\documentclass{article}

\usepackage{arxiv}

\usepackage[utf8]{inputenc} 
\usepackage[T1]{fontenc}    
\usepackage{hyperref}       
\usepackage{url}            
\usepackage{booktabs}       
\usepackage{amsfonts}       
\usepackage{nicefrac}       
\usepackage{microtype}      
\usepackage{lipsum}
\usepackage{graphicx}
\usepackage{float}
\usepackage{lmodern}
\usepackage{enumitem}
\usepackage{amsmath}
\usepackage{nicefrac}
\usepackage{appendix}
\usepackage{dcolumn}
\usepackage{bm}

\title{A Roadmap to Interstellar Flight}

\author{
  Philip Lubin \\
  Department of Physics\\
  University of California - Santa Barbara\\
  Santa Barbara, CA, 93106\\
  \texttt{lubin@ucsb.edu} \\
  \texttt{www.deepspace.ucsb.edu} \\
  \\
  Submitted April 2015, Published February 2016\\
  JBIS Vol 69, Pages 40-72, Feb 2016\\
  \texttt{www.jbis.org.uk/paper/2016.69.40}\\
}

\begin{document}

\maketitle

\begin{abstract}
In the nearly 60 years of spaceflight we have accomplished wonderful feats of exploration
that have shown the incredible spirit of the human drive to explore and understand our universe. Yet in those 60 years we have barely left our solar system with the Voyager 1 spacecraft launched in 1977
finally leaving the solar system after 37 years of flight at a speed of 17 km/s or less than 0.006\% the speed of light. As remarkable as this, to reach even the nearest stars with our current propulsion technology will take 100 millennium. We have to radically rethink our strategy or give up our dreams of reaching the stars, or wait for technology that does not currently exist. While we all dream of human spaceflight to the stars in a way romanticized in books and movies, it is not within our power to do so, nor it is clear that this is the path we should choose. We posit a path forward, that while not simple, it is within our technological reach. We propose a roadmap to a program that will lead to sending relativistic
probes to the nearest stars and will open up a vast array of possibilities of flight both within our solar system and far beyond. Spacecraft from gram level complete spacecraft on a wafer (``WaferSats'') that
reach more than $1/4c$ and reach the nearest star in 20 years to spacecraft with masses more than 105 kg (100 tons) that can reach speeds of greater than 1000 km/s. These systems can be propelled to speeds
currently unimaginable with existing propulsion technologies. To do so requires a fundamental change
in our thinking of both propulsion and in many cases what a spacecraft is. In addition to larger
spacecraft, some capable of transporting humans, we consider functional spacecraft on a wafer, including integrated optical communications, imaging systems, photon thrusters, power and sensors combined with directed energy propulsion. The costs can be amortized over a very large number of
missions beyond relativistic spacecraft as such planetary defense, beamed energy for distant spacecraft,
sending power back to Earth, stand-off composition analysis of solar system targets, long range laser communications, SETI searches and even terra forming. The human factor of exploring the nearest stars and exo-planets would be a profound voyage for humanity, one whose non-scientific implications would be enormous. It is time to begin this inevitable journey far beyond our home. 
\end{abstract}


\newpage
\tableofcontents
\newpage

\section{\label{sec:intro}Introduction}

\begin{figure}
        \centering
        \includegraphics[width=0.6\textwidth]{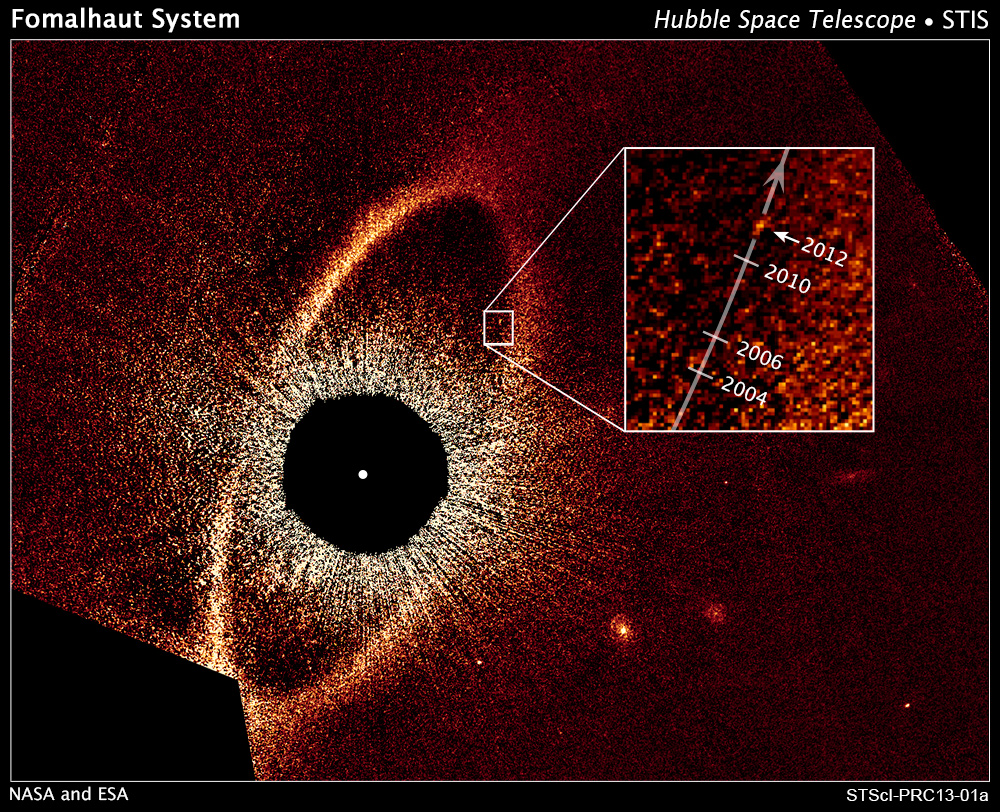}
        \caption{HST image of the star Fomalhaut indicating a possible exoplanet about 25 ly away. This is one possible target that is within reach. Credit: HST Image -- NASA/ESA STScI 13-01a.}
        \label{fig:fomalhaut}
    \end{figure}

\begin{figure}
        \centering
        \includegraphics[width=0.6\textwidth]{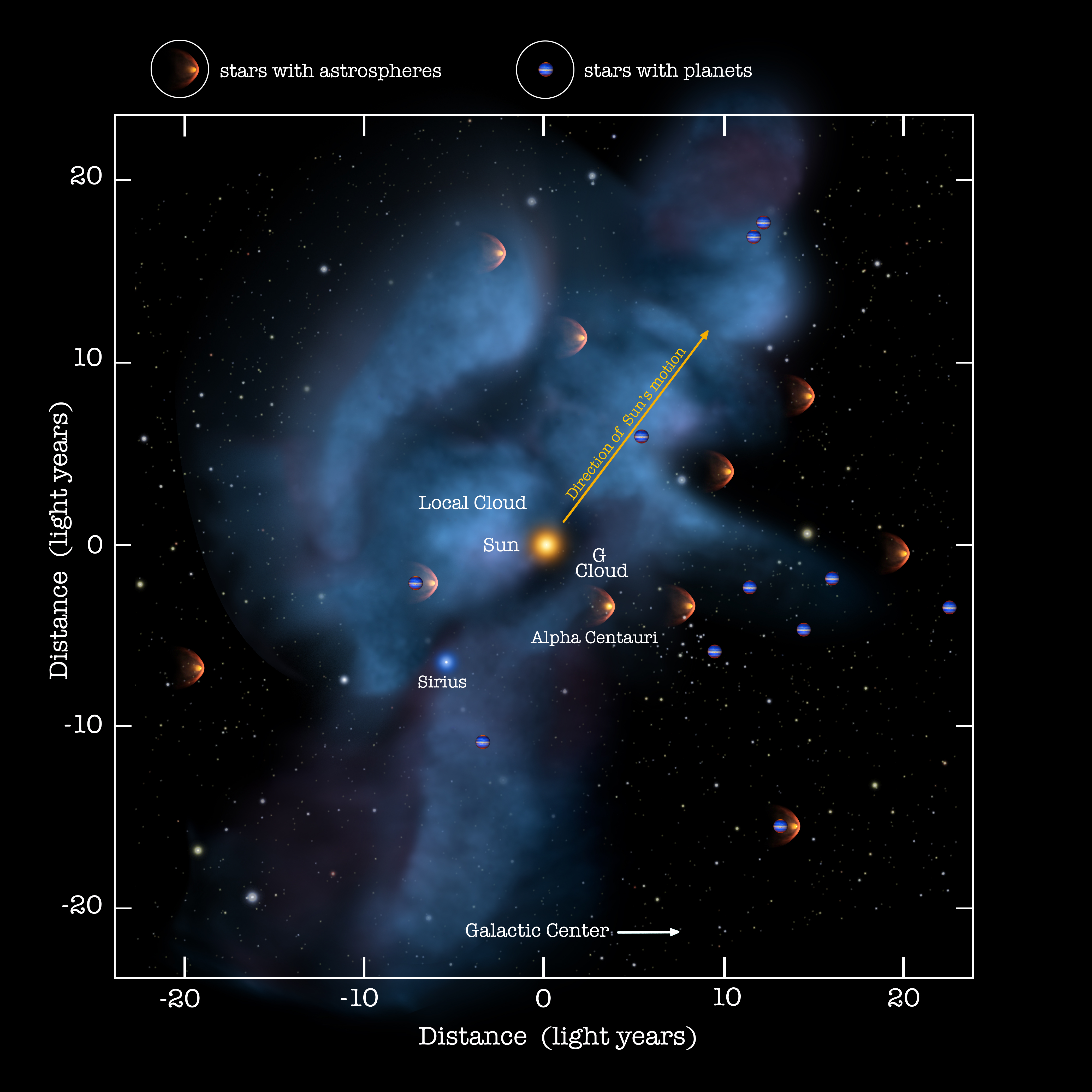}
        \caption{Stars and structures within about 25 ly of the Earth. As indicated some of the stars nearby are already known to contain planets and planetary systems that are potential targets. Credit: NASA Goddard/Adler/U. Chicago/Wesleyan.}
        \label{fig:starsandstructures}
    \end{figure}

    Nearly 50 years ago we set foot on the surface of the moon and in doing so opened up the vision and imaginations of literally billion of people. The number of children ennobled to dream of spaceflight is truly without equal in our history. Many reading this will remember this event or look back at the grainy images with optimism for the future. At the same time we sent robotic probes throughout our solar system and took images of distant galaxies and exoplanet signatures that are forever engrained in our minds. One of humanity’s grand challenges is to explore other planetary systems by remote sensing, sending probes, and eventually life. Within 20 light-years of the Sun, there are over 150 stars and there are known to be a number of planets around at least 12 of these stars and at least 17 stars in 14 star systems appear to be capable of supporting planets in stable orbits within the ``habitable zone.'' This is an incredibly rich environment to explore and shown in Figures \ref{fig:fomalhaut} and \ref{fig:starsandstructures}. Even within the outer reaches of our solar system and into the beginnings of interstellar space lay a profoundly interesting number of objects  we would love to explore if we could. These include the Oort cloud, the heliosheath and heliopause, the solar gravitational lens focus (actually a line) among others.  
    
    Previous ideas for interstellar exploration have included fission, fusion, nuclear weapon driven, fusion using collected interstellar protons, antimatter as well as laser driven \cite{Beals1988}\cite{Bond1978}\cite{Bussard1958}\cite{Forward1984}\cite{Everett1955}. Conceptually many ideas work but the details and practicality are often missing.  Photon propulsion is an old idea going back many years, with some poetic references several hundred years ago. A decade ago what we now propose would have been pure fantasy. \textbf{It is no longer fantasy.} Recent dramatic and poorly-appreciated technological advancements in directed energy have made what we propose possible, though difficult. \textbf{There has been a game change in directed energy technology whose consequences are profound for many applications including photon driven propulsion. This allows for a completely modular and scalable technology with radical consequences.}
    
    We propose a system that will allow us to take the step to interstellar exploration using directed energy propulsion combined with miniature probes including some where we would put an entire spacecraft on a wafer to achieve relativistic flight and allow us to reach nearby stars in a human lifetime. Combined with recent work on wafer scale photonics, we can now envision combining these technologies to allow for a realistic approach of sending probes far outside our solar system and to nearby stars. As a part of our effort we propose a roadmap to allow for staged development that will \textbf{allow us not only to dream but to do.} By leaving the main propulsion system back in Earth orbit (or nearby) and propelling wafer scale highly integrated spacecraft that include cameras, bi-directional optical communications, power and other sensors we can achieve gram scale systems coupled with small laser driven sails to achieve relativistic speeds and traverse the distance to the nearest exoplanets in a human lifetime. \textbf{While this is not the same as sending humans it is a step towards this goal and more importantly allows us to develop the relevant technological base and the ability to build a single ``photon driver" to send out literally millions of low mass probes in a human lifetime.} The key to the system lays in the ability to build both the photon driver and the ultra-low mass probe. Recent developments now make it possible to seriously consider this program and while decades of work will be required, long term consequences will be truly transformative in many areas, not just propulsion \cite{Lubin2013}.
 
\subsection{Electromagnetic Acceleration vs Chemical Acceleration}

    \begin{figure*}
        \centering
        \begin{tabular}{c c}
            \hspace{-17mm}\includegraphics[width=0.425\textwidth]{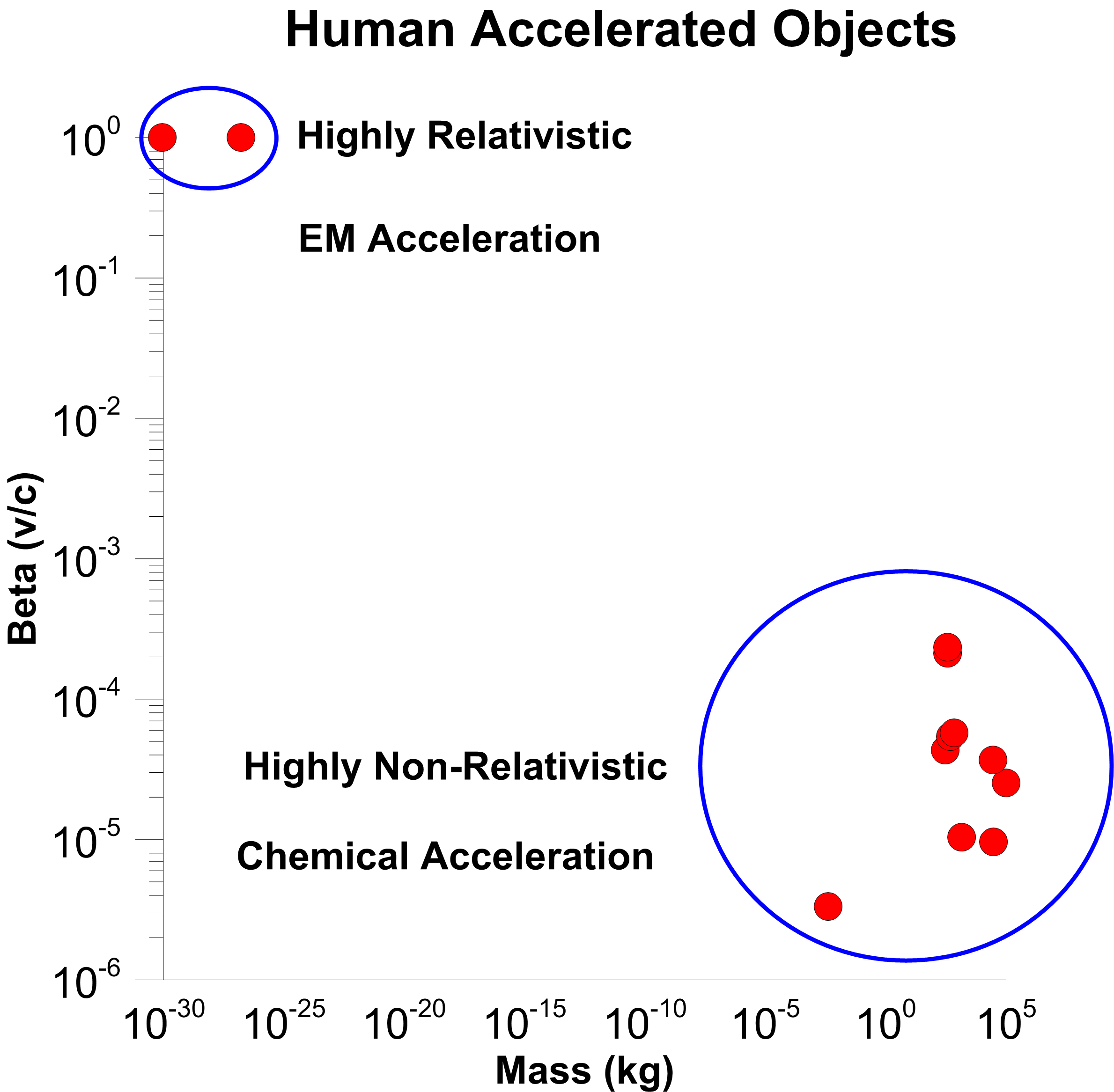} &
            \includegraphics[width=0.45\textwidth]{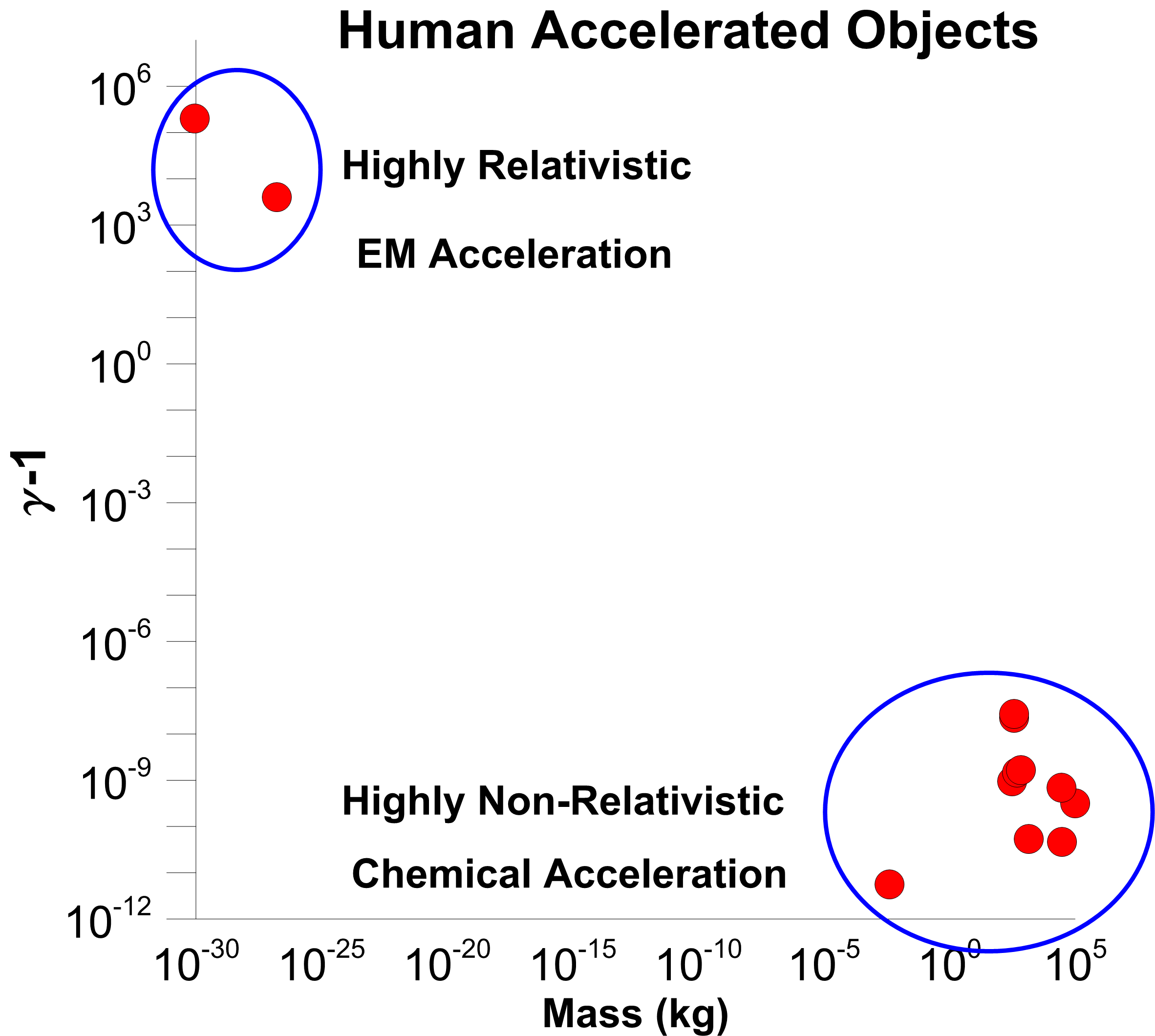} \\
            (a) & (b) \\
        \end{tabular}
        \caption{(a) Fractional speed of light achieved by human accelerated objects vs mass of object from sub-atomic to large macroscopic objects. (b) The same but showing $\gamma-1$ where $\gamma$ is the relativistic ``gamma factor.'' $\gamma-1$ multiplied by the rest mass energy is the kinetic energy of the object.}
        \label{fig:betagamma}
    \end{figure*}

    There is a profound difference in what we have been able to do in accelerating material via chemical means vs electromagnetic means. This difference in speeds achieved is dramatically illustrated if we compare the beta (v/c) and gamma factors achieved (Figure \ref{fig:betagamma}). We clearly have the ability to produce highly relativistic systems but only at the particle level. Practical systems need to be macroscopic as we do not currently have the technological means to self assemble relativistic particle into macroscopic systems. Electromagnetic acceleration is only limited by the speed of light while chemical systems are limited to the energy of chemical processes which are typically of order 1 eV per bond or molecule. To reach relativistic speeds we would need GeV per bond equivalent or about a billion times larger than chemical interactions.  
    
    We propose electromagnetic acceleration to achieve relativistic speeds for macroscopic objects though not using conventional accelerators but using light to directly couple to macroscopic objects. This is simply using a very intense light source to accelerate matter. It has the additional advantage of leaving the propulsion source behind to greatly reduce the spacecraft mass. Of course, this has the disadvantage of reducing or eliminating (depending on the system design) maneuverability once accelerated. For many systems this is not acceptable so hybrid systems are proposed as well as pure photon driven systems. 
    
\begin{figure}
        \centering
        \includegraphics[width=0.5\textwidth]{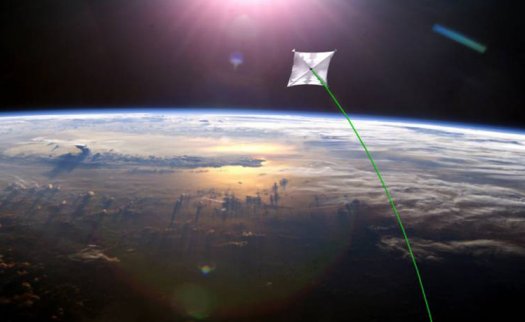}
        \caption{Artistic rendition of a photon driven spacecraft. Pictorial only. Credit: L'Garde.}
        \label{fig:artisticrender}
    \end{figure}

    While the photon drive is not a new concept (solar sails, laser sails, etc. --- Figure \ref{fig:artisticrender}) what is new is that directed energy photonic technology has recently progressed to the point where it is now possible to begin the construction of systems to accelerate macroscopic systems to relativistic speeds. Reaching relativistic speeds with macroscopic systems would be a watershed in our path to the stars. 
    
    \textbf{There has been a game change in directed energy technology whose consequences are profound for many applications including photon driven propulsion. This allows for a completely modular and scalable technology without ``dead ends''} \cite{Hughes2013}\cite{Lubin2014}. This will allow us to take the step to interstellar exploration using directed energy propulsion combined with miniature probes including some where we would put an entire spacecraft on a wafer in some cases to achieve relativistic flight and allow us to reach nearby stars in a human lifetime. 
    
\subsection{Photon Driver --- Laser Phased Array}
    
    The key to the system lays in the ability to build the photon driver. For relativistic flight ($>0.1$ c) development of ultra-low mass probes is also needed. Recent developments now make both of these possible. The photon driver is a laser phased array which eliminates the need to develop one extremely large laser and replaces it with a large number of modest (kW class) laser amplifiers that are inherently phase locked as they are fed by a common seed laser. This approach also eliminates the conventional optics and replaces it with a phased array of small optics that are thin film optical elements. Both of these are a follow on  DARPA and DoD programs and hence there is enormous leverage in this system. The laser array has been described in a series of papers we have published and is called \textbf{DE-STAR} (\textbf{D}irected \textbf{E}nergy \textbf{S}ystem for \textbf{T}argeting of \textbf{A}steroids and \textbf{E}xplo\textbf{R}ation) \cite{Pelton2015}. \textbf{Powered by the solar PV array the same size as the 2D modular array of modest and currently existing kilowatt class Yb fiber-fed lasers and phased-array optics it would be capable of delivering sufficient power to propel a small  scale probe combined with a modest (meter class) laser sail to reach speeds that are relativistic.} DE-STAR units are denoted by numbers referring to the log of the array size in meters (assumed square). Thus DE-STAR-1 is 10 m on a side, -2 is 100 m, etc. Photon recycling (multiple bounces) to increase the thrust is conceivable and has been tested in our lab but it NOT assumed. The modular sub systems (baselined here at 1-4 m in diameter) fit into current launchers such as the upcoming SLS and while deployment of the full system is not our goal in the short term, smaller version could be launched to test proof of concept. \textbf{As an example, on the eventual upper end, a full scale DE-STAR 4 (50-70 GW) will propel a wafer scale spacecraft with a 1 m laser sail to about 26\% the speed of light in about 10 minutes (20 kg$_o$ accel), reach Mars (1 AU) in 30 minutes, pass Voyager I in less than 3 days, pass 1,000 AU in 12 days and reach Alpha Centauri in about 20 years.} The same directed energy driver (DE-STAR 4) can also \textbf{propel a 100 kg payload to about 1\% c and a 10,000 kg payload to more than 1,000 km/s.} While such missions would be truly remarkable, the system is scalable to any level of power and array size where the tradeoff is between the desired mass and speed of the spacecraft. Our roadmap will start with MUCH more modest systems including ground based tests, CubeSat tests, possibly ISS tests and increasingly sophisticated systems. Useful testing can begin at the sub kilowatt level as the system is basically ``self-similar'' with all arrangements being essentially scaled versions of the others. There is no intrinsic barrier to the speed, except the speed of light, and thus unlike other technologies there is no ``dead end.'' On the lower end kilowatt or even 100 W class tests can be conducted to propel very small sub mm payloads to 10 km/s and with larger (but modest) system to $\sim100$ km/s. \textbf{This technology is scalable over an enormous range of mass scales.} The ``laser photon driver'' can propel virtually any mass system with the final speed only dependent on the scale of the driver built. Accelerating small 10 $\mu$m ``grains'' to Mach 100-1000 for hypersonic tests would be of great interest, for example. \textbf{Once built the system can be amortized over a very large range of missions allowing literally hundreds of relativistic wafer scale payload launches per day ($\sim$40,000/yr or one per sq deg on the sky) or 100-10,000 kg payloads for interplanetary missions at much slower rates (few days to weeks).} For reference 40,000 wafers/ and reflectors, enough to send one per square degree on the entire sky, have a total mass of only 80 kg. No other current technology nor any realistically envisioned opens this level of possibility.  While worm holes, antimatter drives and warp drives may someday be reality they are not currently, nor do we have a technological path forward to them. \textbf{Directed energy propulsion allows a path forward to true interstellar probes.} The laser array technology is modular and extensible allowing a logical and well defined roadmap with immediate probe tests to start exploring the interstellar medium on the way to the nearest stars. \textbf{This technology is NOT science fiction. Things have changed.} The deployment is complex and much remains to be done but it is time to begin.  Since the system is modular and scalable the costs to begin are very modest as even small systems are useful. The same system can be used for many other applications as outlined in our papers which amortizes the costs over multiple tasks \cite{Pelton2015}\cite{Hughes2013DE-STAR:Purposes}. 
    
\subsection{Exploring the Interstellar Medium (ISM)}

    On the development path to the nearest stars lay a wealth of information at the edge of and just outside our solar system. It is not ``all or nothing" in going outside of our solar system. We will have many targets, including the solar system plasma and magnetic fields and its interface with the ISM, the heliopause and heliosheath, the Oort cloud outside and Kuiper belt inside, asteroids, KBO's, solar lens focus where the Sun acts as a gravitational lens to magnify distant objects. For example, a more modest  mission at 200 km/s (40 AU/yr) allows many ISM studies. 
    
\subsection{Power Levels and Efficiency}

    The systems can be designed and tested at any power level but it is worth comparing the power levels with conventional flight systems such as the previous Shuttle system. Each Shuttle solid rocket booster (SRB) is 12.5 MN thrust with an exhaust speed of 2.64 km/s and burns for 124 seconds with an ISP of 269 s. The main engine (H$_2$/O$_2$) has 5.25 MN thrust with an exhaust speed of 4.46 km/s and burns for 480 s with an ISP of 455 s. The ``power`` of each SRB is thus about 17 GW. This language is not commonly used but is instructive here. The main engine has a power of about 11 GW and thus at liftoff the Shuttle consumed about 45 GW of chemical power or very similar to the largest systems we have studied (DE-STAR 4) and that are required for interstellar missions. The total energy expended to get the orbiter plus the maximum payload to LEO is about $9.4\times10^{12}$ J = 9.4 TJ. The kinetic energy (KE) of the orbiter + payload at LEO is about 4.0 TJ. The efficiency (KE at LEO/ total launch energy) is about 43\%. Compare this to a 1 g wafer and a 1 g reflector travelling at 26\% c for one of our many missions. This KE is about 3.0 TJ and the integrated efficiency is about 26\% or very similar to the Shuttle at LEO. It is not surprising that the amount of power needed for the laser drive is similar to that required by the Shuttle.  Shuttle  technology cannot get us to the stars but directed energy can.
    
\section{\label{sec:physics}Physics of Directed Energy Propulsion}

\begin{figure}
        \centering
        \includegraphics[width=0.45\textwidth]{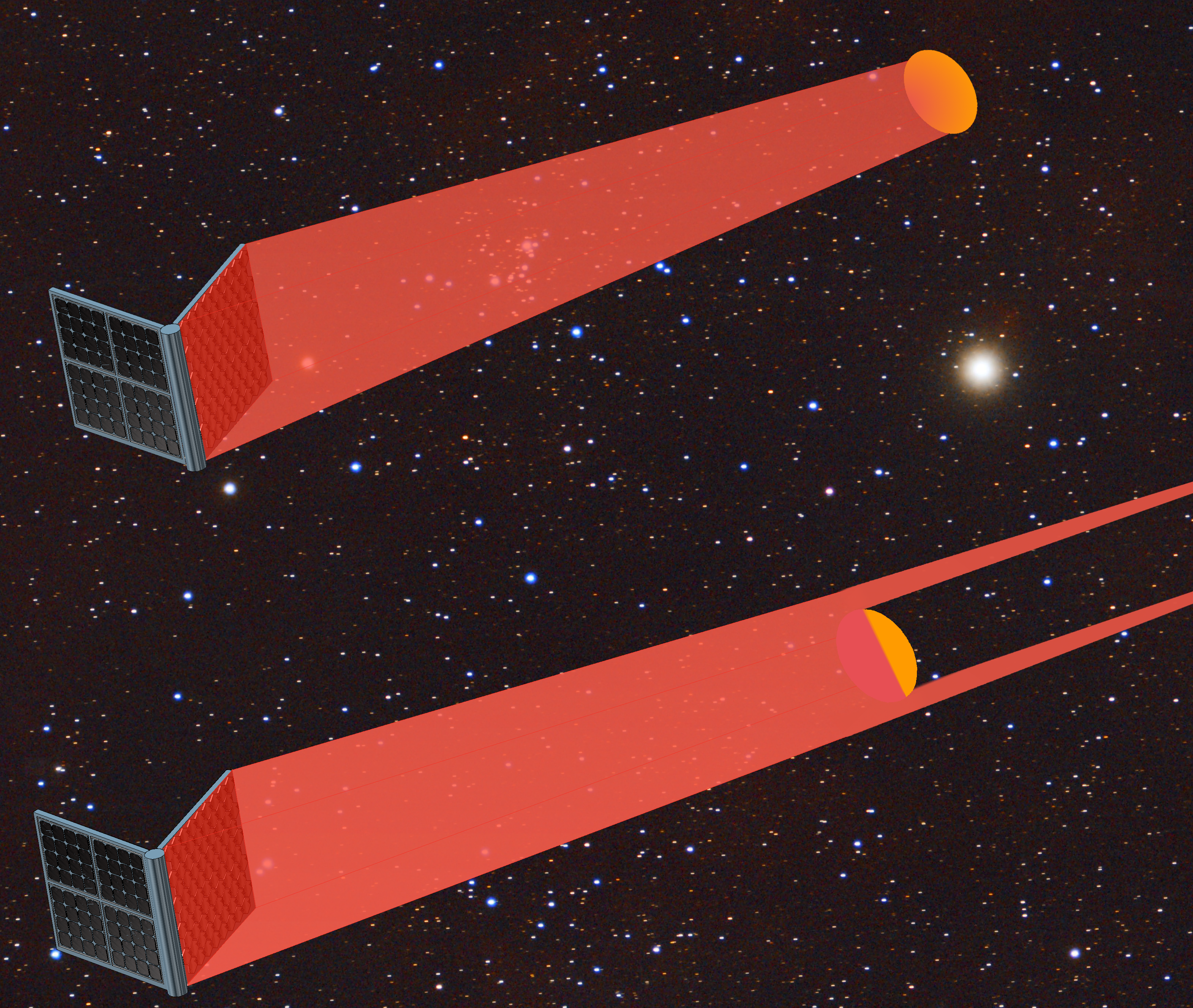}
        \caption{Conceptual drawing showing beam filling the sail and then as distance increases eventually overflowing. With continued illumination speed increases by $\sqrt{2}$.}
        \label{fig:laserpropulsion}
    \end{figure}

    We assume a square diffraction limited phased array of size $d$ illuminating a square payload ``laser'' sail of size $D$. The non-relativistic solution for acceleration, speed and distance to where the laser spot equals the payload sail (Figure \ref{fig:laserpropulsion}) is given below \cite{Bible2013}\cite{Pelton2015}.
    
    We solve for the non-relativistic case here and the relativistic case below. We assume a laser power $P_0$ in the main beam and a total mass (spacecraft + sail) of $m$. The detailed solution is given in the appendix. It is summarized here. We solve for a square or circular array on a square or circular sail in any combination. Assume the DE array has size $d$ and the sail has size $D$. For a circular array of diameter $d_c$ replace the size $d$ with $d=d_c/\alpha$ where $\alpha$ is related to the first minimum of the Bessel function of the first kind $J_1$. In this case $\alpha\sim1.22$. Below $\alpha=1$ for a square array and $\alpha=1.22$ for a circular array. The sail has area $A=\xi D^2$ where $\xi=1$ for a square and $\pi/4$ for a circle. The speed is proportional to $\alpha^{-1/2} \xi^{-1/4}$ for the optimized case (sail mass = bare spacecraft mass). A circular array of the same power and size on a square sail is slower by $\alpha^{-1/2}=(1.22)^{-1/2} = 0.905$ or about 9.5\% slower compared to a square array on a square sail. However for a circular array on a circular sail the speed is only lower by $\alpha^{-1/2} \xi^{-1/4} = (1.22)^{-1/2}(\pi/4)^{-1/4}=0.962$ or 3.8\% slower. Thus there is little difference.
    \begin{equation}
        F=\frac{P_0 (1+\epsilon_r)}{c}
    \end{equation}
    is the laser thrust on payload with laser power $P_0$ and sail reflection $\epsilon_r$ while the laser spot is smaller than the sail. $\epsilon_r=0$ for no reflection (all absorbed) and 1 for complete reflection. For our cases, $\epsilon_r \sim 1$. The resulting acceleration is given by
    \begin{equation}
        a=\frac{F}{m}=\frac{P_0 (1+\epsilon_r)}{mc},
    \end{equation}
    where $m=m_\textrm{sail}+m_0$ is the combined mass of the sail and base payload, $m_0$. The sail mass is given by
    \begin{equation}
        m_\textrm{sail}=\xi D^2 h\rho,
    \end{equation}
    where $D$ is the sail size, $h$ is the sail thickness, and $\rho$ is its density. The sail size $D$ can be written as a function of the total mass $m$ as follows:
    \begin{align}
    \begin{split}
        D(m)=\sqrt{m_\textrm{sail}/\xi h\rho}\sim 31.6\sqrt{\mathop{m_\textrm{sail}(\textrm{kg})}/\xi \mathop{h(\mu)}\mathop{\rho(\textrm{g/cc)}}}.
    \end{split}
    \end{align}
    
    The speed at the point where the laser spot is equal to the sail size is given by
    \begin{equation}
        v_0=\bigg(\frac{P_0 (1+\epsilon_r)dD}{c\lambda\alpha(\xi D^2 h\rho + m_0)}\bigg)^{1/2}.
    \end{equation}
    This occurs at a distance
    \begin{equation}
        L_0=\frac{dD}{2\lambda\xi}.
    \end{equation}
    With continued illumination, the speed increases by $\sqrt{2}$:
    \begin{equation}
        v_\infty=\bigg(\frac{2P_0 (1+\epsilon_r)dD}{c\lambda\alpha(\xi D^2 h\rho + m_0)}\bigg)^{1/2}.
    \end{equation}
    
    We can show that the maximum speed occurs when the sail mass equals the payload mass. In this case,
    \begin{equation}
        D=(m_\textrm{sail}/\xi h\rho)^{1/2}=(m_0/\xi h\rho)^{1/2}
    \end{equation}
    and the speed with continued illumination is
    \begin{align}
       \begin{split}
       v_{\textrm{max}-\infty}=&\bigg(\frac{P_0 (1+\epsilon_r)dD}{c\lambda\alpha m_0}\bigg)^{1/2}\\
       =&\bigg(\frac{P_0 (1+\epsilon_r)d}{c\lambda\alpha D\xi h\rho}\bigg)^{1/2}\\
       =&c\bigg(\frac{P_0 (1+\epsilon_r)}{P_1}\bigg)^{1/2}\bigg(\frac{d}{\alpha\xi D}\bigg)^{1/2}\\ =&\bigg(\frac{P_0 (1+\epsilon_r)d}{c\lambda\alpha}\bigg)^{1/2}(\xi h\rho m_0)^{-1/4},
       \end{split}
    \end{align}
    where 
    \begin{align}
    \begin{split}
        P_1\equiv & c^3 \lambda \xi h\rho\\ \approx & 2.7\times10^{16}(\textrm{W})\cdot\mathop{\lambda(\mu \textrm{m})} \mathop{h(\mu \textrm{m})} \mathop{\rho(\textrm{g/cc})}\\
        = & 27 (\textrm{PW}) \mathop{\lambda(\mu\textrm{m})} \mathop{h(\mu\textrm{m})} \mathop{\rho(\textrm{g/cc})},
    \end{split}
    \end{align}
    or
    \begin{align}
    \begin{split}
        \beta_{\textrm{max}-\infty}=&\bigg(\frac{P_0 (1+\epsilon_r)}{P_1}\bigg)^{1/2}\bigg(\frac{d}{\alpha\xi D}\bigg)^{1/2} \\
        =&\bigg(\frac{2P_0}{P_1}\bigg)^{1/2}\bigg(\frac{d}{\alpha\xi D}\bigg)^{1/2}
    \end{split}
    \end{align}
    for $\epsilon_r=1$ and $m_\textrm{sail}=m_0$.
    
    The time to where the laser spot equals the sail size is (time to $L_0$)
    \begin{equation}
        t_0=\frac{v_0}{a}=\bigg(\frac{cdD(\xi D^2 h\rho+m_0)}{P_0(1+\epsilon_r)\lambda\alpha}\bigg)^{1/2}
    \end{equation}
    and
    \begin{equation}
        L_0=\frac{dD}{2\lambda\alpha}=\frac{d}{2\lambda\alpha}\bigg(\frac{m_\textrm{sail}}{\xi h\rho}\bigg)^{1/2}.
    \end{equation}
    The time to where the laser spot equals the sail size for the case where the sail mass equals the payload mass is
    \begin{align}
    \begin{split}
        t_0=&\bigg(\frac{2cdD^3 \xi h\rho}{P_0(1+\epsilon_r)\lambda\alpha}\bigg)^{1/2}\\
        =&\bigg(\frac{2cd}{P_0(1+\epsilon_r)\lambda\alpha}\bigg)^{1/2}\bigg(\frac{m_0^3}{\xi h\rho}\bigg)^{1/4}\\
        =&m_0 c^2\bigg(\frac{2}{P_0(1+\epsilon_r)P_1}\bigg)^{1/2}\bigg(\frac{d}{\alpha\xi D}\bigg)^{1/2}\\
        \sim&\frac{m_0 c^2}{\sqrt{P_0 P_1}}\bigg(\frac{d}{\alpha\xi D}\bigg)^{1/2}\\
        \sim&1.73\times10^4(\textrm{s}) m_0\bigg(\frac{d}{\alpha\xi D}\bigg)^{1/2}P_0^{-1/2}\\
        \sim&3.08\times10^3(\textrm{s})m_0\bigg(\frac{d\sqrt{\xi h\rho}}{\lambda\alpha\sqrt{m_0}}\bigg)^{1/2} P_0^{-1/2},
    \end{split}    
    \end{align}
    where $[P_0]=\textrm{GW}$ and $[m_0]=\textrm{kg}$.
    The kinetic energy at time $t_0$ is
    \begin{equation}
        \textrm{KE}=\frac{1}{2}mv_0^2=\bigg(\frac{P_0(1+\epsilon_r)dD}{2c\lambda\alpha}\bigg)=FL_0.
    \end{equation}
    Note that this is independent of optimization and overall mass, but does depend on sail size $D$.
    
    While counter-intuitive in the context of solar sails, the highest speed is achieved with the smallest sail and thus smallest payload mass (Figure \ref{fig:distanceandillumination}). The laser has very narrow bandwidth so we can design the reflector with multi-layer dielectric coatings to have $\epsilon_r$ very close to unity. 
    
\subsection{Beam Efficiency}

    The laser power $P_0$ is defined above as the laser power in the main beam that is on the reflector. We define the beam efficiency $\epsilon_\textrm{beam}$ as the fraction of the total generated optical (laser) power $P_\textrm{optical}$ in the main beam as $\epsilon_\textrm{beam} = P_0/P_\textrm{optical}$ or $P_0 = \epsilon_\textrm{beam} P_\textrm{optical}$. The beam efficiency includes all effects that reduce the total optical power generated including losses in the optical chain (fiber losses, lens or telescope losses) as well as diffraction into side lobes, etc. For example, in a phased array the maximum packing fraction for close packed circular apertures is $\pi/4\sim 0.785$. Typical array beam efficiencies are 0.5-0.7. The actual beam distribution will be more complex than a simple top hat, cosine, Bessel or Gaussian distribution, though a Gaussian will generally be a reasonable approximation.
    
\subsection{Energy Required per Launch}

    The energy required per launch is helpful in planning a system design as there may be a need to store the energy needed rather than have a continuous mode. This would have the effect of lowering the capacity of the electrical power system and allow a ``trickle charge mode.'' We define $E_\gamma$ as the photon energy in the main beam (on the sail) to get to the point where the spot size equals the reflector size. Thus, $E_\gamma = P_0 t_0$. In general we will increase the illumination time by a factor of a few times greater than $t_0$ in order to get most of the factor of $2^{1/2}$ increase in speed that comes from continued illumination. There is little  need for additional continued illumination since the speed increase is of diminishing return beyond a few times $t_0$.  
    
    The electrical energy $E_\textrm{elec}$ use over time $t_0$ is $E_\textrm{elec}= E_\gamma/(\epsilon_\textrm{elec}\epsilon_\textrm{beam})=P_0 t_0/(\epsilon_\textrm{elec}\epsilon_\textrm{beam})$ where $\epsilon_\textrm{elec}$ is the total electrical to overall photon conversion efficiency ($\epsilon_\textrm{elec}=P_\textrm{optical}/P_\textrm{electrical}$) and includes all efficiencies such as power supply, laser amplifier, etc. This is a system level efficiency. As an example the current ``wall plug'' efficiency of the Yb laser amplifiers is about 0.42. Note that $P_0=(\epsilon_\textrm{elec}\epsilon_\textrm{beam})P_\textrm{electrical}$.
    
    The total photon energy in the main beam (on the sail) to time $t_0$ is
    \begin{align}
    \begin{split}
        E_\gamma=P_0 t_0=&P_0\bigg(\frac{cdD(\xi D^2 h\rho+m_0)}{P_0(1+\epsilon_r)\lambda\alpha}\bigg)^{1/2}\\
        =&\bigg(\frac{cdD(\xi D^2 h\rho+m_0)P_0}{(1+\epsilon_r)\lambda\alpha}\bigg)^{1/2}\\
        =&m_0 c^2\bigg(\frac{2P_0}{(1+\epsilon_r)P_1}\bigg)^{1/2}\bigg(\frac{d}{\alpha\xi D}\bigg)^{1/2}\\
        \sim&m_0 c^2 \bigg(\frac{P_0}{P_1}\bigg)^{1/2}\bigg(\frac{d}{\alpha\xi D}\bigg)^{1/2}
    \end{split}
    \end{align}
    for the optimized case (i.e. $m_\textrm{sail}=m_0$ and $\epsilon_r=1$ in the last term).
    
    We define the (main beam) launch energy efficiency to time $t_0$ as
    \begin{align}
    \begin{split}
        \epsilon&_\textrm{\tiny{launch}}=\frac{\textrm{KE}(t_0)}{E_\gamma}\\
        =&\bigg(\frac{P_0(1+\epsilon_r)dD}{2c\lambda\alpha}\bigg)\bigg(\frac{cdD(\xi D^2 h\rho+m_0)P_0}{(1+\epsilon_r)\lambda\alpha}\bigg)^{1/2}\\
        =&\frac{(1+\epsilon_r)}{2c}\bigg(\frac{P_0(1+\epsilon_r)dD}{c\lambda\alpha(\xi D^2 h\rho+m_0)}\bigg)^{1/2}\\
        =&\frac{(1+\epsilon_r)}{2c}v_0=\frac{(1+\epsilon_r)}{2}\beta_0
    \end{split}
    \end{align}
    Note the launch efficiency does NOT depend on optimization (sail mass = payload mass) and for $\epsilon_r=1$ then $\epsilon_\textrm{launch}=\beta_0$.
    
    The total electrical energy used to time $t_0$ is
    \begin{equation}
        E_\textrm{elec}=E_\gamma/(\epsilon_\textrm{elec}\epsilon_\textrm{beam}).
    \end{equation}
    
    We can see the relatively simple scaling for photon energy used per launch (to time $t_0$) in terms of the rest mass energy of the spacecraft, the power $P_0$ and the array size $d$ and reflector size $D$. The reason the photon energy scales with the array size as $d^{1/2}$ is due to the fact that a larger array will have a smaller spot and thus a longer illumination time. The reason it scales as $d^{1/2}$ is due to the fact that the distance $L_0$ to where the spot size equals the reflector size is $L_0=dD/2\lambda$ and thus $L_0$ is proportional to the array size $d$. Since the acceleration is constant while the beam is contained within the reflector $L<L_0$ we have $L_0 =\frac{1}{2}a t_0^2$ and thus $t_0 = (2 L_0/a)^{1/2}$ and hence $t_0$ scales as $L_0^{1/2}$ or as $d^{1/2}$.
    
\subsection{Efficiency}

    The instantaneous energy efficiency (power that goes into direct kinetic energy/laser power on reflector) is 
    \begin{align}
    \begin{split}
        \epsilon_p=&\frac{dKE/dt}{P_0}=\frac{mva}{P_0}=\frac{mvP_0(1+\epsilon_r)}{cmP_0}\\
        =&\beta(1+\epsilon_r)=\frac{ma^2 t}{P_0}=\frac{P_0 t (1+\epsilon_r)^2}{mc^2}\\
        \sim&2\beta \sim 4P_0 t/mc^2
    \end{split}
    \end{align}
    for $\epsilon_r \sim1$. The total integrated energy efficiency ($\epsilon_p\sim t$) is   
    \begin{align}
    \begin{split}
        \epsilon_\textrm{total}=&\epsilon_p/2=\beta(1+\epsilon_r)/2=\frac{P_0 t(1+\epsilon_r)^2}{2mc^2}\\
        \sim&\beta\sim\frac{2P_0 t}{mc^2}
    \end{split}
    \end{align}
    for $\epsilon_r\sim1$  where  $m=m_\textrm{sail} + m_0$. See launch efficiency above. Momentum ``efficiency'' $=(1+ \epsilon_r)\sim2$ for $\epsilon_r\sim1$   with $\beta<<1$. The energy transfer efficiency starts out at very low levels and then increases proportional to the speed. The total integrated energy efficiency is just 1/2 that of the instantaneous efficiency at the final speed since the force is constant as long as the laser spot is smaller than or equal to the reflector size and hence the acceleration is constant and hence speed increases proportional to time ($\beta\sim t$) and thus the average $\epsilon_p$ is 1/2 the maximum $\beta$ achieved. This is for the non-relativistic case. For spacecraft accelerated to high speeds the energy efficiency can become quite high.

\subsection{Photon Recycling for Larger Thrust and Efficiency}

    The efficiency of the photon drive can be improved by reusing the photons reflected by the spacecraft reflector in an effective optical cavity mode to get multiple photon reflections. This is known as photon recycling. It is not a new concept and dates back several decades. We will see it greatly complicates the system optics however and is primarily useful at low speeds and short ranges.
    
    In the case of photon recycling the photons bounce back and forth in an optical cavity, one end of which is the spacecraft reflector and the other end is a relatively more massive (referred to here as fixed) mirror \cite{Bae2012}. The total power at the spacecraft mirror sets the force on the spacecraft. The total power on spacecraft mirror is essentially the same as that on the fixed mirror. The combination of the two mirrors forms an optical cavity whose $Q$ factor is defined as $Q = 2\pi E_\textrm{cav}/E_\textrm{loss}\sim \nu/\delta\nu$ where $E\textrm{cav}$ is the energy stored in the cavity, $E_\textrm{loss}$ is the energy lost per cycle, $\nu$ is the optical frequency, and $\delta\nu$ is the FWHM bandwidth of the resonance. One cycle is the round trip travel time of the light or $2L/c$ where $L$ is the distance between the spacecraft and fixed mirror. In general the fixed mirror will be at the laser driver (i.e. near the earth). The energy lost per cycle is due to a variety of effects such as increase of kinetic energy of the spacecraft and fixed mirror per cycle, energy lost to mirror(s) absorption per cycle due to non unity reflection coefficient, diffraction effects as the spacecraft moves away, and mirror misalignments. For the spacecraft close to the laser, optical cavities are possible and do improve efficiency (the effective power on the spacecraft reflector increases by the number of ``bounces''). As the spacecraft begins to move far away diffraction becomes extremely problematic as do mirror alignment issues, surface scattering, side lobes and hence photon recycling has much less practical use at large ranges. In addition there are causality effects and a relativistic effect that is simply the Doppler shifting of the photons off the moving mirrors. These effects complicate the discussion of efficiency \cite{MARX1966}\cite{REDDING1967}\cite{Simmons1993}. Doppler shifts reduce the photon energy and hence momentum on each bounce and ultimately make photon recycling extremely limited at relativistic speeds  even if all other effects are ignored. In general, if we ignore causality and the distance between the laser and spacecraft is small and the spacecraft speed relative to the laser is low, we can replace the power $P_0$ in all the equations by $P_{0r} = N_r (L, v) P_0$ where $N_r (L,v)$ incorporates all the multi-bounce photon recycling effects and is a function of the separation distance $L$ and the speed $v$. The ratio of chemical thrust per unit power (typically 1-few mN/W) to photon (reflection) thrust per unit power (6.6 nN/W) is $\sim10^5$. IF we could develop photon recycling with extremely high efficiency (high finesse) with approx $10^5$ (reflections) we could compete with chemical launches. This would dramatic change launch capability and costs. Since we do not carry the large extra mass of the chemical launch vehicle we do not need as large a number of bounces since there is no ``staging.'' We discuss this further \cite{Kulkarni2018} where we also discuss the relativistic effects that limit the effectiveness of photon recycling as well as parameterize the loss effects.
    
    \begin{figure*}
        \centering
        \begin{tabular}{c c}
            \includegraphics[width=0.45\textwidth]{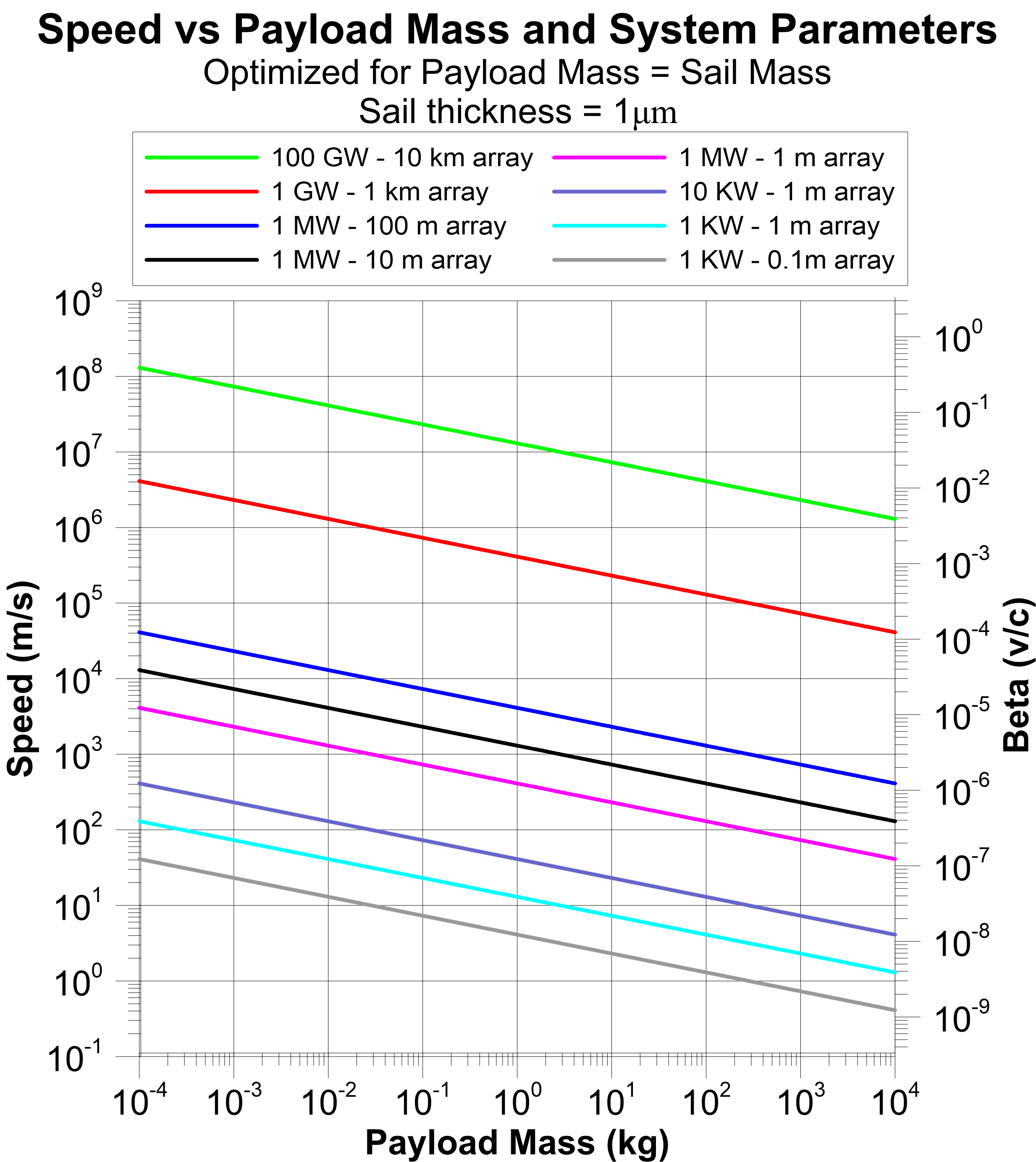} &
            \includegraphics[width=0.45\textwidth]{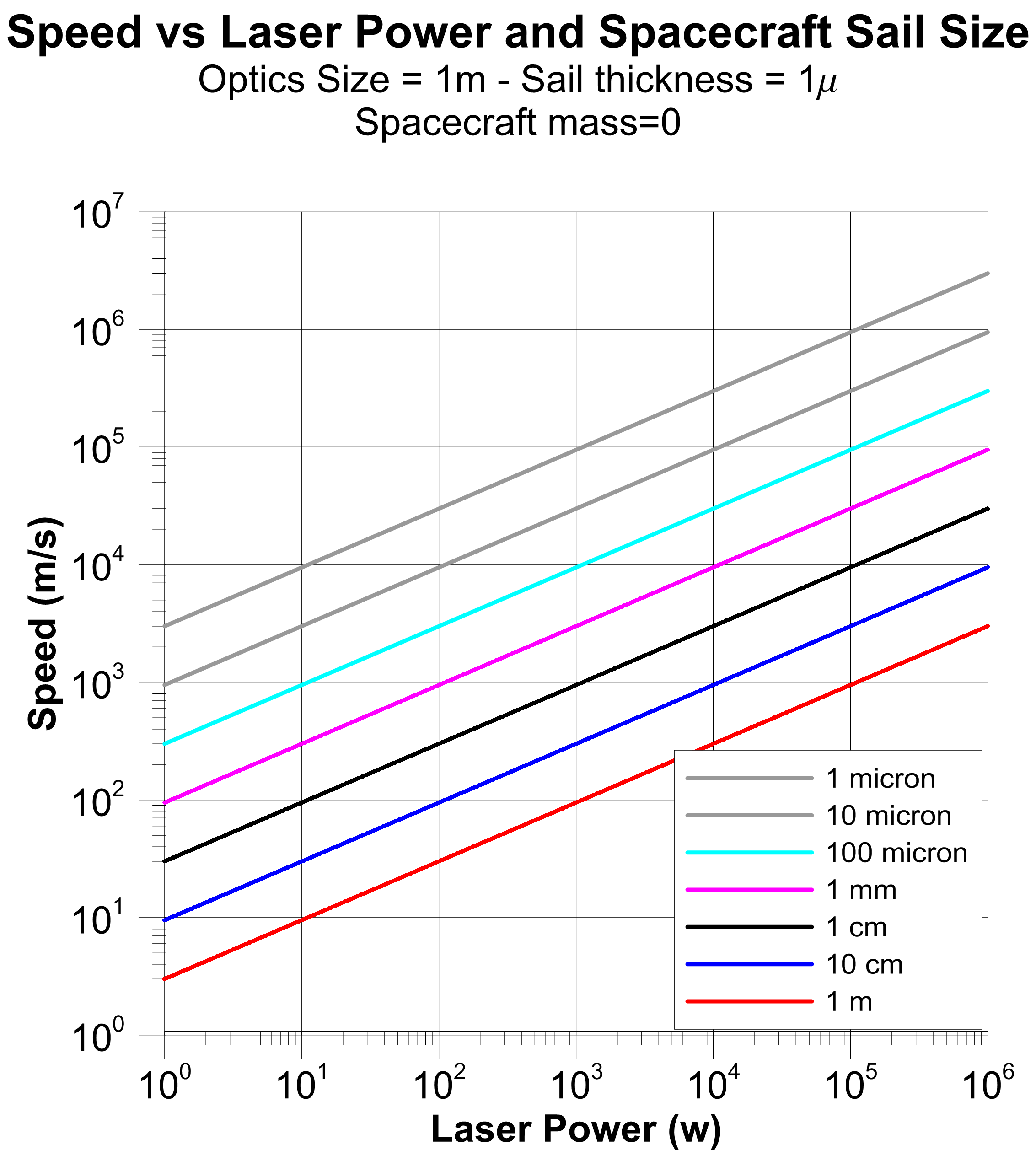} \\
            (a) & (b) \\
        \end{tabular}
        \caption{(a) Speed and beta vs payload mass vs laser array power and size for systems from very small to very large. This range of systems represents a portion of the roadmap. (b) Speed vs laser power for small systems with 1 m optical aperture vs sail size.}
        \label{fig:speed}
    \end{figure*}
    \begin{figure*}
        \centering
        \begin{tabular}{c c}
            \includegraphics[width=0.45\textwidth]{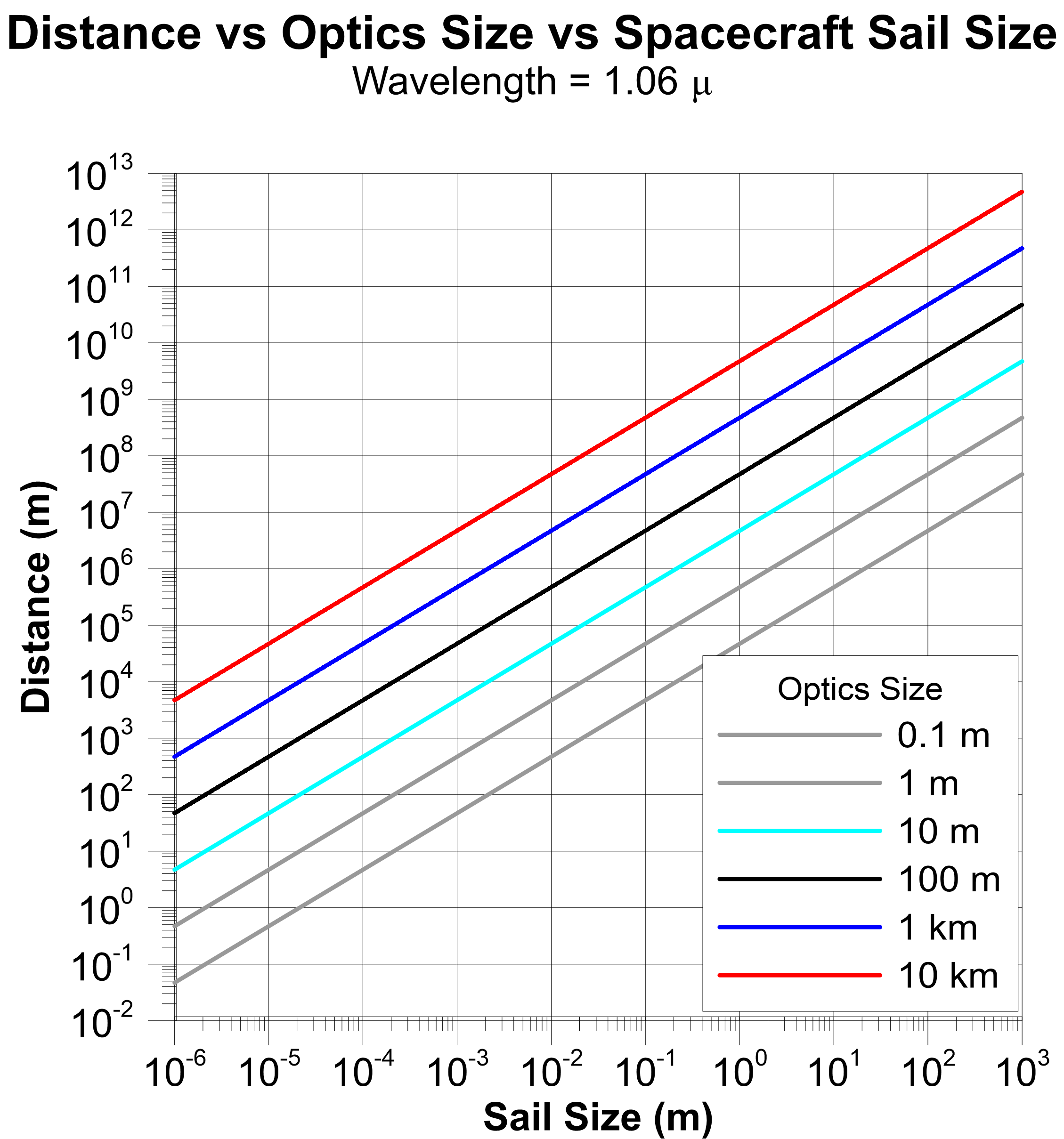} &
            \includegraphics[width=0.475\textwidth]{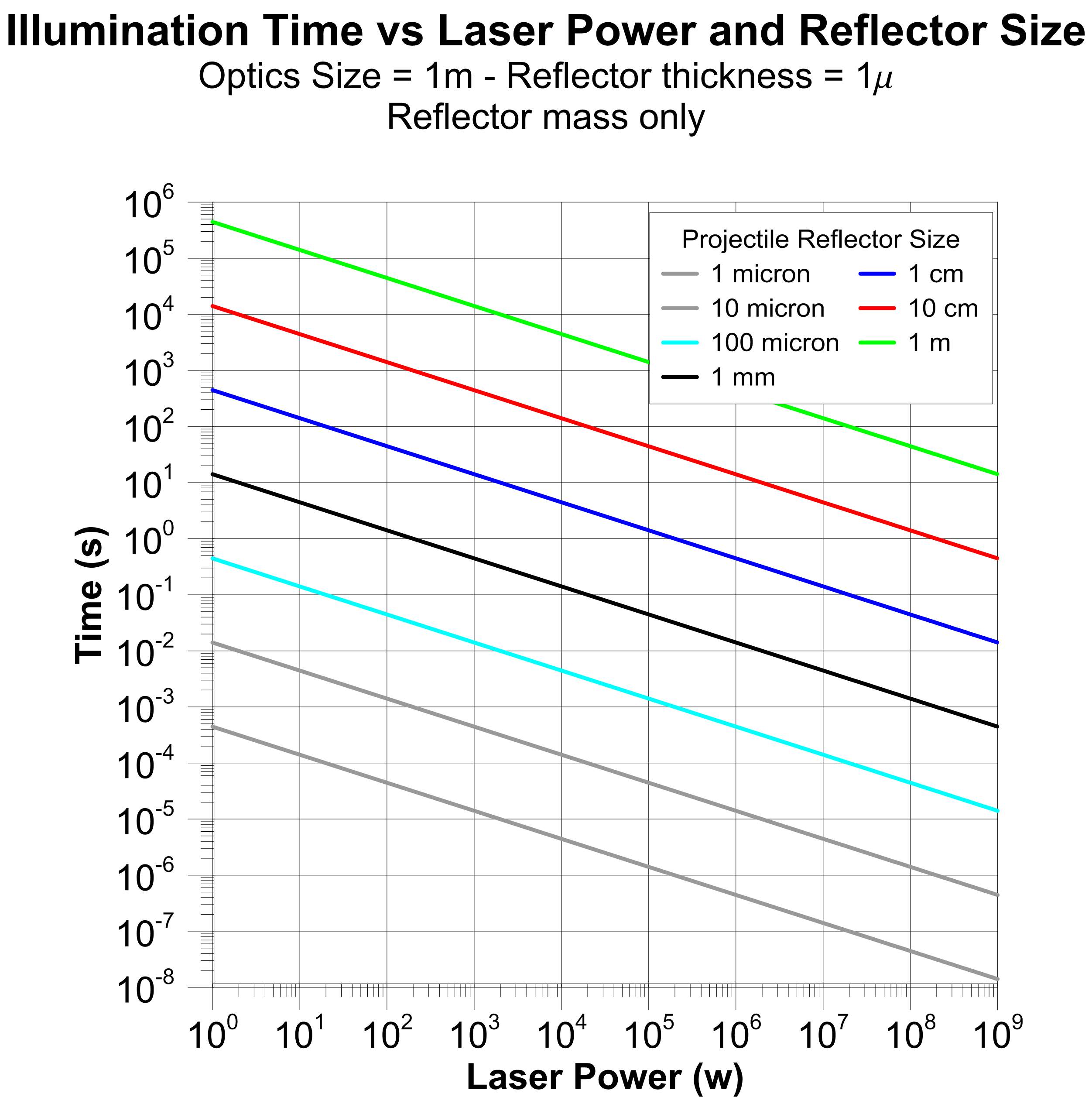} \\
            (a) & (b) \\
        \end{tabular}
        \caption{(a) Distance ($L_0$) within which the laser spot size is within the reflector size vs reflector size and system optical array size. (b) Acceleration time ($t_0$) while the laser spot is within the reflector size vs power and reflector size.}
        \label{fig:distanceandillumination}
    \end{figure*}
    
\subsection{Photon-Assisted Launch}

    Reducing the cost of launch from ground to LEO and beyond is one of the many long term goals of the overall roadmap. Currently all ground to space launches use chemical propulsion. A possible application of the same technology we will develop for the DE driver for relativistic flight is to use a similar system on the ground for laser driven or assisted ground launch but using an ablation drive, though conceivably high $Q$ photon recycling could be used.  In our work on planetary defense \cite{Lubin2013} we show the transition from photon only thrust to ablation thrust, as the flux on ablative targets increases, is typically about 1 MW/m$^2$. In \cite{Lubin2013} we show that for many materials the ratio of the thrust from the ablation case compared to pure photon pressure (assuming complete absorption) is roughly $10^5$ for fluxes well above 1 MW/m$^2$ \cite{Pelton2015}. Comparing ablation thrust per watt of
    \begin{align}
    \begin{split}
        v_\textrm{rel}\frac{dm}{dt}\left(\frac{d}{dt}\frac{1}{2}mv_\textrm{rel}^2\right)=\frac{2}{v_\textrm{rel}}
    \end{split}
    \end{align}
    to photon thrust per watt (for reflection) $= 2P/c/P = 2/c$, we immediately see the thrust ratio is $2/v_\textrm{rel} / 2/c = c/ v_\textrm{rel}$.  Here $v_\textrm{rel}$  is the effective exhaust speed. We note the thrust per watt for ablation assumes a perfectly collimated (zero divergence angle) exhaust plume. In practice the exhaust plume has a significant divergence angle which reduces the net thrust. For many cases in CW ablation the exhaust speed is approximately that of a thermal source. For common high temperature materials that would commonly be used in laser driven ablation engines the materials come to about 3000K, which is similar to the effective plume temperature in a solid rocket engine, and yield about 1-2 km/s for most relevant materials. The Specific Impulse (ISP) is defined in such a way that it is just ISP $= v_\textrm{rel}/g$ where $g$ is the average gravitational acceleration at the surface of the Earth ($\sim 9.8$ m/s$^2$). For $v_\textrm{rel} = 2$ km/s this gives an ISP of about 200, close to the ISP of the SRB on the previous Shuttle system. For photon engines (emission, no reflection) ISP $=3\times10^7$ while for the reflection case it is twice that, or $6\times10^7$. In the idealized case of a laser driven ablation engine we get the thrust ratio, compared to the photon reflection case, of $c/v_\textrm{rel}\sim 1.5-3\times10^5$, consistent with our detailed ablation simulations. It is also possible to use laser heated H$_2$ via heat exchangers to get even higher ISP, due largely to the lower molecular mass, and thus higher exhaust speed for a given temperature and in theory one could approach H$_2$O$_2$ engines which have ISP $\sim450$. For reference ion engines have much larger exhaust speeds (typ $\sim30-60$ km/s) and have an ISP $\sim3000-6000$.
    
    While our work on planetary defense is focused on  asteroids and comets, as well as space debris, the same principal applies to all ablative materials. As no oxidizer is needed this approach offers a number of possible advantages for launch, as well as a number of challenges. There are two regimes of interest. One is the thermal regime and the other is an ionized regime or broadly characterized as the non-thermal regime. The thermal regime is very similar to conventional chemical propellants with a similar thrust per watt of about 1 mN/W with exhaust temperatures being materials limited while the non-thermal regime can have a variety of modes including plasma, ``wake field acceleration'' which is being used in some advanced particle accelerator development and even pulsed relativistic particle production where short high power laser pulses when interacting with thin foils eject relativistic particles. Since the thrust per watt scales roughly as the inverse of the exhaust speed or as $1/v_\textrm{rel}$, the higher speed non-thermal systems have much lower thrust per watt. The advantage of the higher exhaust speeds in the non-thermal case is the higher thrust per unit mass flow which scales as $v_\textrm{rel}$ and thus less launch mass is needed. For ground to orbit the latter is critical. Similarly, laser driven launch using high photon number recycling could be very useful IF we can get very efficient photon recycling with the number of bounces approaching $10^5$. The range is short so this is helpful for photon recycling but there are atmospheric perturbations to contend with and practical orbit insertion generally requires multiple DE driver as the spacecraft moves ``down range.'' In both cases the same basic technology we are developing for the DE-STAR driver could also be used on the ground as the main ``power'' source. This could lead to a major reduction in launch costs leading to a paradigm shift in orbital access. The upcoming SLS uses chemically generated power levels nearly identical to the DE power needed to drive a small probe to relativistic speeds. Even the time scales are similar with a Shuttle or SLS launch requiring about 10 minutes of chemical power to get to LEO and a small (WaferSat) relativistic probe also requiring about 10 minutes of nearly identical power to get to relativistic speeds. Thus a ``boot strap'' approach where a ground DE driver with an ablation booster (or high $Q$ photon recycling) is used for ground launches to enable the deployment of the DE orbital driver using purely photon thrust is one option. This would have significant ramifications for many other programs. This is yet another example of amortization of the technological base for this program \cite{Hughes2013}\cite{Lubin2014}\cite{Pelton2015}\cite{Pelton2015}.
    
\subsection{Relativistic Solution}

    There are a number of issues that modify the classical solution. These are Doppler shift of the laser driver photons on the moving reflector, increase in relativistic mass of the spacecraft and reflector and time dilation. In addition, causality issues related to the time of flight of the photons from the source to the reflector need to be included. The non-relativistic solutions used above allow for reasonable accuracy at modest speeds but a full solution is required at speeds that are a signification fraction of the speed of light. We derive the first order relativistic solutions in our papers \cite{Bible2013}\cite{Pelton2015}. It is give by $t$ vs $\beta$ ($v/c$) and $\gamma=(1-\beta)^{-1/2}$ as follows:
    \begin{align}
    \begin{split}
        t=&\frac{m_0c^2}{2P_0(1+\epsilon_r)}\Bigg[\frac{\beta}{1-\beta^2}+\frac{1}{2}\ln{\left|\frac{1+\beta}{1-\beta}\right|}\Bigg]\\
        =&\frac{m_0c^2}{2P_0(1+\epsilon_r)}\big[\gamma^2\beta + \tanh^{-1}{\beta}\big],
    \end{split}
    \end{align}
    where we define $t_E\equiv m_0c^2/P_0$, and therefore
    \begin{equation}
        t=\frac{t_E}{2(1+\epsilon_r)}\big[\gamma^2\beta + \tanh^{-1}{\beta}\big].
    \end{equation}
    A more complete relativistic solution is given in \cite{Kulkarni2018} which includes efficiency and photon recycling issues.
    
\subsection{Relativistic Effects}

\begin{figure*}
        \centering
        \begin{tabular}{c c}
            \includegraphics[width=0.45\textwidth]{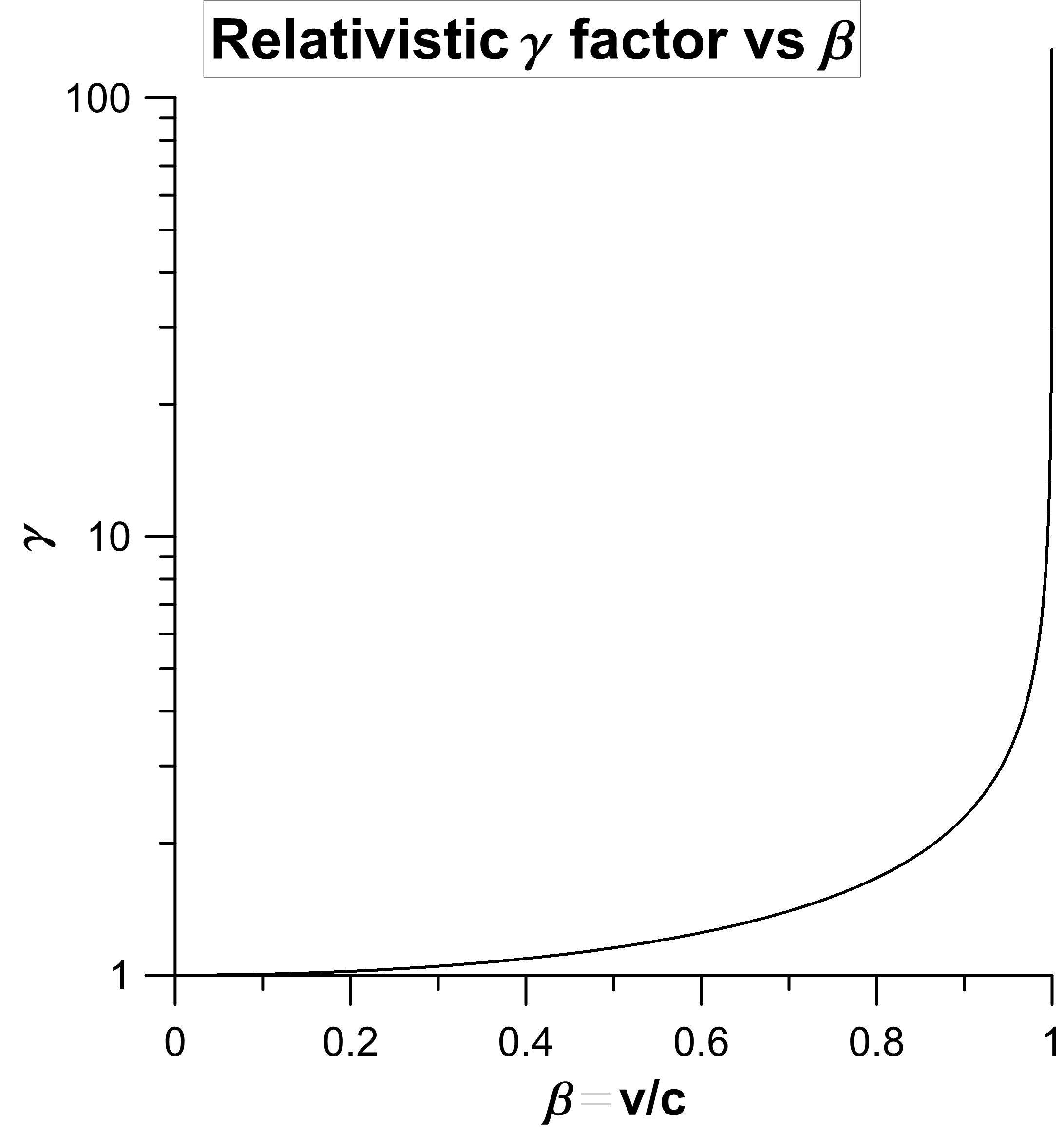} &
            \includegraphics[width=0.46\textwidth]{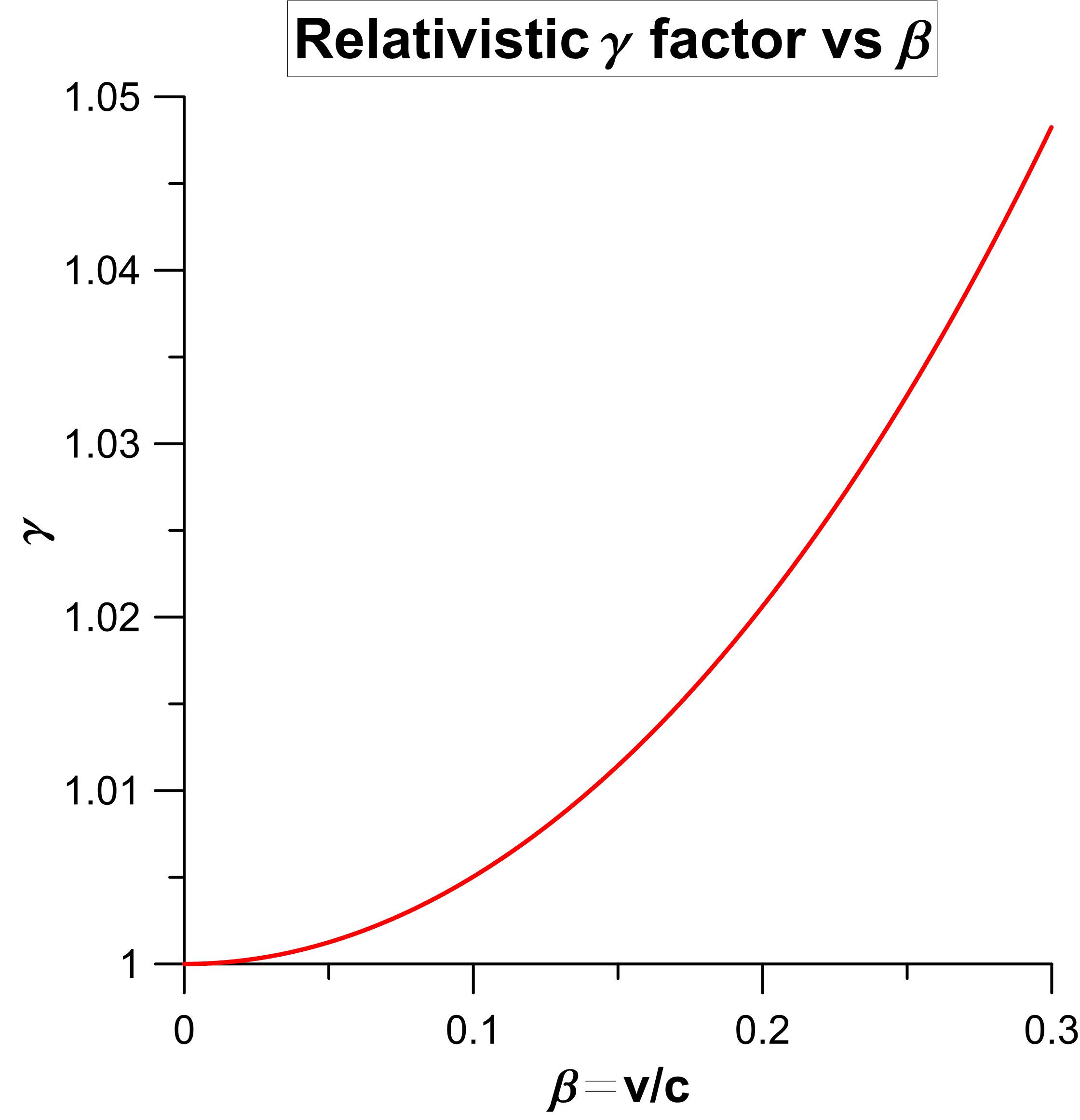} \\
            (a) & (b) \\
        \end{tabular}
        \caption{(a) Relativistic $\gamma$ factor versus $\beta$ factor. $\gamma$ goes to infinity as $\beta$ goes to unity. (b) Range of $\beta$ restricted.}
        \label{fig:relativisticgammabeta}
    \end{figure*}

    Relativistic effects need to be considered for the systems we are proposing. There are a variety of effects to be considered. The critical effects of time dilation, length contraction, wavelength (photon energy) change, and effective mass increase can be parametrized by $\beta$ and $\gamma$ which are defined as:
    \begin{equation}
        \gamma=(1-\beta)^{-1/2}, \hspace{2mm}\textrm{where}\hspace{2mm} \beta=v/c
    \end{equation}
    \begin{align}
    \begin{split}
        \gamma=\frac{1}{\sqrt{1-(v/c)^2}}=&(1-\beta^2)^{-1/2}\\
        \rightarrow& 1+\frac{1}{2}\beta^2,\hspace{1mm} \beta<<1.
    \end{split}
    \end{align}
        In the non-relativistic limit $\beta$ is extremely small and $\gamma$ is very close to unity. As shown the corrections for $\gamma$ differing from unity are relatively small until $\beta$ becomes close to 1 (speed near the speed of light). In the limit as $v\rightarrow c$ ($\beta\rightarrow 1$)  then $\gamma\rightarrow\infty$ and the relativistic effects become extreme. As an example, for a speed of 0.3 c we have $\gamma\sim 1.05$, or a 5\% correction (Figure \ref{fig:relativisticgammabeta}).
    
    \begin{figure*}
        \centering
        \begin{tabular}{c c}
            \includegraphics[width=0.45\textwidth]{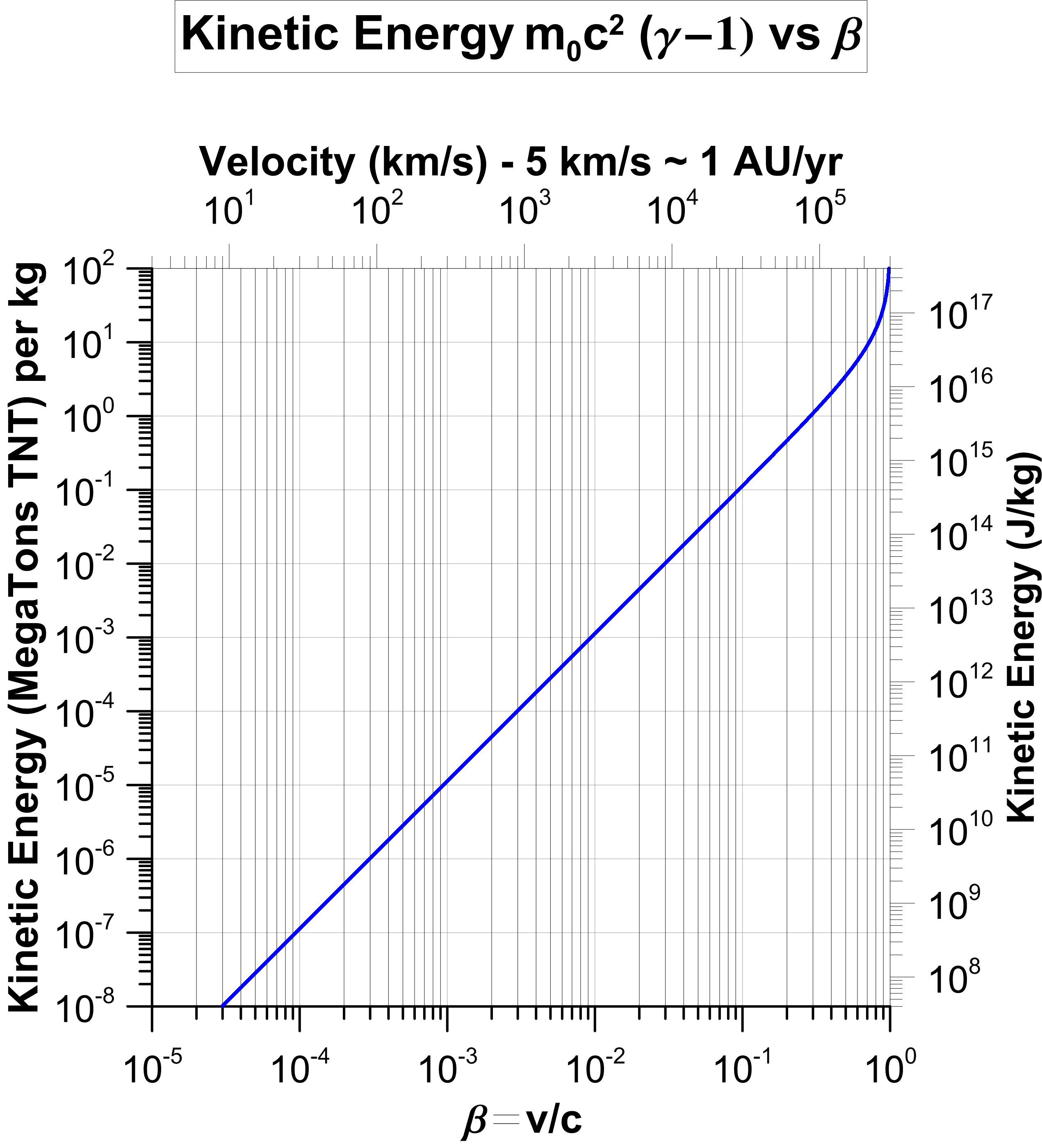} &
            \includegraphics[width=0.45\textwidth]{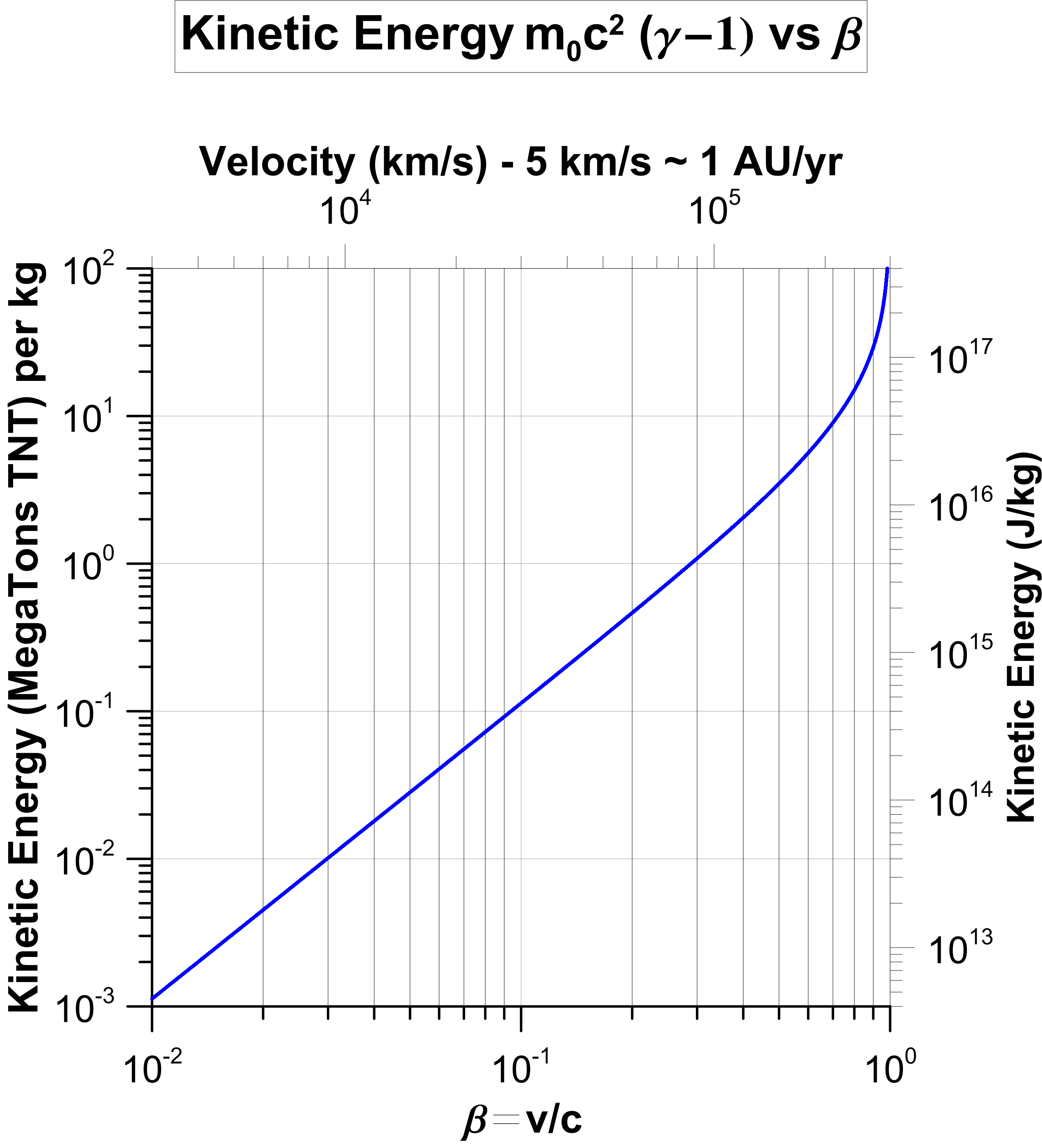} \\
            (a) & (b) \\
        \end{tabular}
        \caption{(a) Kinetic energy vs $\beta$ in units of megatons TNT per kg of payload mass, as well as Joules per kg. (b) Restricted range of $\beta$.}
        \label{fig:kineticenergy}
    \end{figure*}
    
\subsection{Kinetic Energy of Spacecraft}

    The kinetic energy of the spacecraft increases dramatically with speed. The kinetic energy KE$=m_0(\gamma-1)c^2$  where the rest mass is $m_0$ and effective mass $m_\textrm{eff}=m_0 (\gamma-1)$. As the speed approaches c ($\beta\rightarrow1$) the kinetic energy diverges to infinity. Hence the problem of propelling non-zero rest mass objects at the speed of light. We plot the kinetic energy in both J/kg and megaton TNT (MT) equivalent/kg vs speed. At $\beta\sim0.3$ (30\% c) the kinetic energy is about 1 MT/kg. Note that modern thermonuclear weapons have an energy release per unit mass of about 5 MT/ton or about 5 kT/kg for large yield weapons. Small yield weapons are much worse than this. \textbf{Thus a 1 kg spacecraft going at 0.3 c will have an effective ``yield'' of 1 MT or roughly that of a large strategic thermonuclear weapon (Figure \ref{fig:kineticenergy}).} 
    
\subsection{Scaling}
    
    Since the system we propose is not single use but rather scalable to any size it is critical to understand the scaling relations in the section above. In general we use the optimized case of payload mass = sail mass and assume a nearly ideal sail tuned to the laser wavelength so $\epsilon_r=1$. We assume a slightly futuristic sail with thickness of 1 $\mu$m for many cases and 10 $\mu$m (thick even for today's sails). Future advancements in sail thickness down to 0.1 $\mu$m and below can be envisioned but are NOT assumed. They will only make the conclusions even more optimistic. The density of all sails we consider is about the same, namely $\rho\sim 1,400$ kg/m$^3$.  We can then vary power, laser array size, and payload mass as we proceed along the roadmap from small to large systems. The trade-offs between payload mass and speed desired and power and array size required are then explored. We cover this much more in our papers but the basic conclusions are as stated, namely payloads from wafer scale and below to $10^5$ kg and above (human capable) can all be propelled, albeit with different speeds. For $\epsilon_r =1$ we have
    \begin{equation}
        v_{\textrm{max-}\infty}=\bigg(\frac{2P_0 d}{c\lambda\alpha}\bigg)^{1/2}(\xi h\rho m_0)^{-1/4},
    \end{equation}
    which scales as $P_0^{1/2}$, $d^{1/2}$, $\lambda^{-1/2}$, $h^{-1/4}$, $\rho^{-1/4}$, and $m_0^{-1/4}$. The scaling of speed is a mild function of payload mass $\sim m_0^{-1/4}$. This is due to the fact that as the payload mass grows so does the sail. As the sail grows the acceleration distance increases as the laser spot can become larger. These effects tend to mitigate the increased mass. So while a gram scale wafer can be accelerated to relativistic speeds (26\% c in our largest baseline case - DE-STAR 4), the same laser array that allows this also allows propelling a 100 kg craft (Voyager class) to about 1.5\% c or nearly 300 times faster than Voyager achieved after 37 years. A 100 kg craft of this time would reach 1AU ($\sim$Mars) in about a day while a Shuttle class vehicle with a mass of 105 kg ($\sim$100 tons) would reach 0.26\% c or about 780 km/s or 46 times faster than Voyager. This exceeds the galactic escape speed for example (depending on the Dark Matter distribution). While the numbers may be mind numbing they need to be kept in context. We are NOT proposing we should immediately build the largest system but rather begin the roadmap to do so.
    
\subsection{Preliminary System Design}
    
\begin{figure}
        \centering
        \includegraphics[width=0.7\textwidth]{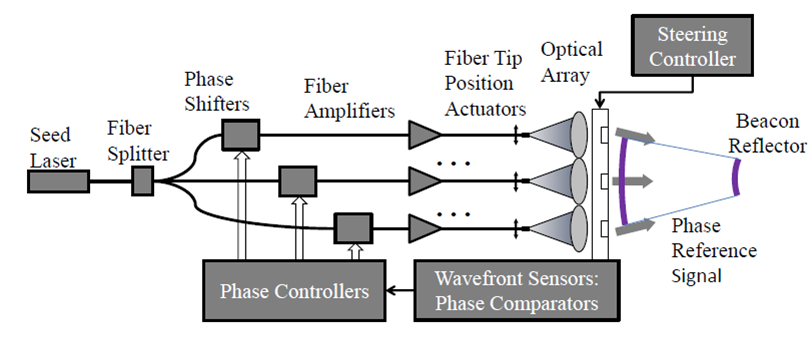}
        \caption{Schematic design of a phased array laser driver. Wavefront (phase) sensing from both local and extended systems combined with the system metrology are critical to forming the final beam.}
        \label{fig:phasecontrol}
    \end{figure}

    Directed energy systems are ubiquitous, used throughout science and industry to melt or vaporize solid objects, such as for laser welding and cutting, as well as in defense. Recent advances in photonics now allow for a 2D array of phase locked laser amplifiers fed by a common low power seed laser that have already achieved near 50\% wall plug conversion efficiency. It is known as a MOPA (Master Oscillator Power Amplifier) design (Figure \ref{fig:phasecontrol}). 
    
\begin{figure*}
        \centering
        \begin{tabular}{c c c}
            \includegraphics[width=0.2\textwidth]{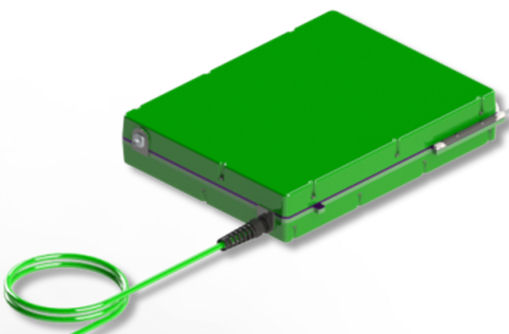} &
            \includegraphics[width=0.34\textwidth]{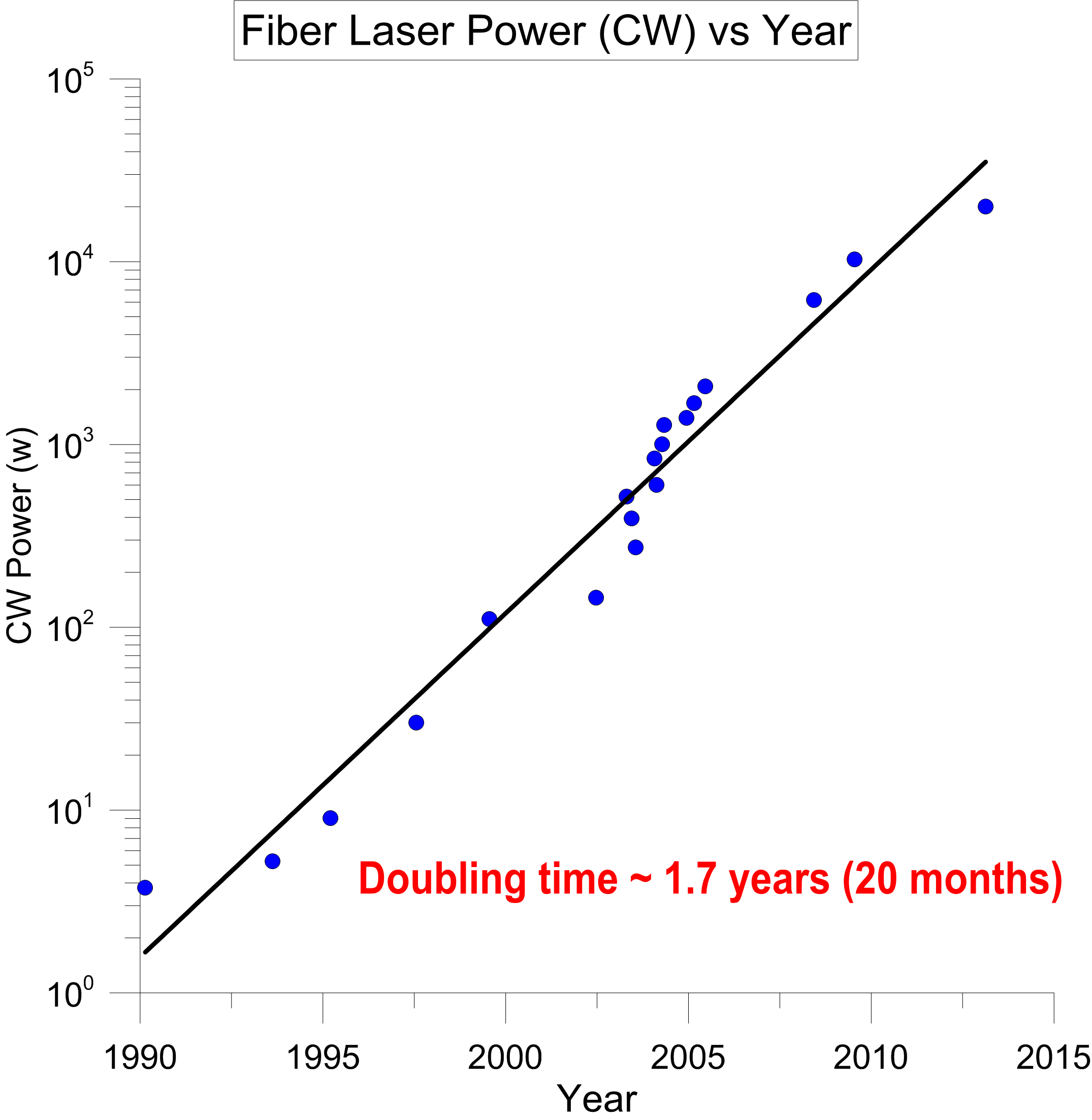} &
            \includegraphics[width=0.36\textwidth]{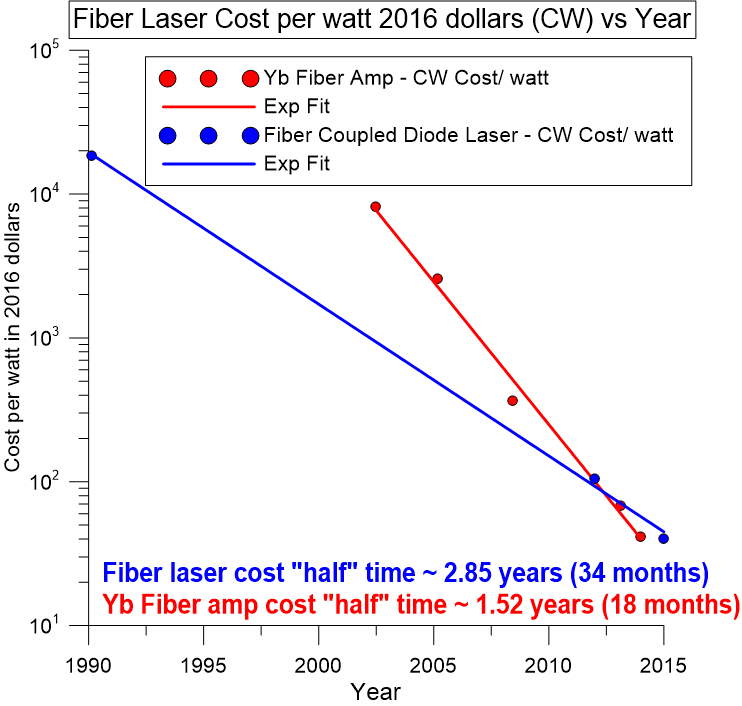} \\
            (a) & (b) & (c) \\
        \end{tabular}
        \caption{(a) Picture of current 1-3 kW class Yb laser amplifier which forms the baseline approach for our design. Fiber output is shown at lower left. Mass is approx 5 kg and size is approximately that of this page. This will evolve rapidly, but is already sufficient to begin. Courtesy Nufern. (b) CW fiber laser power vs year over 25 years showing a ``Moore's Law'' like progression with a doubling time of about 20 months. (c) CW fiber lasers and Yb fiber laser amplifiers (baselined in this paper) cost/watt with an inflation index correction to bring it to 2016 dollars. Note the excellent fit to an exponential with a cost ``halving'' time of 18 months. }
        \label{fig:amplifiers}
    \end{figure*}

    The technology is proceeding on a ``Moore's Law'' like pace with power per mass at 5 kg/kW with the size of a 1 kW amplifier not much larger than a textbook (Figure \ref{fig:amplifiers}). There is already a roadmap to reduce this to 1 kg/kW in the next 5 years and discussions for advancing key aspects of the technology to higher TRL are beginning. These devices are revolutionizing directed energy applications and have the potential to revolutionize many related applications. Due to the phased array technology the system can simultaneous send out multiple beams and thus is inherently capable of simultaneous multitasking  as well as multi-modal. 
    
    The laser system can be built and tested at any level from desktop to extremely large platforms. \textbf{This is radically different than the older "use a huge laser" approach to photon propulsion.} This is the equivalent to modern parallel processing vs an older single processor supercomputer. 
    
    \begin{figure}
        \centering
        \includegraphics[width=0.7\textwidth]{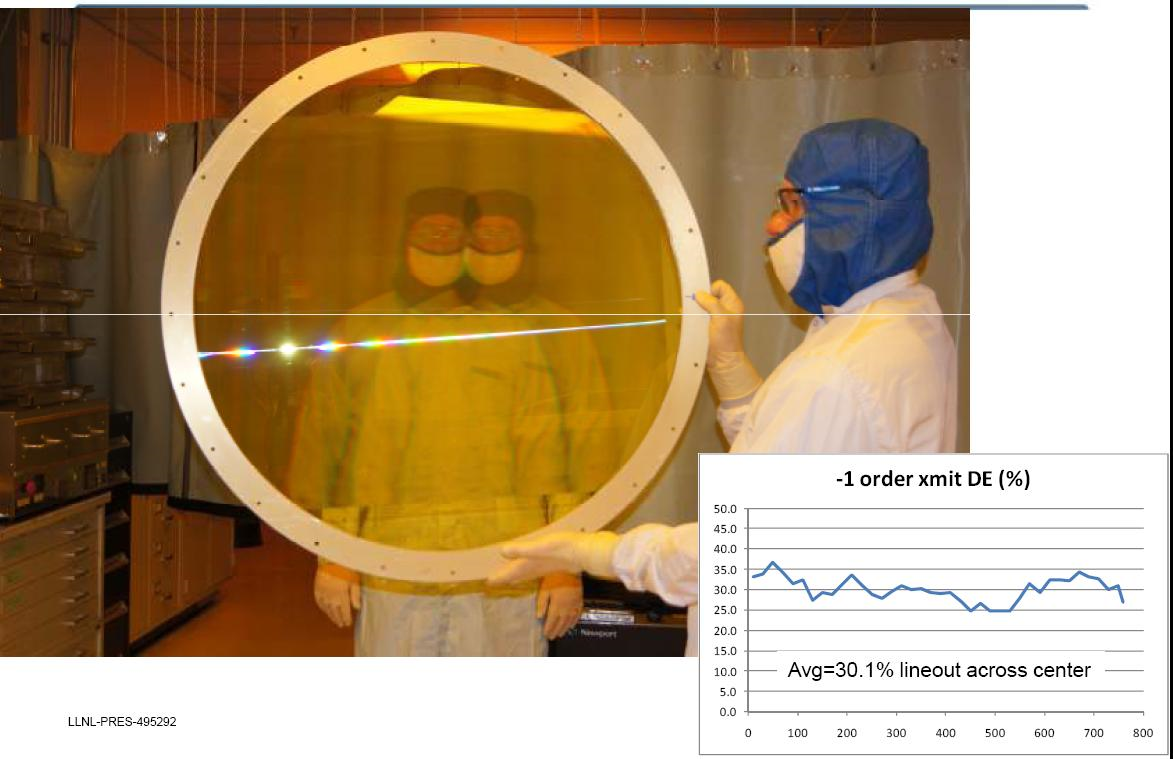}
        \caption{Thin film replicated optics from DARPA MOIRE Ball-LLNL program. The thin ($\sim30$ micron) plastic film is replicated from an etched mandrel. Areal mass of the film is approximately 60 g/m$^2$ with actual optical mass dominated by mounting ring and not the optic itself. Thin glass and other materials are other possibilities for these lightweight replicated optics. In our system all optical elements are identical leading to mass production replicated techniques being optimal. Credit: DARPA.}
        \label{fig:thinfilm}
    \end{figure}
    
    The more modest size systems can be completely tested on the ground as well as sub-orbital flight tested on balloons or possibly sounding rocket. While the largest sized systems (km scale) are required for interstellar missions, small systems have immediate use for roadmap development and applications such as sending small probes into the solar system and then working our way outward as larger laser arrays are built. The laser array is modular, leading to mass production, so that a larger array can be built by adding elements to a smaller array. \textbf{Array testing and propulsion tests are feasible at all levels allowing for roadmap development rather than ``all or nothing''.}  Small arrays can also be used for orbital debris removal, long range laser communication, power beaming, ISS defense from space debris as well as stand-on systems for planetary defense so again there is a use at practically every level and funding is well amortized over multiple uses. This allows practical justification for construction. In addition there is an enormous leveraging of DoD and DARPA funds that dramatically lowers the overall costs (Figure \ref{fig:thinfilm}).
    
\subsection{Phase-Lockable Lasers and Current PV Performance}

    New fiber-fed lasers at 1 $\mu$m have efficiencies near 40\% (DARPA Excalibur program). We assume incremental efficiency increases to 70\%, though current efficiencies are already good enough to start the program. It is conceivable that power density could increase to 10 kW/kg in a decade given the current pace.  Current space multi-junction PV has an efficiency of nearing 40\% with deployable mass per power of less than 7 kg/kW (ATK Megaflex as baselined for DE-STARLITE). Multi-junction devices with efficiency in excess of 50\% are on the horizon with current laboratory work exploring PV at efficiencies up to 70\% over the next decade. We anticipate over a 20 year period PV efficiency will rise significantly, though it is NOT necessary for the roadmap to proceed. The roadmap is relatively ``fault tolerant'' in technology development. Array level metrology as a part of the multi-level servo feedback system is a critical element and one where recent advances in low cost nanometer-level metrology for space applications is another key technology \cite{Hughes2014}. One surprising area that needs significant work is the simple radiators that radiate excess heat. Currently this is the largest mass sub system at 25 kg/kW (radiated). The increase in laser efficiency reduces the radiator mass as does the possibility to run the lasers well above 300 K. Radiation hardening/resistance and the TRL levels needed for orbital use are another area we are currently exploring.
    
\section{\label{sec:waferscalespacecraft}Wafer-Scale Spacecraft}

\begin{figure}
        \centering
        \includegraphics[width=0.45\textwidth]{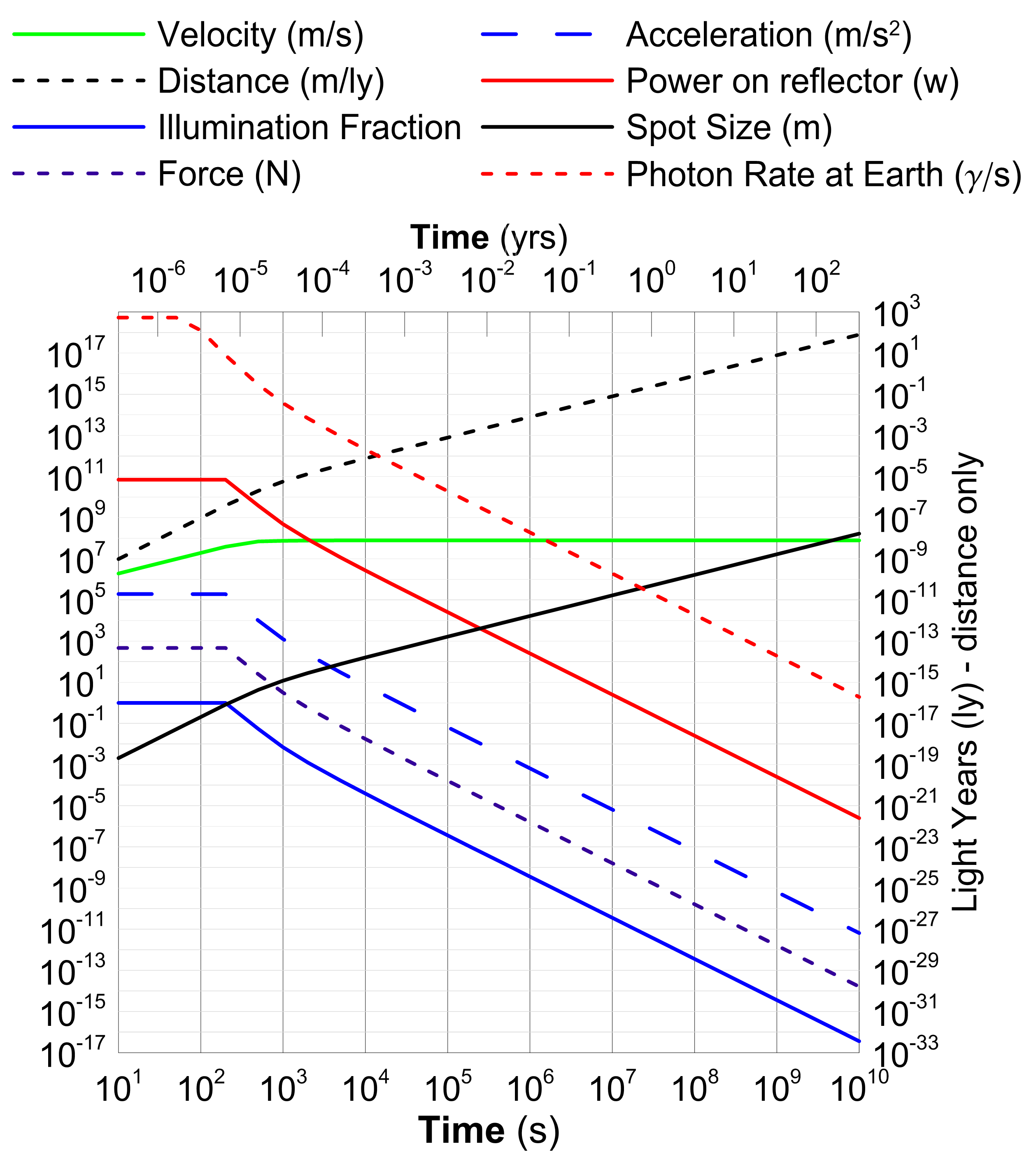}
        \caption{Parameters for full class 4 system with 1 gram wafer SC and 1 m sail. Craft achieves $\sim c/4$ in a few minutes and takes about 20 years to get to Alpha Centauri. Communications rate assumes class 4 drive array is also used for reception with a 1 watt short burst from a 100 mm wafer SC. Here the only optical system on the spacecraft is assumed to the 100 mm wafer. No external optics is assumed for the laser communications system at the wafer. Received photon rate at Earth is during the burst which yields about 650 $\gamma$/s at the Earth in the class 4 receive array with the wafer at Alpha Centauri . Average received rate must be scaled by the burst fraction which for the small 1 g wafer is 0.2\% limited by the RTG and assumed low 7\% thermal to electrical conversion efficiency. One goal is to store enough energy onboard from the RTG to allow a long term ``burst rate'' during critical periods (image download, etc.) and this allows the possibility to achieve the full 1 watt burst rate during these critical times. If we can achieve 1 photon/bit encoding the data rate (bits/sec) would be essentially the photon rate ($\gamma$/s).}
        \label{fig:class4}
    \end{figure}

\begin{figure}
        \centering
        \includegraphics[width=0.6\textwidth]{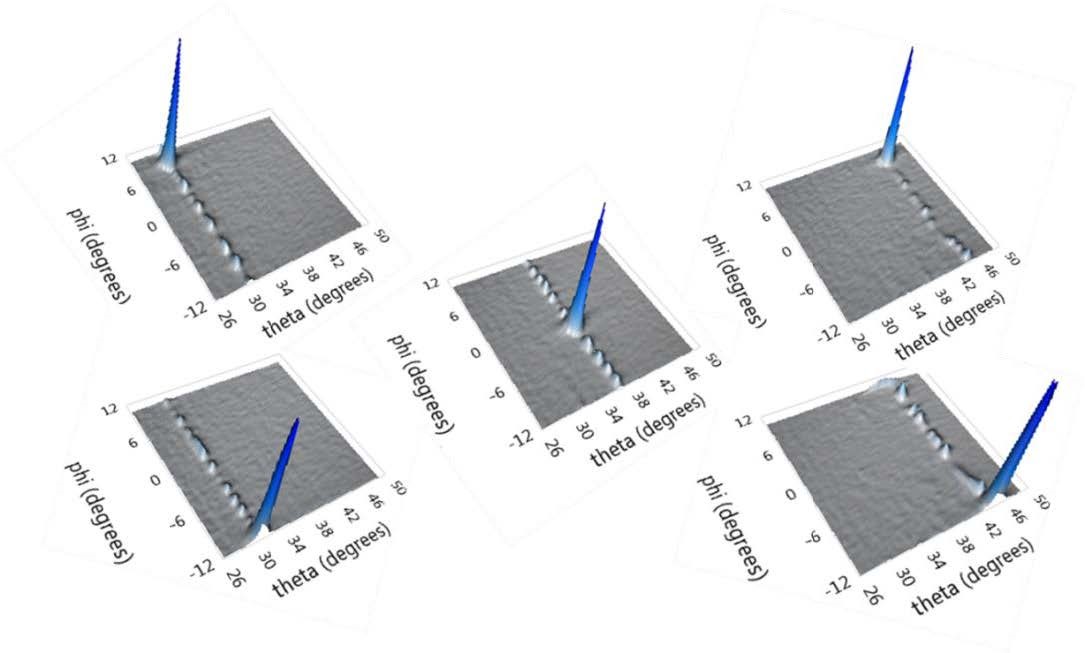}
        \caption{UCSB phased array for chip level laser communication with no external optics showing electronic beam steering \cite{Hulme2014}. Another option we are considering is a single laser with MEMS steering combined with a thin film optic. While not as elegant as a wafer scale phased array it is far simpler in some ways but lacks the full 2D nature of the wafer phased array. The thin film optic in this case adds complexity. Though extremely low mass it must be deployed after launch. More work is needed on these trade studies.}
        \label{fig:chiplaserarray}
    \end{figure}

    Recent work at UCSB on Si photonics now allows us to design and build a ``spacecraft on a wafer.'' The recent (UCSB) work in phased array lasers on a wafer for ground-based optical communications combined with the ability to combine optical arrays (CMOS imagers for example) and MEMS accelerometers and gyros as well as many other sensors and computational abilities allows for extremely complex and novel systems. Traditional spacecraft are still largely built so that the mass is dominated by the packaging and interconnects rather than the fundamental limits on sensors. Our approach is similar to comparing a laptop of today to a supercomputer with similar power of 20 years ago and even a laptop is dominated by the human interface (screen and keyboard) rather than the processor and memory. Combining nano-photonics, MEMS, and electronics with recent UCSB work on Si nano-wire thermal \cite{Hulme2014}\cite{Curtin2014} converters allows us to design a wafer that also has an embedded RTG or beta converter power source (recent LMCO work on thin film beta converters as an example) that can power the system over the many decades required in space. Combined with small photon thrusters (embedded LEDs/lasers for nN thrust steering on the wafer gives a functional spacecraft).  While not suitable for every spacecraft design by any means this approach opens up radically new possibilities including mass production of very low cost wafer-scale spacecraft. In addition, the power from the laser itself can add significant power to the spacecraft even at large distances. We have run link margin calculations including Zodi, CIB, galaxy, and optical emission for such a wafer-scale system run in a hibernate/burst mode using the DE-STAR array as both the transmitter of power to propel and communicate with the spacecraft as well as to receive the very weak signal from the spacecraft and conclude it is feasible to receive data (albeit at low rate) at light year distances (Figures \ref{fig:class4} and \ref{fig:chiplaserarray}). For pointing we use the onboard camera as star tracker and/or lock to DE-STAR laser as a beacon \cite{Brashears2015}.
    
    Using the reflector for laser communications would also help greatly. We have NOT assumed this in the link margin calculations for Figure \ref{fig:class4} but do so in Fig \ref{fig:class4parameters}. There are many challenges here and ones that will require considerable effort, but the rewards will be used in not only interstellar probes but also planetary, space, and terrestrial remote sensing, medical, security, etc. The list of possible uses for self-contained autonomous system is endless.
    
\subsection{Laser Sail}

    The laser sail is both similar to and fundamentally different than a solar sail. For small laser sails, even with low powers, the flux can easily exceed 100 MW/m$^2$ or 105 Suns. This requires a very different approach to the sail design. Fortunately the laser line is very narrow so we can tune the laser sail reflectivity to be extremely high and the absorption to be extremely low using multi-layer dielectric coatings. The relativistic aspects of the highest speed missions present another problem as the laser wavelength is shifted at the reflector.  Laser coatings on glass already can achieve 99.999\% reflectivity or absorption of less than 10$^{-5}$. We have started working with industrial partners and have designed a ``roll to roll'' process that is a multi-layer dielectric on plastic that achieves 99.995\% reflectivity (in design). This looks good enough for most cases except the extreme flux of the true interstellar probes which use small ($\sim1$ m size) reflectors. For the small reflectors we propose using a pure dielectric reflection coating on ultra-thin glass or other material. Spherical (bubbles) sails are an option for lab and orbital testing. The loss in fiber optic quality glasses allows loss in the ppt(10-12)/$\mu$m (of thickness) which is already better than we need. This is an area we need to explore much more. For example, the flux at the tip of high power kilowatt class single mode fiber optic exceeds 10 TW/m$^2$, higher than we need. Rather than the typical 1/4 reduced wavelength anti-reflective (AR) dielectric coating, we will need to design a 1/2 wave reflection coating for the sail.
    
\subsection{Flux on Sail}
    
    The scaling of flux on the reflector is
    \begin{equation}
        F=P_0/D^2,
    \end{equation}
    assuming optimized case where sail mas = payload mas ($m_0=m_\textrm{sail}=\xi D^2 h\rho$).
    \begin{equation}
        D=\bigg(\frac{m_0}{\xi h \rho}\bigg)^{1/2} \rightarrow F=\frac{P_0}{\xi D^2}=\frac{P_0 h\rho}{m_0}.
    \end{equation}
    Note the flux is proportional to the thickness and density (smaller sail) and inversely proportional to the mass (larger sail). \textbf{This means lower mass payloads have high flux requirements on the sail.} We consider two cases: 1) where light is either reflected or absorbed but none is transmitted through the sail (appropriate to dielectric and metal coatings) and 2) where some light is transmitted through the reflector (appropriate to dielectric only coatings).
    
    1) Case of NO transmission trhough reflector (all is reflected or absorbed):
    \begin{equation}
       F=\frac{dp}{dt}=\frac{2P\epsilon_r}{c}+\frac{P(1-\epsilon_r)}{c}=\frac{P(1+\epsilon_r)}{c},
    \end{equation}
    where $P$ is the laser power at the reflector, $\epsilon_r$ is the refelctor reflection coefficient (1 for perfect reflection, 0 for perfect absorption).
    
    2) Case of some transmission of light through the reflector:
    \begin{equation}
        F=\frac{dp}{dt}=\frac{2P_r}{c}+\frac{P_A}{c}=\frac{P}{c}(2\epsilon_r+(1-\epsilon_r)\alpha),
    \end{equation}
    where $\alpha$ is the absorption coefficient. Note that if $\alpha=1$ (complete absorption inside the reflector of the part not reflected), then $F=P(1+\epsilon_r)/c$, as in case 1. Similarly, $\alpha=0$ implies no absorption of light inside the reflector. Above, $P_r=P\epsilon_r$ is the laser power reflected at the first surface, $P_A=P(1-\epsilon_r)\alpha$ is the laser power absorbed inside the reflector. $P_T=P-P_r-P_A$ is the laser power transmitted through the reflector such that $P=P_r+P_A+P_T$. Note that for metalized reflectors $\alpha\approx1$ since radiation not reflected is absorbed. The general case would then replace $P(1+\epsilon_r)$ with $P(2\epsilon_r+(1-\epsilon_r)\alpha)$.\\
    
\subsection{Sail Temperature}

    We can compute the approximate sail temperature $T_\textrm{sail}$ based on the absorption coefficient and the effective emissivity of the front and back of the sail. Let $\epsilon_f$ be the convolved emissivity of the front of the sail, $\epsilon_b$ be the convolved emissivity of the back of the sail, and $A_\textrm{sail}$ be the sail area.
    \begin{equation}
        \frac{P_A}{A_\textrm{sail}}=\frac{\alpha P_0(1-\epsilon_r)}{A_\textrm{sail}}=\sigma T^4(\epsilon_f+\epsilon_b),
    \end{equation}
    where $P_0$ is the power in the main beam on the sail. Therefore,
    \begin{equation}
        T_\textrm{sail}=\bigg[\frac{\alpha P_0(1-\epsilon_r)}{\sigma(\epsilon_f+\epsilon_b)A_\textrm{sail}}\bigg]^{1/4}.
    \end{equation}
    
\subsection{Reflector Mass}

    Current solar sail reflectors have thicknesses in the 1-10 $\mu$m range. Future technologies may allow us to greatly reduce this and thus extend the speed and distance of our probes. In this paper we assume current technologies for reflectors with some modest improvements. It is important to consider the implications of future advances in this area, with new technologies such as graphene and carbon nano tubes. We do not assume any of these advanced technologies in our baseline design but they will likely become extremely important in the future and will only help us.
    
    Assuming $\epsilon_r=1$, recall the (non-relativistic) scaling of speed for the case of reflector mass given in Equation 25:
    \begin{align}
    \begin{split}
        v_{\textrm{max-}\infty}=&\bigg(\frac{2P_0 d}{c\lambda\alpha}\bigg)^{1/2}(\xi h\rho m_0)^{-1/4}\\
        =&\bigg(\frac{2P_0 d}{c\lambda\alpha}\bigg)^{1/2}(\xi \sigma m_0)^{-1/4}.
    \end{split}
    \end{align}
    The scaling of speed with the thickness of the sail $h$ is
    \begin{equation}
        v_{\textrm{max-}\infty} \propto (h\rho m_0)^{-1/4}.
    \end{equation}
    As the sail thickness and density decreases the speed will increase. It is often useful to consider the areal density of the reflector. The areal density is $\sigma(\textrm{kg/m$^2$})=h\rho$ (Figure 15). As an example, $h=1\mu$m and $\rho=1.4$ g/cc gives $\sigma=1.4$ g/m$^2$.
    
    \begin{figure*}
        \centering
        \begin{tabular}{c c c}
            \includegraphics[width=0.32\textwidth]{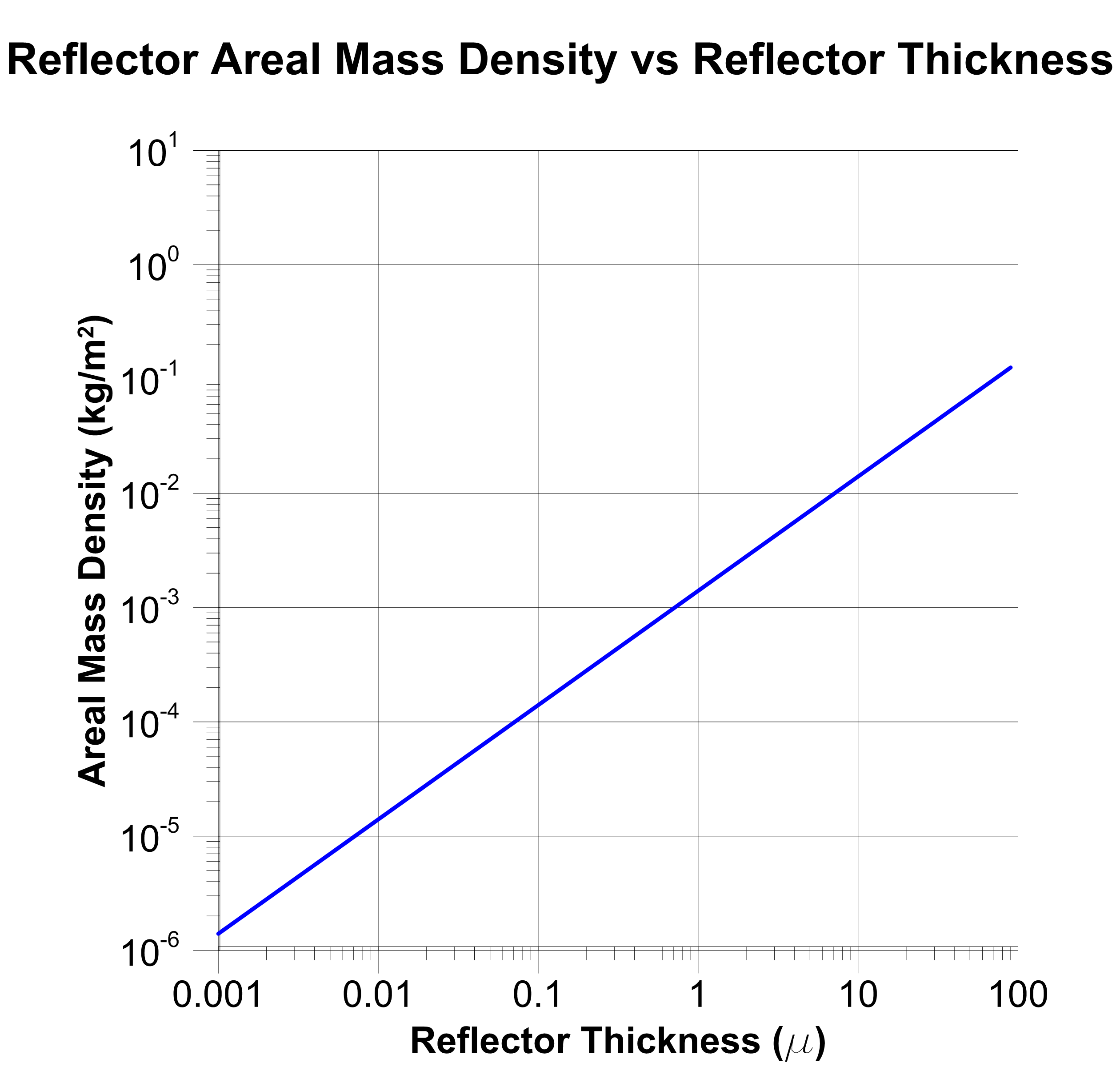} &
            \includegraphics[width=0.28\textwidth]{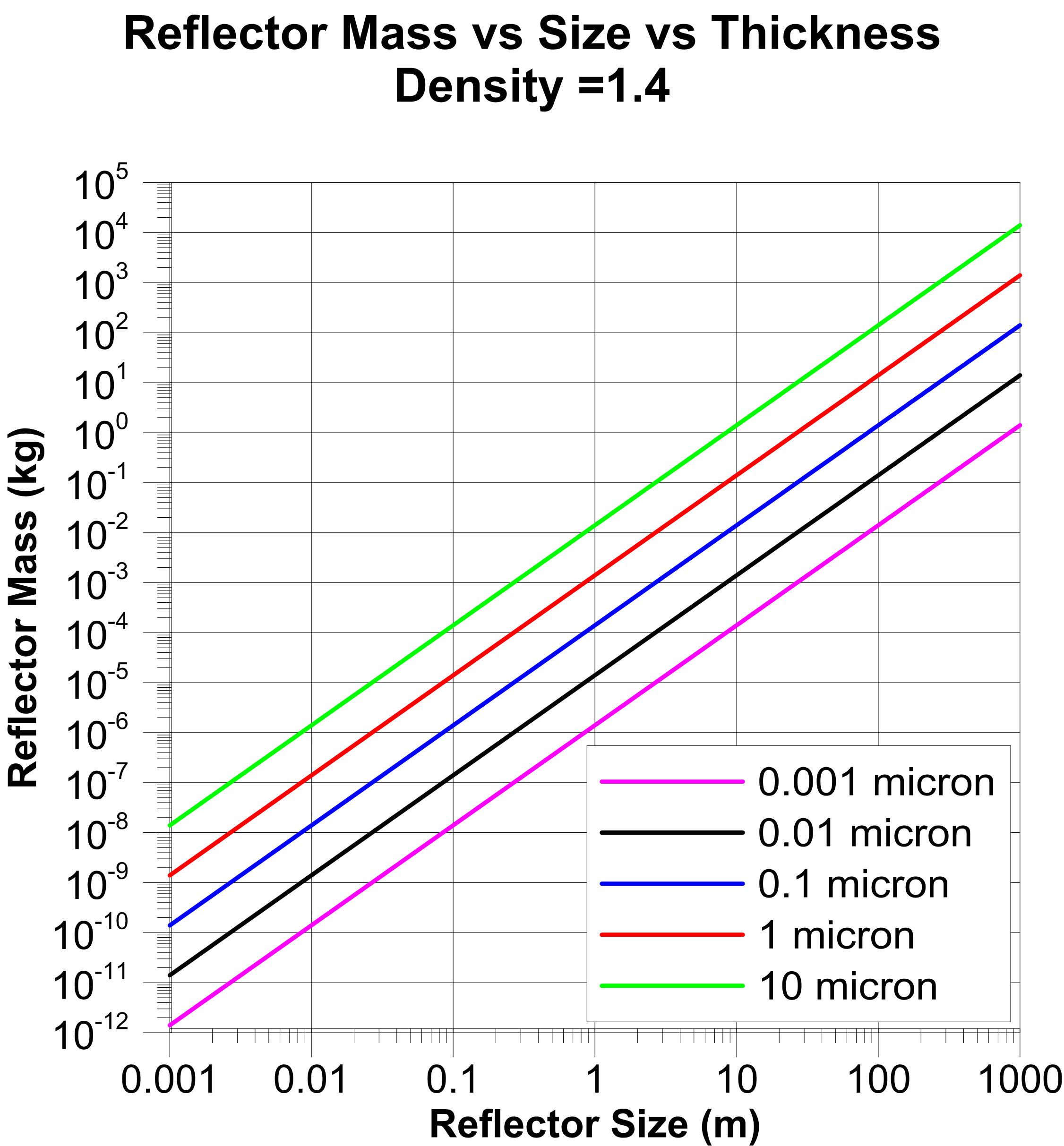} &
            \includegraphics[width=0.31\textwidth]{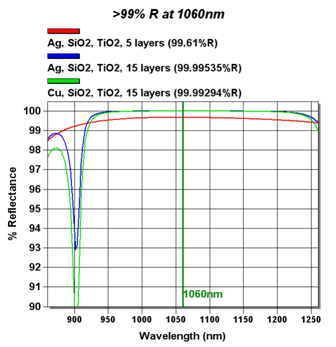} \\
            (a) & (b) & (c) \\
            \includegraphics[width=0.32\textwidth]{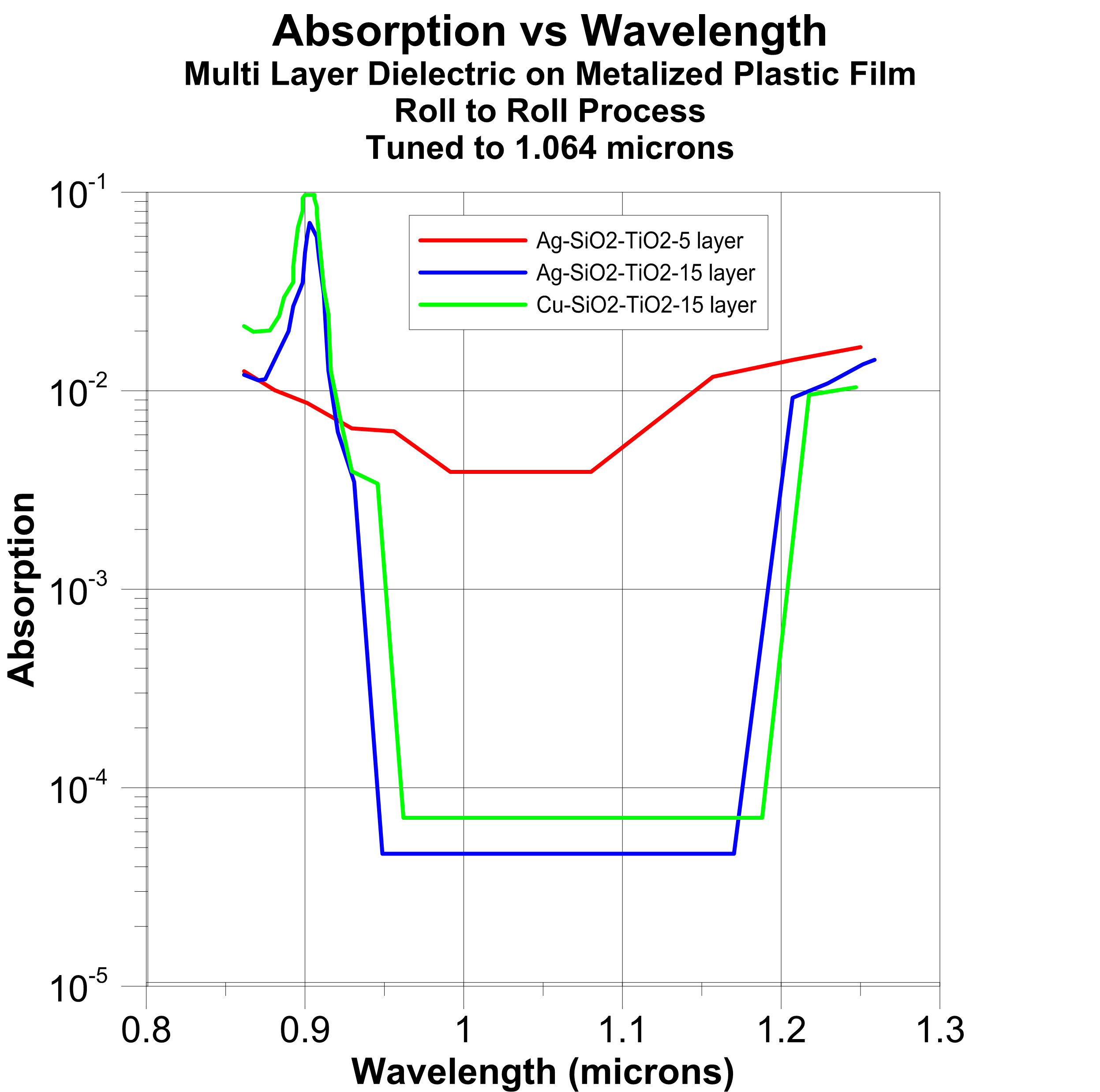} &
            \includegraphics[width=0.3\textwidth]{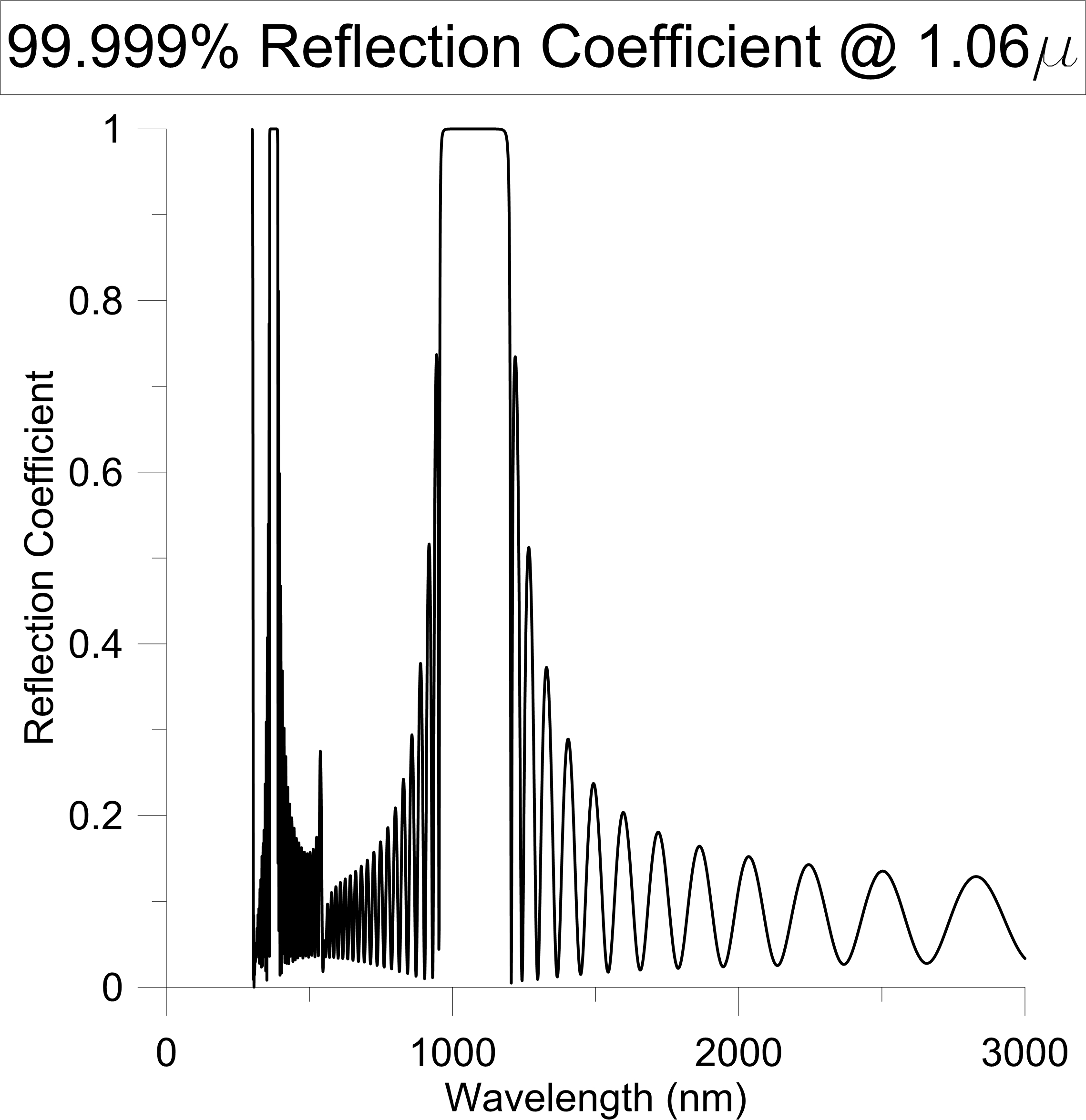} &
            \includegraphics[width=0.31\textwidth]{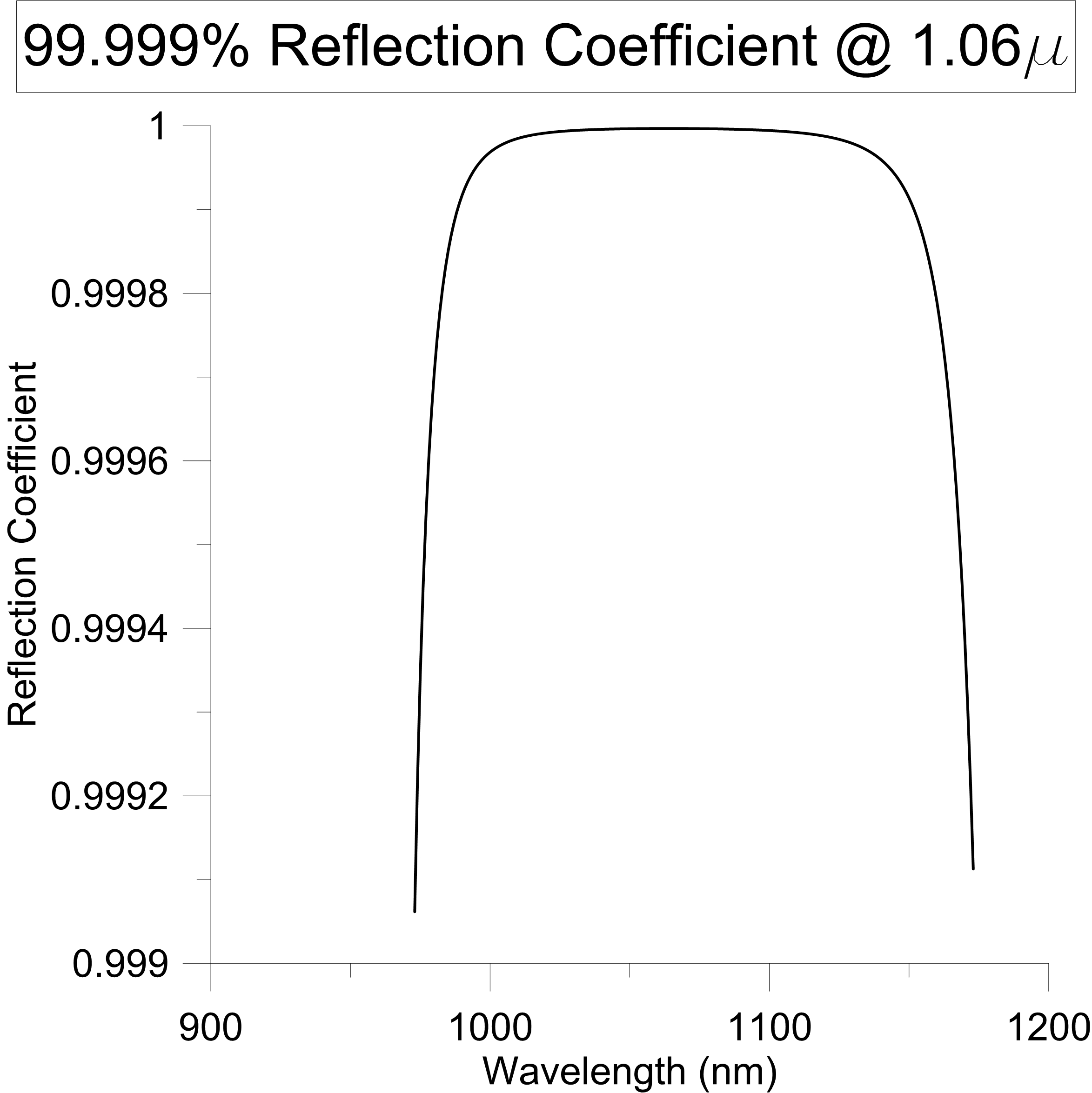} \\
            (d) & (e) & (f) \\
        \end{tabular}
        \caption{(a) Reflector areal mass (kg/m$^2$) vs reflector thickness for density = 1.4 g/cc typical for plastic. Glass films would have a density about 1.8 times larger. (b) Mass of the reflector vs size (assumed to be square) vs reflector thickness. Current reflectors are between 1 (cutting edge) and 10 (conservative) microns thick. Future technologies may allow us to make significantly thinner reflectors. A part of our roadmap to reduce the reflector mass, though we assume in our system analysis a 1 micron thick reflector. Note that the final speed only depends on the system mass to the inverse $1/4$ power (m$^{-1/4}$) for the case of payload mass equaling reflector mass. Hence increasing the reflector thickness has only a mild effect on the speed. (c) Multilayer dielectric deposited on metalized plastic film. The reflectivity is tuned to be maximum at the laser wavelength. Reflectivity vs wavelength for several models of the reflector with varying dielectric layers and compositions. (d) Same but absorption is plotted. Note minimum at the laser wavelength of 1.06 microns with an absorption of about 50 ppm.  This reflector was designed to be part of a mass production ``roll to roll'' process applicable to very large reflectors. For our smaller ``WaferSat'' missions the reflector is relatively small in diameter (meter scale) and hence a roll to roll process is not needed. Roll to roll processing is needed for larger payload masses, which require larger reflectors. (e) Reflectivity vs wavelength over a large wavelength range showing the tuning of maximum reflection at the laser line. It is thus possible to design a reflector that is extremely reflective at the laser wavelength but highly absorptive at longer wavelengths and thus good as a thermal radiator (extreme example of a ``selective coating''). (f) Same but over the narrow region near the laser line (1064 nm). This corresponds to an absorption of about 10 ppm. For thin glass films this could be adopted to a roll to roll process if needed for large reflectors, though it is not currently designed this way.\\}
        \label{fig:reflector}
    \end{figure*}

\subsection{Multi-Layer Dielectric on Metalized Plastic Film}

    Metalized thin film plastic films with multi-layer dielectric coatings can achieve very high reflectivity. We have designed a 99.995\% reflective system suitable for large scale ``roll to roll'' production (Figure \ref{fig:reflector}). Note the reflectivity is tuned to the narrow laser line and that these reflectors are NOT suitable for solar sails which use the broad spectrum of the sunlight to propel them. We illustrate this below with a putative design for our Yb baseline 1064 nm laser case. For large sails ($>10$ m diameter) this is a suitable choice. For example a 30 meter square sail on 10 $\mu$m thick plastic film will have a mass about 13 kg while a more advanced thin film of 1 $\mu$m thickness would have a mass of about 1.3 kg \cite{Bible2013}. Note that the ``bandwidth'' of the tuned reflection in this case is relatively broad, being about 20\%. This is important when we begin to include Doppler shifts for the relativistic cases of the very high speed low mass payloads. Higher bandwidth is possible with additional layers.

\subsection{Multi-Layer Dielectric on Metalized Glass}

    The use of tuned multi-layer dielectric films on metalized glass substrates currently achieves reflections of ``five 9's'' or 99.999\% . This means absorption of less than 10 ppm (parts per million). This is shown for the case of our Yb 1064 nm baseline in Figure \ref{fig:reflector}. Again the high reflectivity is tuned to the laser line. Note that a bandwidth of approximately 20\% is achieved for this case.
    
    We note that even better performance has been achieved for the 30 cm, 40 kg LIGO multi-layer dielectric on glass mirrors with about 0.5 ppm absorption. In this case the mirror does not have metal coatings but the mirrors are quite thick (multiple cm thick).

\subsection{Multi-Layer Dielectric on Glass with no Metal}

\begin{figure}
        \centering
        \includegraphics[width=0.45\textwidth]{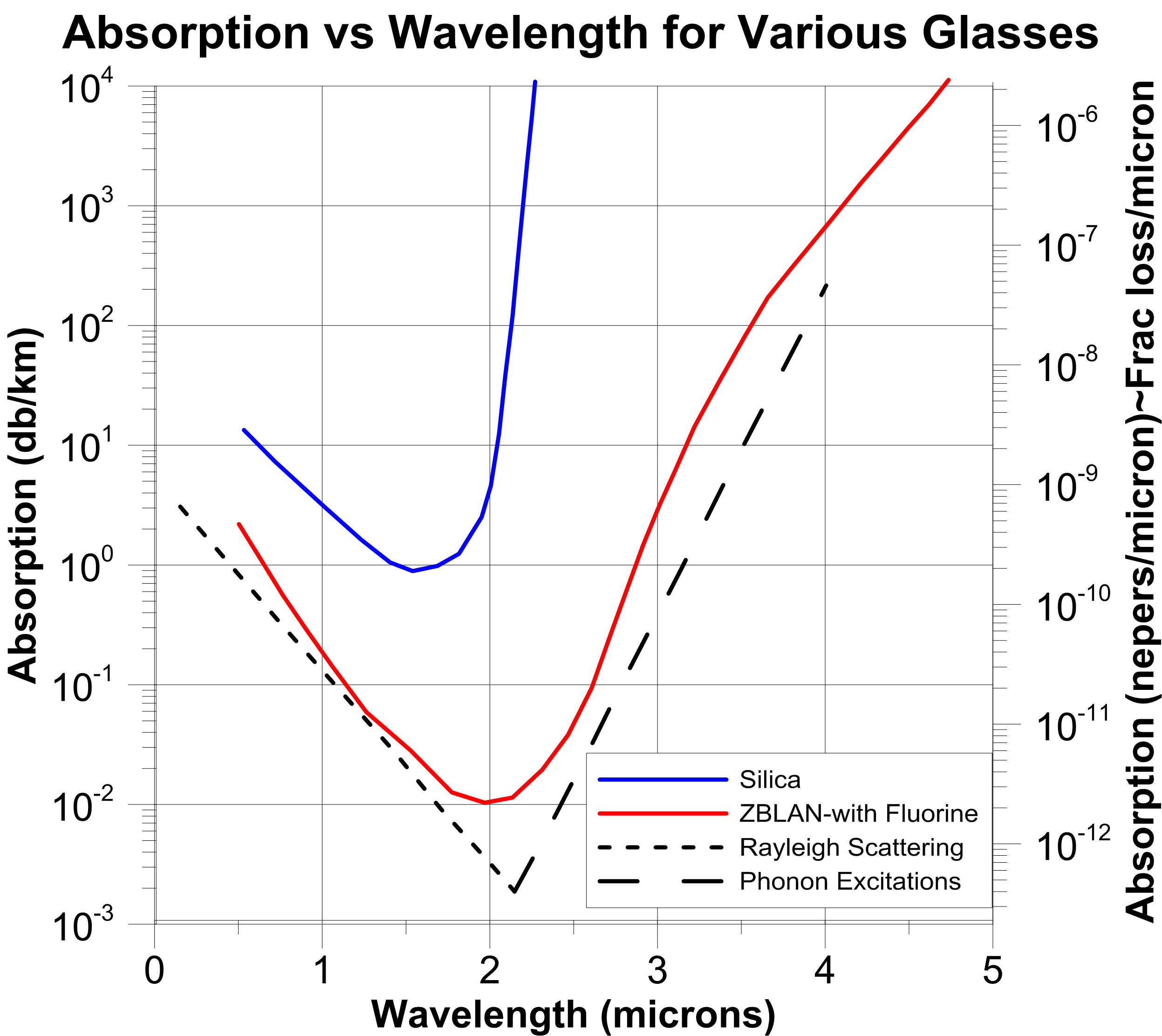}
        \caption{Absorption for optimized glass designed for fiber optics. This is extremely low OH glass and has excellent performance for extremely high flux applications such as those which our small reflector relativistic designs require. Note the absorption at our laser wavelength of 1.06 microns is about 20 ppt/micron reflector thickness. What is not shown is the optimized reflectivity for the multi-layer dielectric coating. In this system there are no metal layers which allows us to achieve such low absorption.}
        \label{fig:glassabsorption}
    \end{figure}

    For the small very high flux sails, such as the  relativistic WaferSat probes, the flux on the reflector becomes so large that metalized reflectors, even with multi-layer dielectric coatings become extremely difficult to make. The issue is that the metalized sub-structure is not reflective enough so the thermal management becomes a critical problem (sails vaporize). One solution is to remove the metals completely and use a fully dielectric reflector. This is what we propose for the extreme cases of high flux small sails. Glasses designed for fiber optics and other photonic communication applications have extremely low absorption coefficients with ppt (parts per trillion) per micron thickness currently achieved. While the reflection coefficient will not be as large as it is for the metalized cases in general, they are sufficient. Note in this case the absorption takes place in the bulk of the glass of the reflector and dielectric coating while in the metalized plastic and glass case it is absorption in the metal film that is dominant. We plot the absorption coefficients for modern fiber optic glasses to illustrate this (Figure \ref{fig:glassabsorption}). As a comparison a 1 cm ($10^4$ micron) thick piece of ZBLAN glass with 20 ppt/micron absorption coefficient would have a total absorption of $20\times10^{-12}\times 10^4 = 2
    \times10^{-7}$ or 0.2 ppm at a wavelength of 1.06 microns.
    
\subsection{Reflector Stability and Shaping}

    A critical issue will be the stability of the sail. There are a number of perturbative effects. These include laser instabilities and laser mode issues, differential forces on the sail and mechanical modes in the sail, heating of the sail and laser pointing instability. This is a complex set of issues that requires a significant amount of research and development. This will not be trivial. Some ways to mitigate these issues are spinning the sail (especially if circular) and shaping of the reflector into a modified conic (similar to a reentry vehicle). Feedback from the sail to the laser will help but the time of flight will lower the effective servo bandwidth for this. Ideally a self stabilizing system is desired. We see this as one of the most critical issues to overcome.
    
\subsection{Beam Shaping and Sail Stability}

    Since we are using a phased array to form the target spot we can shape the beam profile by adjusting the phase of the array. Normally we would produce a Gaussian beam pattern but there are other options that would increase sail stability. One is to form a minimum (null) in the middle of the beam which would damp small perturbations and self stabilize the sail.
    
\subsection{Effect of Reflector Thickness}

\begin{figure}
        \centering
        \begin{tabular}{c c}
            \includegraphics[width=0.422\textwidth]{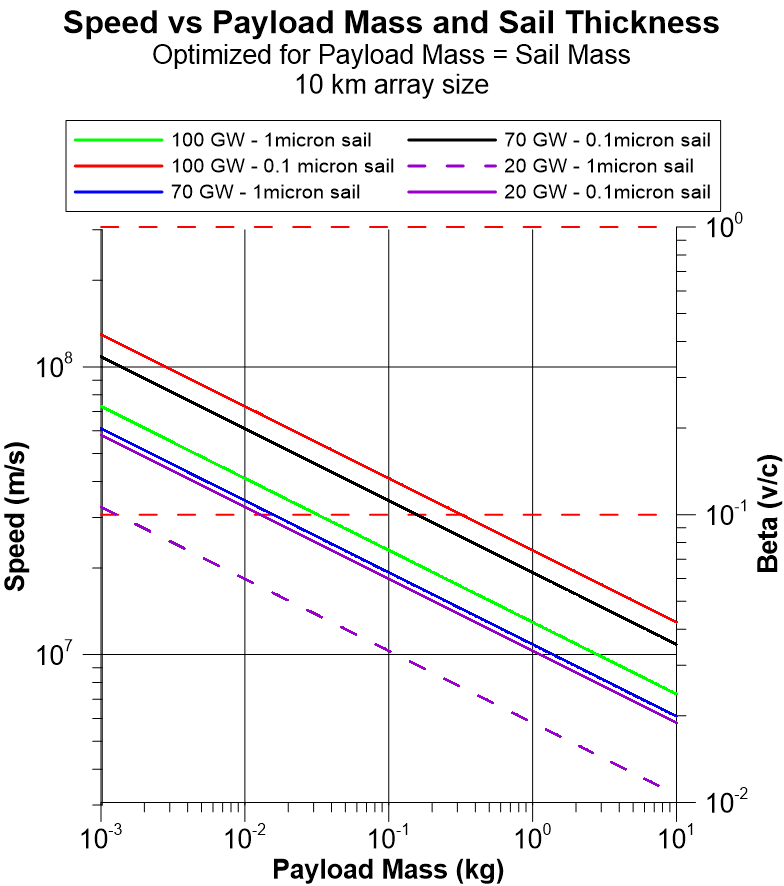} & \includegraphics[width=0.422\textwidth]{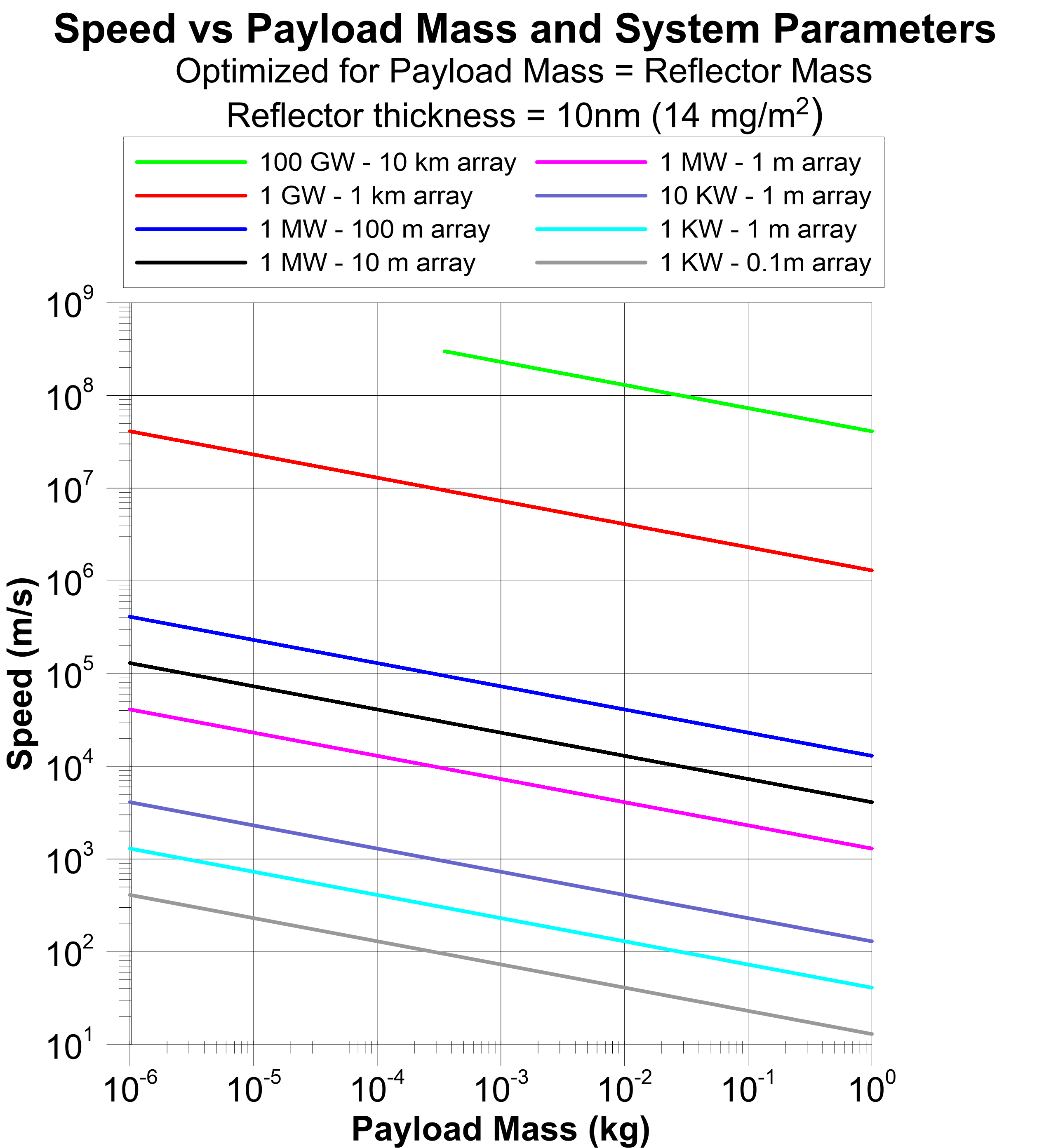}\\ 
            (a) & (b) \\
        \end{tabular}
        \caption{Speed vs payload mass and power for sail thicknesses of 0.1 and 1 micron, assuming a class 4 drive system.  Note that for a factor of 10 decrease in reflector thickness (1 $\mu$m to 0.1 $\mu$m) the speed increases by a factor of $10^{-1/4}\sim1.8$ in the non-relativistic limit. It is possible that even thinner sails could be produced in the future.  Note that a 1 gram payload (with a 1 gram sail with 1 $\mu$m thickness) achieves 0.1c with a 20 GW 10 km array operating at 1 $\mu$m wavelength. (b) Very thin sail (10 nm nominal) with 14 mg/m$^2$ areal density as an example of future (possible) much thinner sails with much lower areal density. Feasibility of these are not established yet. All arrays at 1 $\mu$m wavelength. Reflector density is 1.4 g/cc in these plots.}
        \label{fig:speedvsmass}
    \end{figure}

    In the future we can anticipate thinner sail materials as increased nano-fabrication of ultra thin films become available. Graphene may be one such material that might be possible to coat for good reflectivity. Here we show the effect of using reflectors of varying thickness. Since their speed with continued illumination scales as the sail thickness $h^{-1/4}$ for the optimized case of sail mass = payload mass we show a plot of the effect for a few selected sail thicknesses (Figure \ref{fig:speedvsmass}).
    \begin{equation}
        v_{\textrm{max-}\infty}=\bigg(\frac{P_0(1+\epsilon_r) d}{c\lambda\alpha}\bigg)^{1/2}(\xi h\rho m_0)^{-1/4}.
    \end{equation}
    
\subsection{Effect of Changing Array Wavelength}

    One can imagine using a variety of wavelengths for the system. If this is a ground based system the atmosphere effectively blocks use from about 1 mm to about 30 $\mu$m.  Here we show the effect of using systems of differing wavelengths. Since the speed with continued illumination scales as the $\lambda^{-1/2}$ we plot the ``speed multiplier'' relative to the baseline 1$\mu$m laser case where all other parameters for the array and payload remain fixed (Figure 20). This is true whether the payload is optimized (sail mass = payload mass) or not. To obtain the speed relative to the 1 $\mu$m laser case, multiply the speed by the ``speed multiplier.'' For example, at a wavelength of 1 cm (30 GHz) (about twice the highest frequency of the proposed SKA) the speed multiplier is $10^-2$ and hence the speed with a 1 cm wavelength array is 1\% of the speed of a 1 $\mu$m array, all other parameter being the same. As another example, at a wavelength of 3 cm (10 GHz) the speed multiplier is $5.8\times10^{-3}$.
    
    \begin{figure}
        \centering
        \hspace{-5mm}\includegraphics[width=0.45\textwidth]{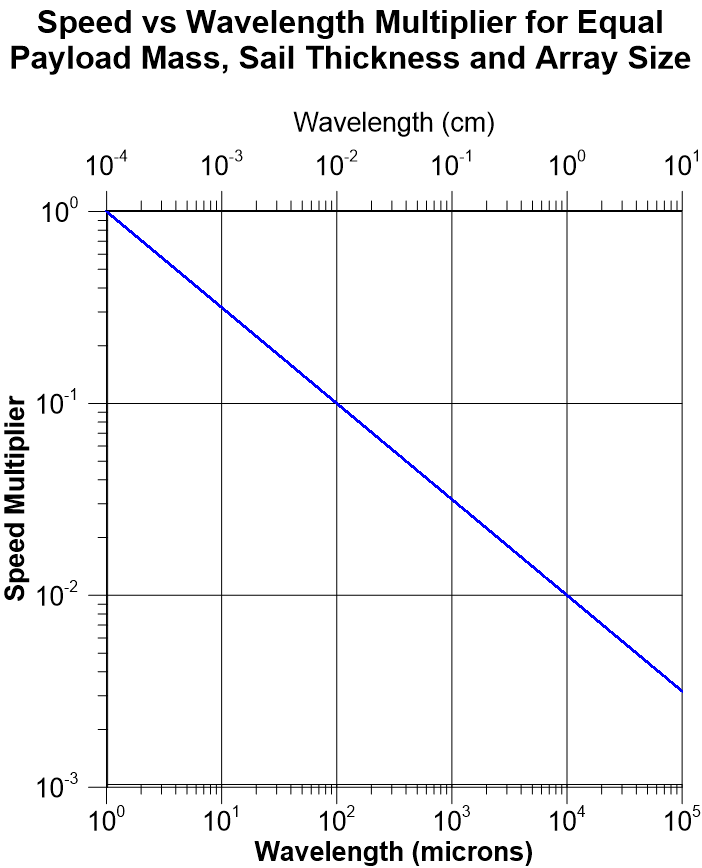}
        \caption{Speed multiplier vs array wavelength relative to the baseline operation at 1 $\mu$m. For operation at a different wavelength, with all other parameters the same, multiply the speed for the baseline 1 $\mu$m case by the speed multiplier to get the new speed at the different wavelength. If the new wavelength uses absorption on the sail (as is the case for some microwave systems) rather than reflection then the speed is reduced by an additional factor of $\sqrt{2}$.}
        \label{fig:wavelengthmultiplier}
    \end{figure}
    \begin{figure}
        \centering
        \includegraphics[width=0.45\textwidth]{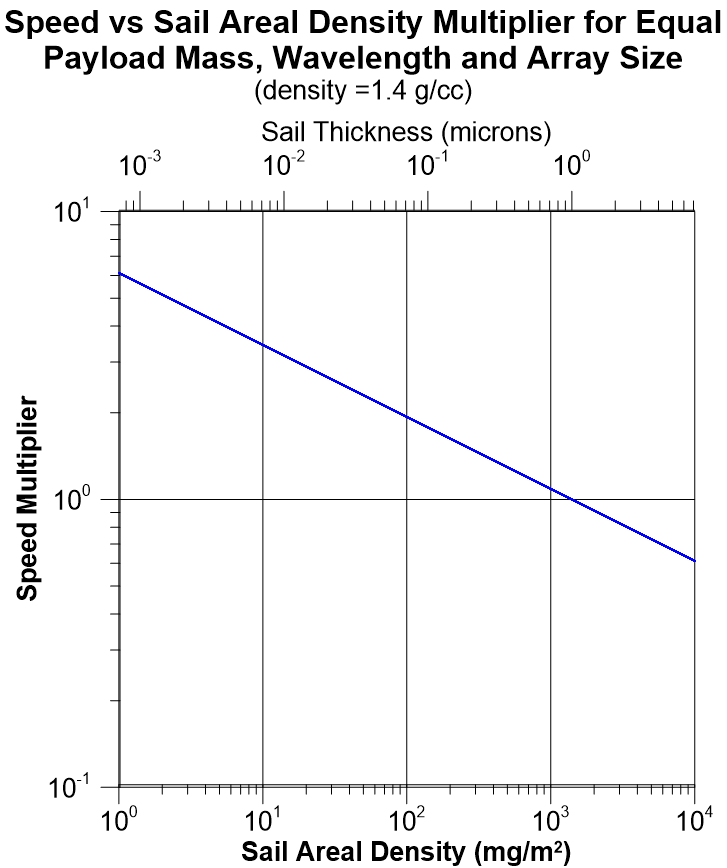}
        \caption{Speed multiplier vs sail areal density ($\sigma$) relative to the baseline operation of $\sigma=1.4$ g/m$^2$ = 1400 mg/m$^2$ (density $\sigma=1.4$ g/cc, sail thickness h=1 $\mu$m). For operation at a different areal sail density, with all other parameters the same, multiply the speed for the baseline case by the speed multiplier to get the new speed at the different areal density. The sail thickness is also given assuming a density of 1.4 g/cc. For example, the speed for an areal density of 14 mg/m$^2$ (corresponding to a thickness of 10 nm) would be higher by a factor of about 3.16 = speed multiplier. This would be an areal density 100 lower than a 1 micron thick film. As another example, for a film at 30 mg/m$^2$ (e.g. a carbon nanotube sheet) the speed multiplier would be 2.61.  If the sail is used in absorption rather than reflection the speed multiplier is reduced by $\sqrt{2}$. For example, in the case of a 30 mg/m$^2$ sail IF the sail were used as an absorber rather than a reflector then the effective speed multiplier would be $2.61/\sqrt{2}\sim1.85$.}
        \label{fig:speedvsdensity}
    \end{figure}
    
    Conversely to obtain the same speed as the 1$\mu$m case you would multiply the array power $P_0$ or the array size $d$ by the inverse square of the speed multiplier (which is the wavelength ratio). \textbf{Thus an array operating at 1 cm would need to have a power $10^4$ larger ($\sim1000$ TW = 1 PW) or a size $10^4$ larger ($10^5$ km or about 10 times the size of the Earth) compared to the baseline 1$\mu$m wavelength 10 km array laser case operating at 100 GW. Note some microwave system use absorption rather than reflection on the sail and in this case the speed would be lower by $\sqrt{2}\sim1.41$.}
    \begin{equation}
        v_{\textrm{max-}\infty}=\bigg(\frac{P_0(1+\epsilon_r) d}{c\lambda\alpha}\bigg)^{1/2}(\xi h\rho m_0)^{-1/4}.
    \end{equation}
    
\subsection{Effect of Changing Sail Areal Density}

    The speed is proportional to the sail areal density $\sigma=h\rho$ to the -1/4 power where $\sigma$ has units of kg/m$^2$, $h$ is the sail thickness and $\rho$ is the sail material average density. Hence $v\sim\sigma^{-1/4}$. Our baseline case is for $h=1$ $\mu$m and $\rho=1.4$ g/cc. This yields a sail areal surface density of $\sigma=1.4$ g/m$^2$. We plot a speed multiplier vs areal density (Figure 21) to make it easy to think about the effects of sail areal density on speed relative to the nominal case of $\sigma=1.4$ g/m$^2$ = 1400 mg/m$^2$.

\subsection{Solving for Array Size}

    We can solve for the array size $d$ needed for a given speed and spacecraft mass, power, etc. as follows:
    \begin{equation}
        v_\infty=\bigg(\frac{2P_0(1+\epsilon_r)dD}{c\lambda\alpha(\xi D^2 h\rho+m_0)}\bigg)^{1/2}
    \end{equation}
    
    \begin{equation}
        d=\frac{v_\infty^2 c\lambda\alpha(\xi D^2 h\rho+m_0)}{2P_0(1+\epsilon_r)D}\propto v_\infty^2.
    \end{equation}
    \\From this we see that $d \propto m_0$ IF $m_0 >> m_\textrm{sail} = D^2 h\rho$, but in order to get to speed $v_\infty$ we need to increase the sail size $D$ if the power is kept constant. This eventually casuses the sail mass to dominate and thus the assumption of $m_0>>m_\textrm{sail}$ becomes invalid. For the optimum (maximum) speed case when the sail mass = payload mass,
    \begin{equation}
        v_{\textrm{max-}\infty}=\bigg(\frac{P_0(1+\epsilon_r) d}{c\lambda\alpha}\bigg)^{1/2}(\xi h\rho m_0)^{-1/4}
    \end{equation}
    \begin{equation}
        P_0=v_{\textrm{max-}\infty}^2\frac{c\lambda\alpha}{d(1+\epsilon_r)}(\xi h\rho m_0)^{-1/4}.
    \end{equation}
    In this optimized case we see a different scaling relationship with $P_0 \propto m_0^{1/2}$. Thus the power for the optimum case scales as the (spacecraft mass)$^{1/2}$. This is good in the sense that increasing the mass by 100 ``only'' increases the power by 10 to get to the same speed with the same array size. However, it also means that the array size only shrinks by a factor of 10 if we reduce the mass by 100 for the same array size. Notice that the power is proportional to the wavelength $\lambda$ for the same speed, spacecraft mass, array size, etc.
    
\subsection{Trading Array Size and Power}

    If we fix the speed requirement for a given spacecraft mass, sail thickness, and density and fixed wavelength we can trade power $P_0$ for array size $d$ since it is the product that matters. For example if we want to shrink the size of the array by a factor of 10 we would have to increase the power by a factor of 10 to compensate.
    
\section{\label{sec:orbits}Orbital Trajectory Simulation}

\begin{figure*}
        \centering
        \begin{tabular}{c c}
             \hspace{-6mm}\includegraphics[width=0.45\textwidth]{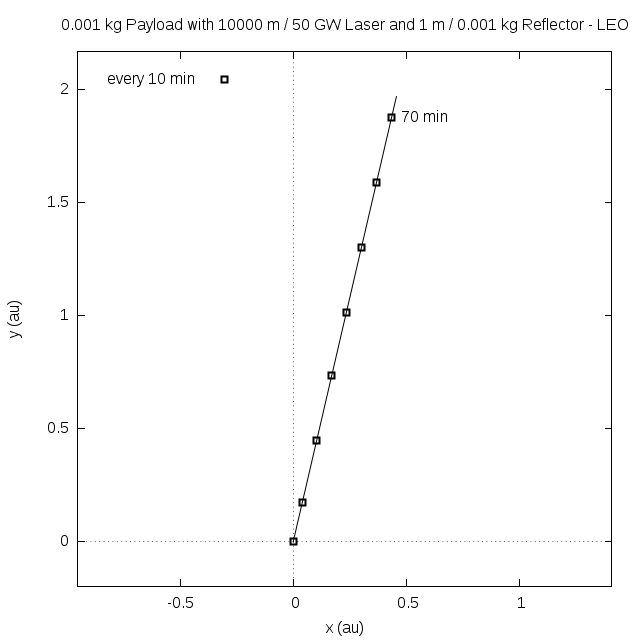} & 
             \includegraphics[width=0.325\textwidth]{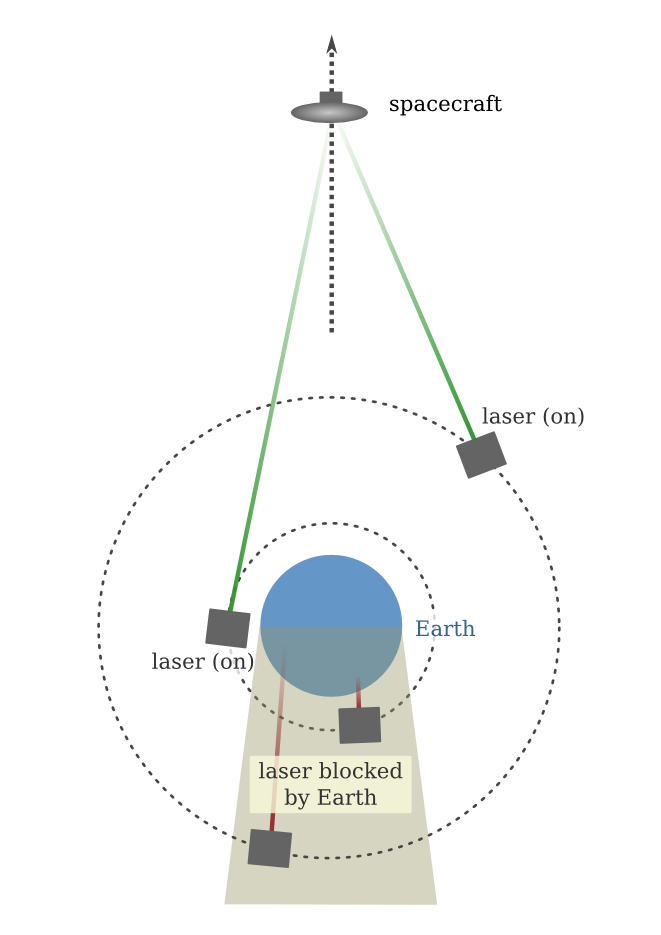} \\
             (a) & (b) \\
        \end{tabular}
        \caption{(a) x-y solar system trajectory coordinates every 10 minutes for numerical simulations of 1 g payload with Class 4 driver. (b) LEO and GEO based laser driver showing the blockage of the Earth during laser illumination of spacecraft.}
        \label{fig:trajectory}
    \end{figure*}

\begin{figure}
        \centering
        \includegraphics[width=0.45\textwidth]{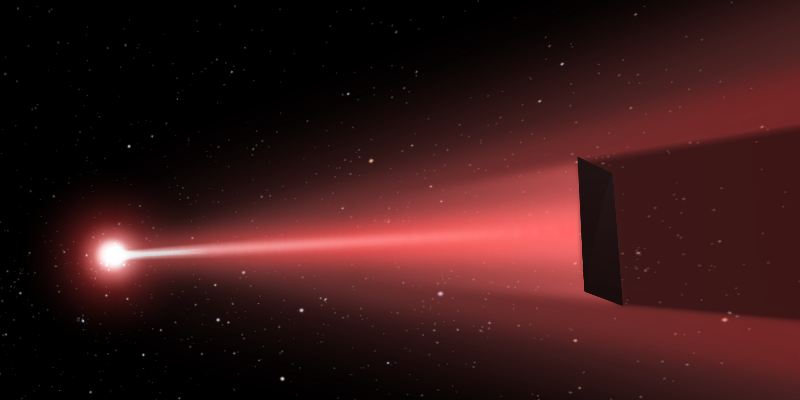}
        \caption{Artistic rendering of a laser driven reflector. Credit: Q. Zhang -- UC Santa Barbara.}
        \label{fig:deepin}
    \end{figure}

    We have worked out the orbital trajectories assuming a Sun synchronous LEO based DE-STAR (it can be placed in other orbits, or moon, etc.) to keep it fully illuminated except during eclipses. For the very low mass payloads the time to near maximum speed is so short that the spacecraft travels in nearly a straight trajectory as the acceleration time is small compared to a LEO orbital time while for the heavier payloads this is not true and the path is more complex  (Figures \ref{fig:trajectory} and \ref{fig:deepin}) \cite{Zhang2015}\cite{Zhang2015a}.
    
\subsection{Multiple Independent Re-Entry Vehicles Option}

    An option to be considered is to use a large number of wafer scale spacecraft contained in a larger (but still relatively low mass) ``mothership.'' Upon entry near the target star the mother ship would eject the wafers which would then interconnect with each other and the mothership via an optical link. The larger mothership would then transmit the collected data back to Earth. We envision several hundred ``WaferSats,'' each with their own power (RTG and PV) as well as their own optical communications and photon thruster attitude control. Each wafer has a mass of about 1 gram and hence a 1 kg mothership could carry perhaps 500 of these and disperse them in a roughly 1 AU spacing on a $20\times20$ array upon entry to allow a much more thorough exploration. Since command and control back to Earth is not feasible due to TOF issues the system would have to be autonomous. The disadvantage of this is the mothership has lower speed due to its larger mass.
    
\subsection{Braking to Enter Orbit on Arrival}

    A very difficult challenge is to slow the spacecraft to typical planetary orbital speeds to enable orbital capture once arriving. This task is difficult as the initial entry speeds are so high ($\sim$c) and the orbital speeds are so low ($\sim 10^{-4}$ c). Dissipating this much energy is challenging. We have considered using the star's photon pressure, the stellar wind (assuming it is like our own solar system), using the magnetic coupling to the exo-solar system plasma. None of these techniques appears to be obviously able to accomplish this task and much more work and simulation is needed. A simple fly-by mission is clearly the first type of mission to explore in any case to assess the environment in a given system to design (if possible) an optimized braking strategy. High mass payloads that travel within our solar system could be slowed by either a ping-pong system using a second laser at the destination (Mars in some very distant scenario for example) or by ejecting a reflector that is then used as a braking system (similar to thrust reversal on jets) but this only works if the payload is still within illumination range of the primary laser system. 
    
\section{\label{sec:communications}Communications}

    Another use of the DE-STAR system would be for long range interstellar communications to and from the spacecraft. This is a critical issue for long range interstellar probes in the future. Can we get high speed data back?
    
\subsection{DE-STAR to Spacecraft Data Rate}

    Consider an optical link calculation with DE-STAR 4 which emits about 50 GW at 1.06 $\mu$m or about $3\times10^{29}$ $\gamma$/s, with a divergence half angle of 
    \begin{equation}
        \theta\approx\frac{\lambda}{D}\approx10^{-10}\textrm{rad}.
    \end{equation}
    \\At a distance of $L = 1$ ly ($\sim10^{16}$ m) the spot size (diameter) is about $D_s\sim2\times10^6$ m. For the case of the 100 kg robotic craft and with a very conservative 10$\mu$m thick, 30 m diameter reflector this gives a spacecraft received photon rate of $3\times10^{29}\times(30/2\times10^6)^2\sim7\times10^{19}$ $\gamma$/s. If we assume it takes 40 photons per bit (which is very conservative) this yields data rate of about $2\times10^18$ bits/s, clearly an enormous rate. Below we use more optimistic values approaching 1 ph/bit. Less than 2 ph/bit has been achieved in long range (but not interstellar distances). See figure 24 for a summary of the 100 kg probe with a 30 m reflector case.
    
\subsection{Spacecraft to DE-STAR Data Rate}

\begin{figure}
        \centering
        \includegraphics[width=0.475\textwidth]{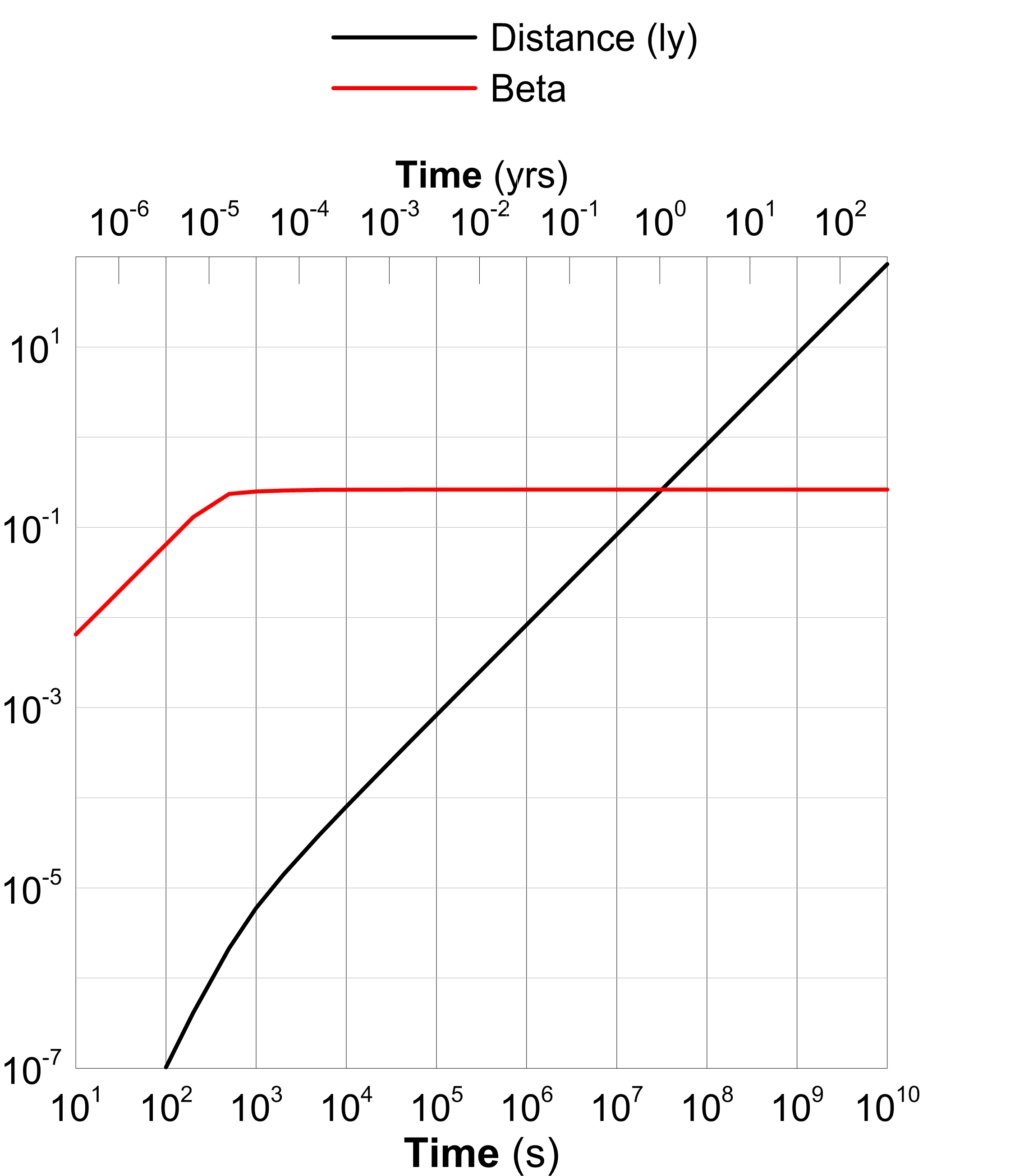}
        \caption{WaferSat case with same DE-STAR 4 driver (70 GW -- 10 km array) as in Figure \ref{fig:wavelengthmultiplier} but for wafer scale spacecraft driven by 1 meter sized reflector (1 micron thick), Wafer mass is 1 g and reflector mass is 1.4 g. Total mass is 2.4g. Parameters shown are relativistic factor $\beta$ and distance vs time. Note that Alpha Centauri (4.3 ly) is achieved in about 20 years with 10 ly achieved in about 45 years.}
        \label{fig:distancebeta}
    \end{figure}

    Assuming the same spacecraft has a 10 W transmitter onboard (an RTG for example) for an optical link at the same basic wavelength $\sim1.06$ $\mu$m (slightly different to allow full duplex communications if needed) and that it uses the same 30 m reflector as for the photon drive, but this time it uses it as the communications transmitter antenna (mirror). We do the same basic analysis as above.  10 W at 1.06 $\mu$m or about $5\times10^{19}$ $\gamma$/s, with a divergence half angle of
    \begin{equation}
        \theta\approx\frac{\lambda}{D}\approx3.5\times10^{-8}\textrm{rad}.
    \end{equation}
    At a distance of $L = 1$ ly ($\sim 10^{16}$ m) the spot size (diameter) is about $D_s\sim3.5\times10^8$ m. For the case of the 100 kg robotic craft and with a 30 m diameter reflector transmitting BACK to a DE-STAR 4 which acts as the receiver this gives a received (by the DE-STAR) photon rate of $5\times10^{19}\times(10^4/3.5\times10^8)^2\sim4\times10^{10}$ $\gamma$/s. If we assume it takes the same 40 photons per bit this yields a received (at Earth or wherever the DE-STAR system is located) data rate of about $1\times10^9$ bits/s or 1 Gbps. At the nearest star (Proxima Centauri) at a distance of about 4 ly the data rate at Earth from the spacecraft is about 70 Mbps. \textbf{Live streaming $>$HD video looks feasible all the way to our nearby interstellar neighbors IF we use the drive reflector for data transmission and IF we had this large a reflector.} A critical element in long range laser communications is the very narrow bandwidth nature of the laser. This fact greatly reduces the background ``noise'' received enabling a large SNR. This is summarized in Figure 24. For the wafer-scale case (below) the data rate is much lower due to the lower wafer transmit power and the smaller transmit optic size (Figures \ref{fig:distancebeta} and \ref{fig:class4parameters}). 
    
    \begin{figure}
        \centering
        \includegraphics[width=0.45\textwidth]{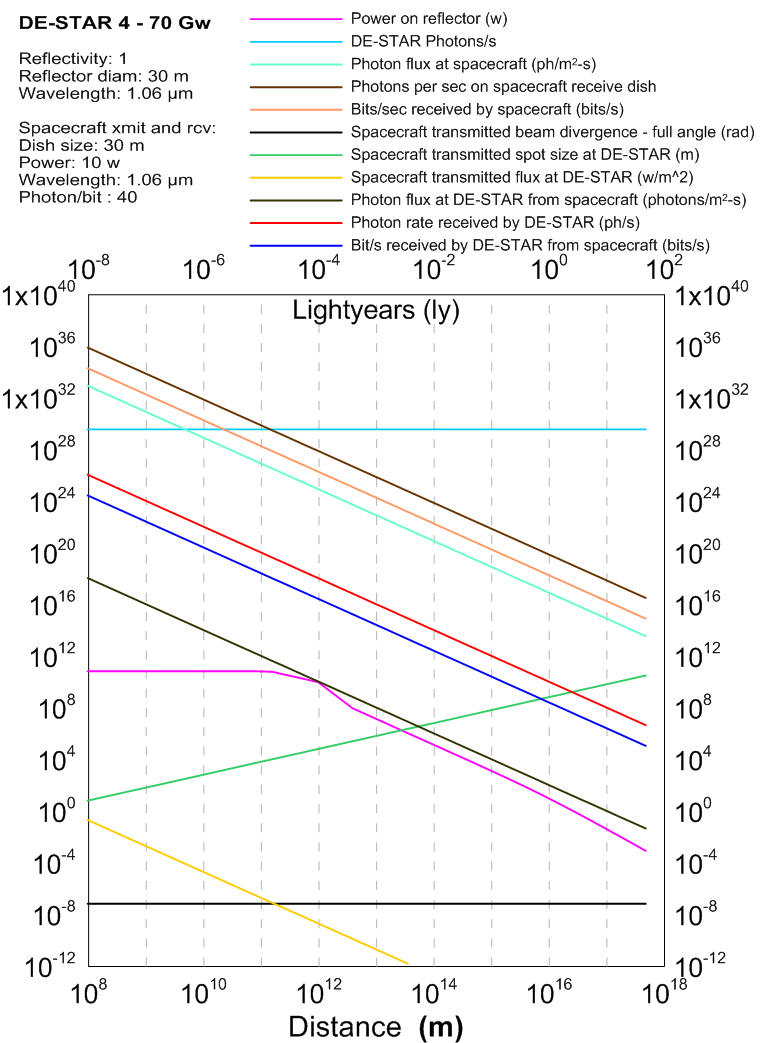}
        \caption{Flight parameters including data rates for a 100 kg spacecraft with a conservative 10 $\mu$m thick and small 30 m reflector made from the ``roll to roll'' multi-layer dielectric process as described above.  Laser driver is 10 km (DE-STAR 4) array with an optical output of 70 GW power. This is NOT an optimized design (sail mass does NOT equal spacecraft mass) and is shown as an example of one of many possible variations.}
        \label{fig:destar4}
    \end{figure}
    \begin{figure}
        \centering
        \includegraphics[width=0.45\textwidth]{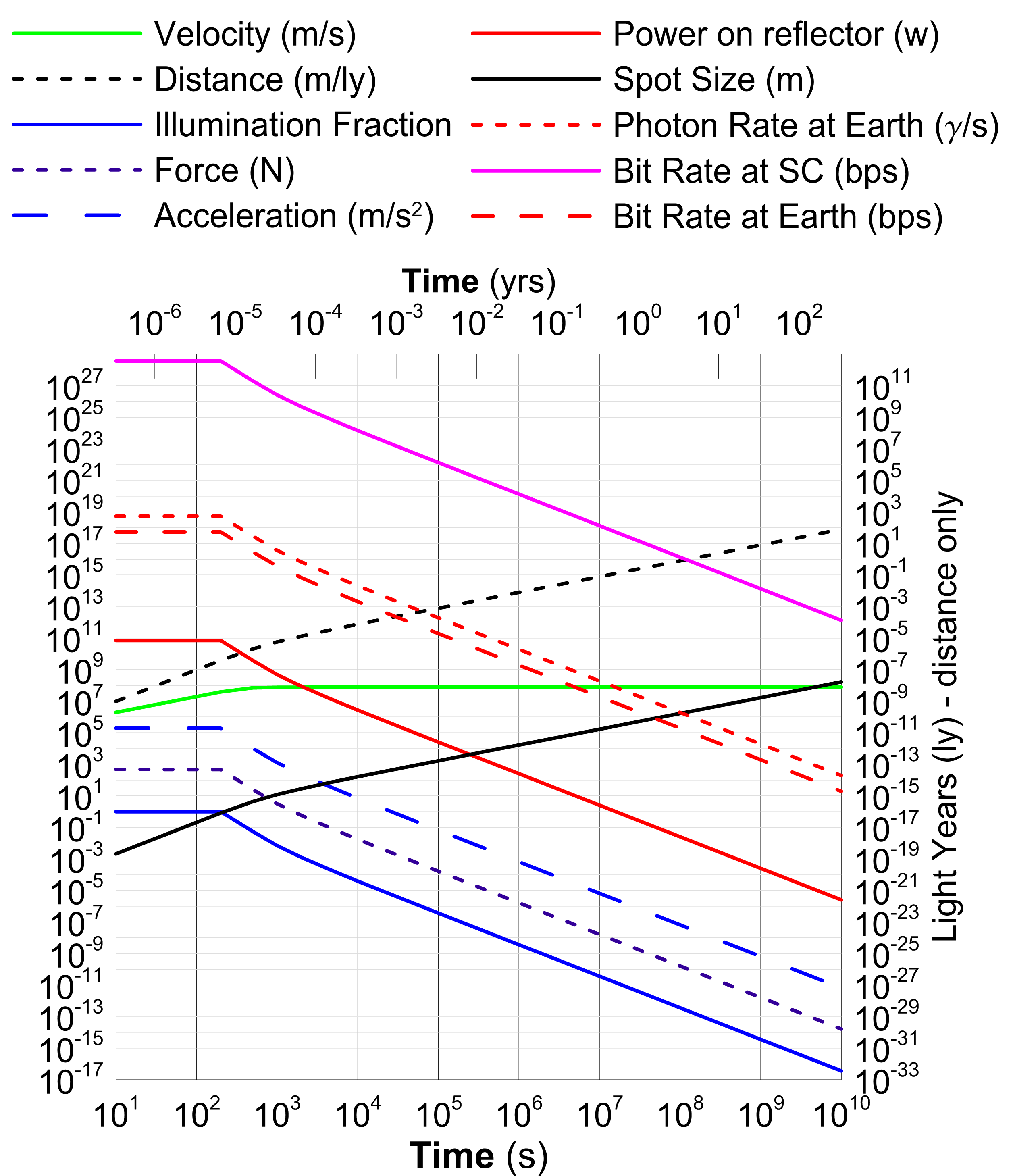}
        \caption{Parameters for full class 4 system with 1 g wafer SC and 1 m sail. Craft achieves 0.2 c in about 10 minutes (assuming an extended illumination) and takes about 20 years to get to Alpha Centauri. Communications rate assumes class 4 drive array is also used for reception with a 1 watt short burst from a 100 mm wafer SC. Here we use the 1 meter drive reflector as the transmit and receive optical system on the spacecraft. We also assume a photon/bit ratio near unity. In this case we get a data rate at Andromeda of about 65 kb/s. In the previous figure for the same wafer scale spacecraft the \textbf{only optical system on the spacecraft was the 100 mm wafer.} The data rate received at the Earth from Alpha Centauri is about 0.65 kb/s during the burst assuming we can use the DE-STAR 4 driver as the receiver and only the wafer itself for the transmission optic. The plot above shows a much more conservative photon/bit ratio of 40 while unity has been achieved but never over the extremely long distances discussed here.}
        \label{fig:class4parameters}
    \end{figure}
    
\subsection{Backgrounds}

    The relevant backgrounds at 1 $\mu$m wavelengths are optical emission from the telescope/array, zodiacal emission from our solar system dust both scattering sunlight and emitting greybody thermal radiation (Zodi), and the Cosmic Infrared Background (CIB). It is assumed that the latter is the sum of all unresolved galaxies in the universe in the field of view \cite{Brashears2015}\cite{Bible2013}. The Cosmic Microwave Background (CMB) is not relevant and light from our galaxy is relatively small unless looking directly at a star. If searches/communications are done inside our atmosphere then the situation is more complex due to emission from our atmosphere. 
    
\begin{figure}
        \centering
        \includegraphics[width=0.45\textwidth]{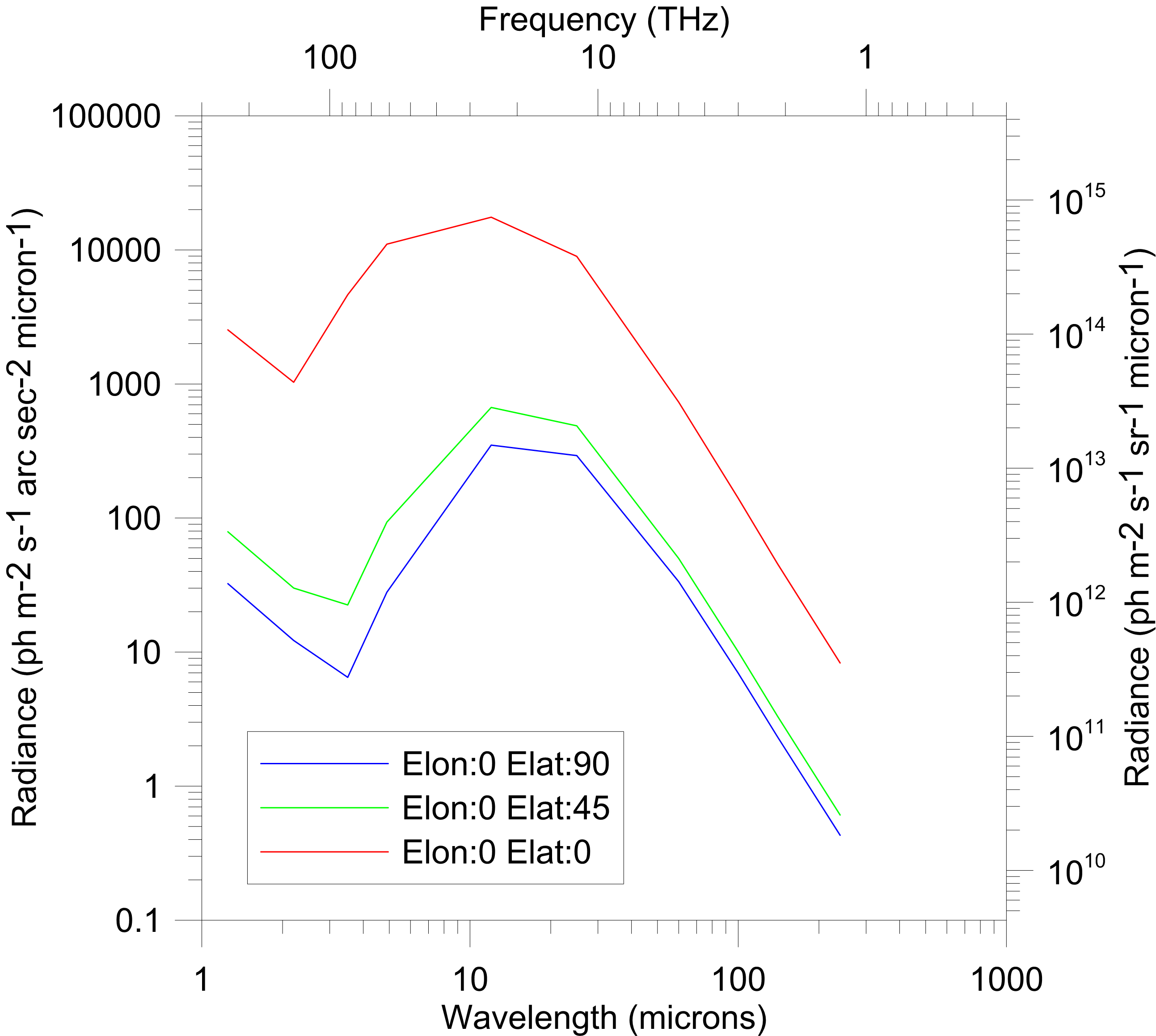}
        \caption{Zodiacal emission from the COBE DIRBE measurements showing day 100 of the COBE mission at observation from along the ecliptic plane to 45$^{\circ}$ to 90$^{\circ}$ (ecliptic pole). William Reach private communication (2012). Note the radiated peak from solar system dust around 15 $\mu$m due to the dust temperature being about 200 K and the scattered rise near 1 $\mu$m due to the Zodi dust scattered sunlight.}
        \label{fig:zodi}
    \end{figure}

\begin{figure}
        \centering
        \includegraphics[width=0.45\textwidth]{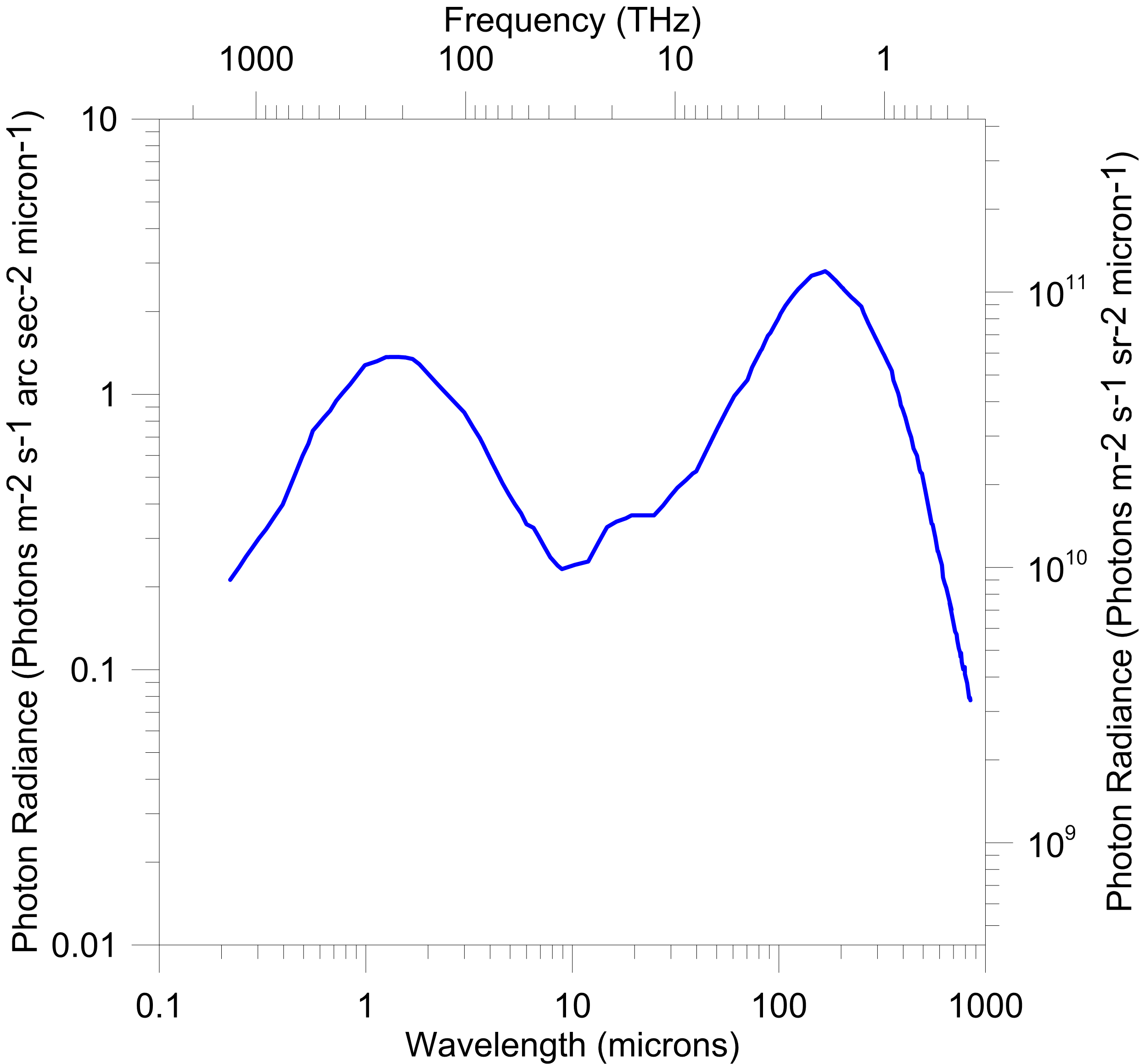}
        \caption{Cosmic Infrared Background (CIB) radiation from COBE DIRBE and subsequent measurements.}
        \label{fig:cib}
    \end{figure}

    The Zodiacal light is highly anisotropic and also time dependent as well as dependent on the location of the Earth in the orbit around the sun. We model this from the DIRBE instrument on COBE. The CIB is far more isotropic on modest angular scales and becomes largely point-like on very small scales. Again we model this from the DIRBE data on COBE and subsequent measurements. We also model the optics at various temperatures and the Earth's atmospheric emission for inside the atmosphere measurement but will focus here on orbital programs \cite{Denny2014}\cite{Suen2015}. The Zodi and CIB are shown in Figures \ref{fig:zodi} and \ref{fig:cib}. 
    
\subsection{The Phased Array Telescope --- Using DE Driver Array as a Communications Receiver}

\begin{figure}
        \centering
        \includegraphics[width=0.6\textwidth]{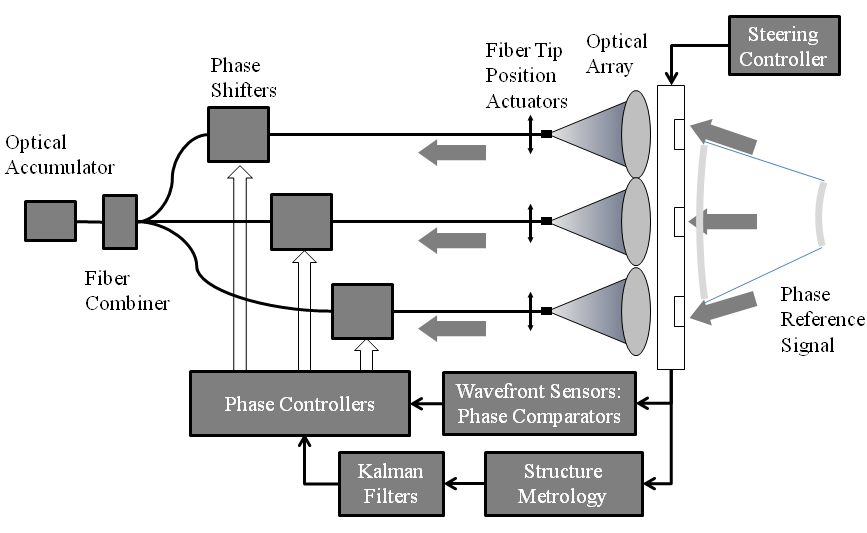}
        \caption{Phased array telescope (PAT) receive mode. The laser phased array is used in reverse as a phased array receiver. The system topology and elements are essentially the same in transmit (propulsion) and receive modes. The phase reference signal for the receive mode can be from the target star or from a beacon. Inside the atmosphere the same PAT mode is used for adaptive optics as every element is phase controlled.}
        \label{fig:arraytelescope}
    \end{figure}

    In addition to its use as the propulsion system for the payload, the main phased array is also used as the laser communications receiver (Figure \ref{fig:arraytelescope}). This is critical in reception of the long range data from the probes. In this case the system is essentially time reverse symmetric except that the main laser amplifiers used for drive are not on during reception due to possible large cross talk between transmit and receive modes. The current baseline approach is to add at least one and possibly a fiber bundle at each optical sub element to be used for reception. All the receive fibers are brought back to the receive detector which could be either an APD (avalanche photo diode array) or a superconducting sensor. Both can offer single photon counting capability. Since the transmit and receive fibers are slightly displaced spatially in the focal plane, we can either change the phase of the receive fibers (each receive fiber has a phase shifter) or use the fiber tip actuator to move the fiber over or act as an additional phase shifter to allow the beams to properly point at the spacecraft. In this mode the entire array acts as a phased array telescope (PAT). In the PAT mode the synthetic beam is basically equivalent to the transmit beam. Since the signal from the spacecraft is in a very narrow band we do not have to do absolute fiber optical length matching but only relative phase matching (i.e. modulo $2\pi$). This makes it much easier to ``phase up'' than white light broad band phase matching that requires optical line length matching. There is another possibility which is the light bucket telescope (LBT) mode. In the LBT mode there is no phase alignment needed during reception and all the optical sub elements act as independent telescope. The light from all the receive fibers is then combined before detection and each sub-element will have a larger beam pointing at the target. The advantage of the LBT approach is that it is much simpler as no phase alignment is needed. The disadvantage is that the background (noise) becomes much larger since the largely isotropic noise from the Zodi and CIB will be present at the same level in each sub-element fiber as the background light detected in each fiber, which is the product of the background brightness $B_\lambda$ and $A\Omega\beta$, where $A$ is the sub-element telescope area, $\Omega$ is the sub-element solid angle, and $\beta$ is the detection bandwidth. In general we want to match the laser communication linewidth to the detection bandwidth $\beta$ to minimize the background and thus maximize the SNR. Assuming a diffraction limited telescope for each sub-element we have $A\Omega=\lambda^2$, independent of the size of the sub-element. If there are $N$ sub elements in the system then the signal in each sub-element is reduced by $1/N$ compared to that received by the full system. When the entire telescope acts as a phased array, we have the same background in the total system as in one-sub element, namely 
    \begin{equation}
        B_\lambda A\Omega\beta=B_\lambda \lambda^2\beta,
    \end{equation}
   but the signal is increased by a factor of $N$ relative to that received by each sub-element. Since the signal in each fiber will add while the background noise from all the fibers will add in quadrature there is a signal to noise (SNR) advantage of between $N^{1/2}$ and $N$ for using the PAT mode vs the LBT mode. \textbf{The PAT is also a possible path forward to extremely large telescopes for astronomy and laser communications (optical DSN)  purposes.} While the bandwidth is relatively narrow, such a telescope topology offers a path for specialized spectroscopy systems such high redshift follow up to upcoming 30 m class telescopes as one example and could also possibly be used for exoplanet imaging where the phased array allows deep nulls in the response pattern to reduce the light from the parent star.
    
\subsection{Backgrounds for Full Phased Array vs a Non-Phased Array or Light Bucket Array}

    It is interesting to compare the background for a fully phased array (fully synthesized) vs a non-phased array or light bucket approach. As we discussed we can imagine we have $N$ sub elements in the array which has size $d$. Each sub element then has size $d/N^{1/2}$ and area
    \begin{equation}
        A_\textrm{sub}=\frac{d^2}{N}=\frac{A_\textrm{full}}{N},
    \end{equation}
    where $A_\textrm{full}$ is the full array area. Each sub element has beam size
    \begin{equation}
        \Theta_\textrm{sub}=\frac{2\lambda/d}{N^{1/2}}=\frac{2N^{1/2}\lambda}{d}=N^{1/2}\Theta_\textrm{full},
    \end{equation}
    where $\Theta_\textrm{full}$ is the whole synthesized (full) beam size. This is for a square array. For a circular array we replace $d$ with $d/\lambda$ with $\alpha\sim1.22$. There is also a packing fraction to be considered. For circular sub apertures the maximum packing fraction is $\pi/4$. The sub-element solid angle is
    \begin{equation}
        \Omega_\textrm{sub}=\Theta_\textrm{sub}^2=\frac{N4\lambda^2}{d^2}=N\Omega_\textrm{full}.
    \end{equation}
    
    Assuming that each sub-element is diffraction limited, then the entendue of each sub-element is $A_\textrm{sub}\Omega_\textrm{sub}=\lambda^2$ which is the same as the entendue of the full synthesized array since the isotropic backgrounds all contribute proportional to the entendue.  However, if we run the telescope as a non-phased array (light bucket) we have $N$ of these sub-elements and each contributes a background contribution the same as the fully synthesized array but there are $N$ of these sub-elements \textbf{so there is now $N$ times the background.} This has the effect of greatly reducing the overall SNR as mentioned above. For the case of a large class 4 array needed for relativistic propulsion $N$ can be quite large. If we assume that each sub-element is just large enough to fit into a current heavy lift launcher fairing (many sub-elements go into each launcher) then the max sub-element diameter is about 3-4 meters. This would yield about 10 m$^2$ per sub-element or $N=10^7$ for a class 4 system. This would increase the total background by $N$ or by $10^7$ which would be extremely damaging to the SNR for many applications and thus a fully synthesized receive mode is critical.
    
\subsection{Laser Communication Bandwidth}

    Lasers for communications purposes can have very narrow intrinsic linewidth but with data modulation the effective bandwidth is often dominated by the data bandwidth not the intrinsic laser linewidth. We will assume a nominal laser communications wavelength of near $\lambda=1.0$ $\mu$m ($\sim300$ THz or 1.2 eV/$\gamma$) to take advantage of the transmit array optics. Since we quote the various backgrounds in per unit wavelength whereas a more natural unit of laser line is in Hz it is convenient to convert between the two. This is simply done since the fractional wavelength bandwidth and the fractional frequency bandwidth are equal: $\Delta\lambda/\lambda=\Delta\nu/\nu$. The intrinsic laser linewidth can be extremely narrow with specialized lasers being under 1 Hz bandwidth. More common laser diodes can have intrinsic bandwidths of about 1 KHz but since the data modulation is often much larger than the intrinsic linewidth, the effective laser linewidth (including modulation) can exceed 10 GHz for high data rate laser communication systems. We will consider cases for data rates from 1 KHz to 10 GHz or $10^3$ - $10^{10}$ Hz which corresponds to an equivalent wavelength bandwidth of $3\times10^{-12}$ to $3\times10^{-5}$ $\mu$m. This corresponds to a spectrometer resolving power $R=\lambda/\Delta\lambda$ at $\lambda=1.0$ $\mu$m of $3\times10^{11}$ to $3\times10^4$.
    
\subsection{Doppler Shifts}
    
    There are a variety of Doppler shifts that are significant relative to the intrinsic laser linewidth and even for data modulation. We solve for the general case at arbitrary speed. For an emitter of frequency $\nu_0$ in its rest frame moving away from an observer at $\beta$ ($=v/c$) the observer will measure a frequency
    \begin{equation}
        \nu=\gamma\nu_0(1-\beta)=\nu_0[(1-\beta)/(1+\beta)]^{1/2},
    \end{equation}
    or 
    \begin{equation}
        \lambda=\lambda_0[\gamma(1-\beta)]^{-1}=\lambda_0[(1+\beta)/(1-\beta)]^{1/2}.
    \end{equation}
    For low speeds ($\beta<<1$) $\gamma$ is very close to unity so the Doppler shifted wavelength is approximately $\lambda=\lambda_0(1+\beta)$ or $\Delta\lambda=\lambda-\lambda_0=\lambda_0\beta$. Some examples are instructive. At the equator the Earth's rotation speed is about 460 m/s or a $\beta\sim1.5\times10^{-6}$. In LEO the orbital speed is about 7.8 km/s or $\beta\sim2.6\times10^{-5}$. The Earth's orbital speed around the Sun is about 30 km/s or $\beta\sim1\times10^{-4}$. Note that all of these are larger than the typical laser linewidth and thus need to be considered in any truly matched filter and then a matched filter needs to be adaptive for Doppler effects, let alone the speed of the spacecraft that are discussed here where $\beta\sim0.2$ for the small wafer-scale spacecraft.
    
\subsection{Optimized Receiver Filtering}

    Ideally we would match the receiver filter to pass the laser communications signal but block as much of the background as possible. As discussed this is ideally an adaptive filter that adjusts to the spacecraft speed and any local motions, especially at low data rates where the effective laser linewidth can be quite small (KHz range for example). This presents very significant challenges to filter designs. For example narrow band multilayer dielectric interference filters are difficult to make with an $R$ greater than $10^3$ and spectrometers typically have $R$ less than $10^5$. Specialized resonant ring filters can have $R\sim10^7$ but the best solution would likely be a super heterodyne detector with a wide enough IF bandwidth or tunable local oscillator to track the laser line and an IF that can adjust to varying data rates. This is not trivial to do but is becoming feasible.
    
\subsection{Laser Communications Flux}

    To compute the SNR for the laser communications system we need to know the signal level at the Earth from the distant spacecraft. The waferscale case is by far the most difficult. At a distance $L$ the spot size at the Earth for a square wafer of size $D$ is $2L\lambda/D$ assuming a diffraction limited system with a square aperture. For a circular aperture we replace $D$ with $D/\alpha$ with $\alpha=1.22$. In practice there will be an efficiency factor that takes into account to optical efficiency and side lobes (similar to a Strehl ratio). For a laser communications power of $P_{lc}$ we get a flux at the Earth of 
    \begin{equation}
        F_{lc}=\frac{P_{lc} D^2}{4L^2 \lambda^2}.
    \end{equation}
    For the case of the burst mode wafer with $P_{lc}=1$ watt ($\sim5\times10^{18}$ $\gamma$/s for $\lambda=1.0$ $\mu$m) with $D=0.1$m (10 cm wafer) and a distance of 1 ly ($\sim10^{16}$ m) we get $F_{lc}$($L=1$ ly)$\sim1.3\times10^{-4}$ $\gamma$/s-m$^2$. This seems like a small flux but if we assume we can use the whole laser drive array as the receiver we get a collecting area $A$ for a class 4 system of $10^8$ m$^2$ and then total collected signal is then $AF_{lc}\sim1.3\times10^4$ $\gamma$/s again at 1 ly. At $\alpha$ Centauri $L\sim4.4$ ly ($4.4\times10^{16}$ m) and the signal drops by $(4.4)^2\sim20$ and hence the signal is then about 650  $\gamma$/s which would only enable an average data rate of less than 1 kbs during the burst assuming a photon/bit ratio near unity (currently achieved in much shorter range free space laser communications). While a higher power burst (say 10 W) would increase the burst rate it would not increase the average data rate. Since the 1 watt burst fraction is only 0.2\% of the time (limited by the onboard RTG power) the effective ``total average data rate'' would only be about 1 bps but the data transmission only needs to be during mission critical phases, such as image transmission. As an example, a modest quality image would be (assuming optimized compression) around 0.1-1 MB. At 0.65 kbs (1 ph/bit) for the 1 watt with 10 cm optics case, a 0.1 MB (800 kb) image would take about 1200 s and use (at 50\% eff) about 2400 J of energy for a 1 watt burst mode. If we could achieve onboard energy storage densities of current Li batteries ($\sim$0.2 W-hr/g $\sim720$ J/g), but suitable for long term and low temperature use here it would require about 3 g to store 2400 J. We would need to improve energy storage density by a factor of at least 10 to be useful for this purpose. \textbf{Using the proposed wafer scale RTG the 2400 J data burst could only be done about once per week.} A simple alternative would be to burst less contiguous data more frequently as this requires less onboard energy storage. It is important to optimize the burst time width given the data rate even with a fixed burst fraction. Hence bursting 1/500 s every second at 650 $\gamma$/s received is probably not the best way to optimize the SNR. Bursting for 1 second every 500 secondd is probably more efficient for data collection SNR but requires more energy storage onboard. An optimization of burst power, energy storage, and SNR needs to be done. The waferscale case is clearly the most challenging. Since the wafer size (and reflector size) vs payload mass ($m_0$) scales as $D^2\sim m_0$ and \textbf{hence the flux $F_{lc}=P_{lc}D^2/(4L^2\lambda^2)$ and data rate at the Earth scales as $P_{lc}D^2$}. For a circular aperture replace $D$ by $D/\alpha$ where $\alpha=1.22$. \textbf{$P_{lc}$ is linear in payload mass $m_0$}, assuming a fixed fraction of the payload is in the onboard power (RTG mass) AND \textbf{$D^2$ is linear in payload mass so the data rate is proportional to the SQUARE of the payload mass} in general. As the speed only scales as $m_0^{-1/4}$ there is relatively little speed penalty and hence time to target with increased payload mass.  IF we could use the actual reflector as a part of the optical communication link we could dramatically increase the data rate, which for the case of the small 10 cm wafer would be a factor of 100 increase in data rate. As another example a 1 kg payload (with the same fixed fraction of RTG power) would be about 5.6 times slower in speed but would have a data rate that could be $10^6$ higher. This gives us a large amount of flexibility in mission designs with a tradeoff between mass, speed, system capability, and data rate.
    
    For the receive side (laser drive array) in either the PAT or LBT modes we should be able to (ideally) get this level of signal. The difference between the PAT and LBT is not the collecting area but the effective solid angle which then affects the background and hence the SNR.
    
\subsection{Optics Background}

    The warm optics of the array radiate into the fibers and the fibers themselves radiate and this constitutes a background. Since the plan is to use the same array elements used to transmit the optical emission of the laser illuminator, it is necessary to compute the optical emission rate into the detector.  The optics emission can be modeled as a grey body of temperature $T$ with emission rate
    \begin{equation}
        B_\lambda=\frac{2hc^2}{\lambda^5(e^{hc/\lambda kT}-1)},
    \end{equation}
    where $B_\lambda$ (W/m$^2\cdot$sr$\cdot$m) is the Planck function brightness of a blackbody. This needs to be modified by the emissivity $\epsilon_\lambda$ of the optical system. For the moment we assume the worst case of a blackbody whereas typical optical systems will have emissivities closer to 0.1. The optics are assumed to be at roughly 300 K for simplicity (this could be changed in some scenarios), giving a brightness of about $1\times10^6$ ph/s$\cdot$m$^2\cdot$sr$\cdot\mu$m for unity emissivity at the baseline wavelength of 1.0 $\mu$m. Under the assumption of a diffraction limited system, the entendue of the optics is such that $A\Omega=\lambda^2$ (single mode) $\sim10^{-12}$ m$^2\cdot$sr where $A$ is the effective receiving area and $\Omega$ is the received solid angle. The bandwidth of reception must also be included. Here a matched filter spectrometer or heterodyning is assumed  with a bandwidth equal to the laser linewidth and modulation bandwidth and to match Doppler effects from both the spacecraft motion as well as the motion of the receiver around the Earth, relative to the Sun etc. This is typically $10^3$ - $10^{10}$ Hz or approximately $3\times10^{-12}$ to $3\times10^{-5}$ $\mu$m.  The total per sub-element is thus an emission of about $3\times10^{-18}$ to $3\times10^{-11}$ ph/s. Comparing the optics emission of $1\times10^6$ ph/s$\cdot$m$^2\cdot$sr$\cdot\mu$m at 1.0 $\mu$m to the CIB and Zodiacal light shows the CIB and Zodiacal light are both much larger than the optics emission.  For comparison, note that when looking directly at the Sun the brightness of the solar surface is $\sim5\times10^{25}$ ph/s$\cdot$m$^2\cdot$sr$\cdot\mu$m at 1.0 $\mu$m.  Assuming a diffraction-limited system, the resulting photon rate would be about $2\times10^1$ to $2\times10^8$ ph/s for laser (receiver) bandwidths from $10^3$ to $10^{10}$ Hz ($3\times10^{-12}$ to $3\times10^{-5}$ $\mu$m) as above when looking at the Sun with the beam contained completely within the Sun as one example.
    
\subsection{Zodiacal Light Background}

    As shown above the Zodiacal light background has a brightness at 1.0 $\mu$m of approximately $2\times10^{12}$ ph/s$\cdot$m$^2\cdot$sr$\cdot\mu$m (90 -- perpendicular to ecliptic plane), $4\times10^{12}$ ph/s$\cdot$m$^2\cdot$sr$\cdot\mu$m (45 degrees to ecliptic) and $2\times10^{14}$ ph/s$\cdot$m$^2\cdot$sr$\cdot\mu$m (0 -- in ecliptic plane). This gives a diffraction-limited system resultant photon rate (independent of aperture size), assuming $\lambda=1.0$ $\mu$m, of about $6\times10^{-12}$ to $6\times10^{-5}$ ph/s (90 -- perpendicular to ecliptic plane), $1\times10^{-11}$ to $1\times10^{-4}$ ph/s (45 degrees to ecliptic) and $6\times10^{-10}$ to $6\times10^{-3}$ ph/s (0 -- in ecliptic plane) for receiver bandwidths from $10^3$ to $10^{10}$ Hz.
    
\subsection{CIB Background}

    As shown above the CIB has a brightness at 1.0 $\mu$m of approximately $2\times10^{10}$ ph/s$\cdot$m$^2\cdot$sr$\cdot\mu$m. This gives a diffraction-limited system resultant photon rate (independent of aperture size), assuming $\lambda=1.0$ $\mu$m, of about $6\times10^{-14}$ to $6\times10^{-7}$ ph/s for laser (receiver) bandwidths from $10^3$ to $10^{10}$ Hz.
    
\subsection{Stellar Target Background}

    The planetary system we may wish to visit will generally have a close star and its background needs to be considered. For a class 4 system, with a $d=10$ km baseline, the beam divergence is $2\lambda/d\sim2\times10^{-10}$ radians (full width) at $\lambda=1.06$ $\mu$m. At $\alpha$ Centauri, $L\sim4.4$ ly ($4.4\times10^{16}$ m), the fully synthesized (PAT mode) receive spot size would be $2L\lambda/d\sim10^7$ m or about that of the radius of the Earth. Thus the synthesized receiver beam size would be much smaller than the star. The question of the host starlight in the receive beam is one of the side lobe response. As discussed above for the example of the Sun, assuming a diffraction-limited system (fully synthesized beam), the resulting photon rate would be about $2\times10^1$ to $2\times10^8$ ph/s for laser (receiver) bandwidths from $10^3$ to $10^{10}$ Hz when looking directly at the host star, assuming it is like our Sun and having the beam contained completely within the star. Since the main beam in this example is small even compared to the size of our Sun (or the host star) and since the host star signal is potentially quite large compared to the thermal, Zodiac, and CIB backgrounds, care needs to be taken to insure that the host star background does not dominate the SNR budget. The fully synthesized beam for a class 4 system has a spot size at $\alpha$ Centauri of about $10^7$ m or about $7\times10^{-5}$ AU. Thus at 1 AU from the host star we are about 15,000 beams away from the host star. Thus we would be firmly into the far side lobes for such a beam and the rejection at 15,000 main beams can be quite large (typ $>30$ dB rejection). At low data rates (say 1 KHz) \textbf{even if the main beam were pointed directly at the host star} the data rate would only be about 20 ph/s while even the wafer-scale spacecraft would have a photon rate of about $10^3$ ph/s AND we would NOT be pointed in general at the host star.
    
\section{\label{sec:SETI}Implications of the Technology for SETI}

    \begin{figure*}
        \centering
        \begin{tabular}{c c}
            \hspace{-10mm}\includegraphics[width=0.45\textwidth]{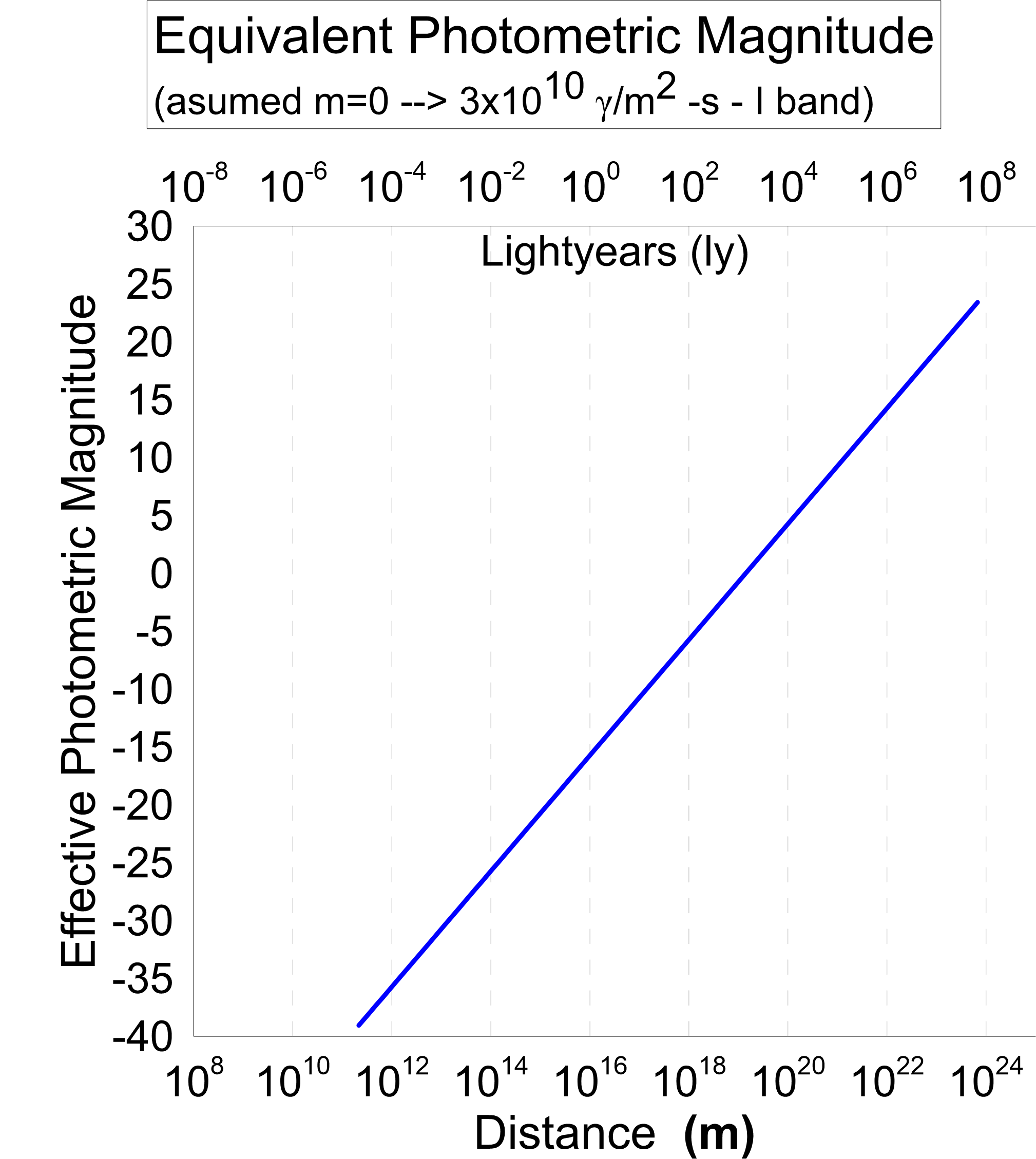} & \includegraphics[width=0.425\textwidth]{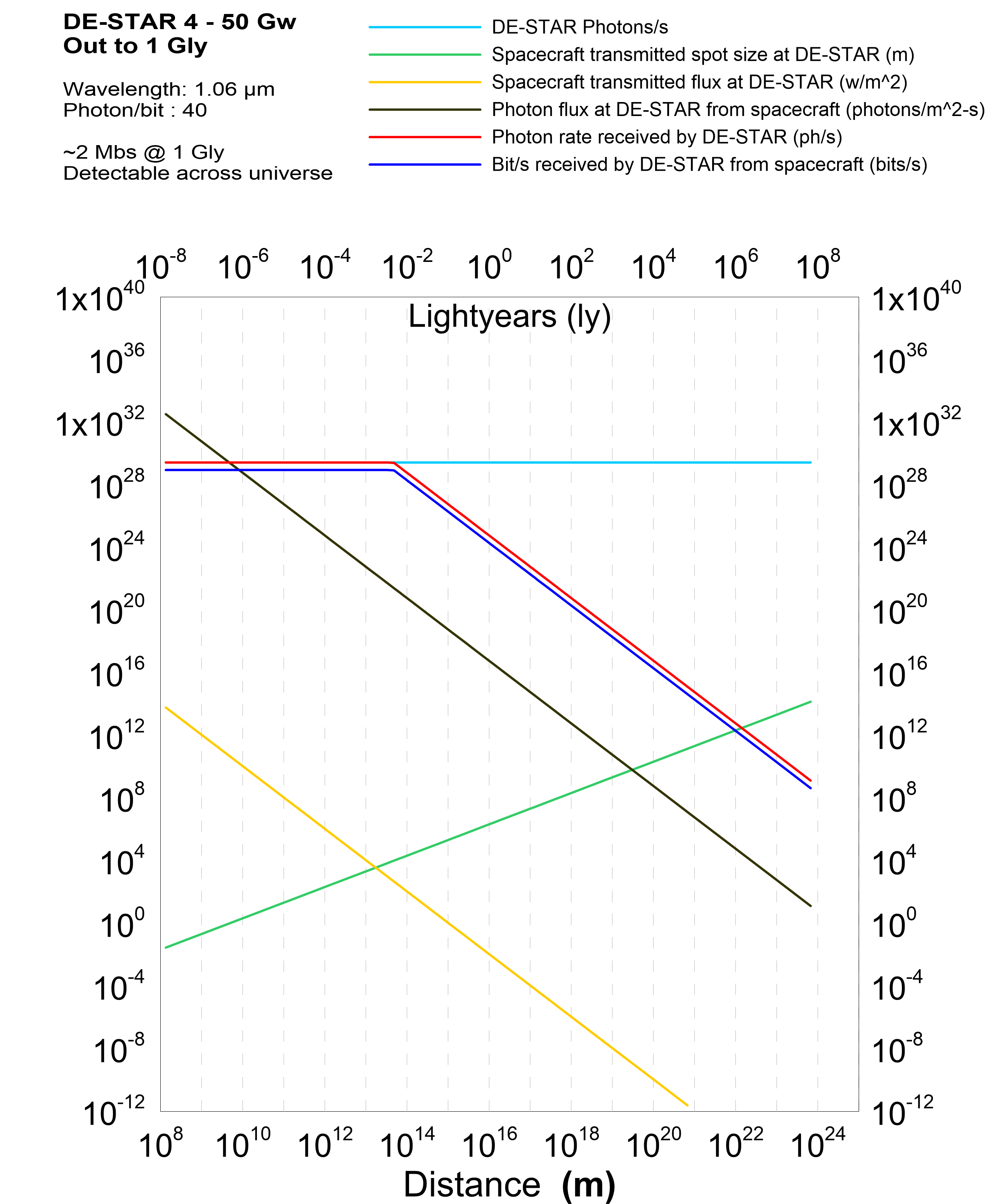} \\
             (a) & (b) \\ 
        \end{tabular}
        \caption{(a) Equivalent I band photometric magnitude of DE-STAR laser driver as observed vs distance of observation. Even at 1 Gly the laser driver would be visible in a modest telescope by another civilization. Note that this is the equivalent photometric magnitude which uses a broadband filter ($R\sim3$) and that an optimally tuned filter would have vastly better SNR as it would exclude most of  the background light \cite{Lubin2016}. We assume a worst case of an I band photometric filter. Even with this extremely pessimistic design the laser driver is visible across the entire horizon. (b) Photon and data rate for two civilizations with comparable DE-STAR 4 systems facing each other with an optimally tuned filter (resolution bandwidth to laser to minimize background) where each unit is used both for transmission and reception. Even at 1 Gly distances data rates are high.}
        \label{fig:datarates}
    \end{figure*}

    The same laser used to drive the probe can also be used as a beacon to ``signal'' other planetary systems as well as establish extremely long range ``communication'' systems with data delays modulo the speed of light. We discuss this in detail in detail in \cite{Lubin2015DirectedCommunications}\cite{Lubin2016}. The implications for SETI searches are quite profound. A similarly ``advanced'' civilization like ours would be visible across the entire visible universe (horizon). This implies that optical SETI searches can not only search nearby planetary systems but could search the entire universe for similar or more advanced civilizations. If the current Kepler statistics ($\sim1$ planet/star) on the abundance of suitable planets is scalable to the entire universe, this would imply of order $10^{22}$ planets within our horizon. We show in \cite{Lubin2016} that even ground based searches using modest existing apertures can detect civilizations with technology at our current technology level (though not yet deployed) at relatively high redshifts (Figure \ref{fig:datarates}).  Pondering the number of possible planets this allows us to search for has profound implications \cite{Lubin2016}.
    
\section{\label{sec:dust}Interstellar and Interplanetary Dust Impacts}

    The issue of dust grain and gas (molecular and atomic) as well as electron and proton impacts with a relativistic craft are important. We will consider interstellar and interplanetary impacts separately. The latter is only important during the acceleration phase but may also be indicative of the dust impacts upon arrival at a distant solar system, though the interplanetary dust around other stars is poorly understood currently. We distinguish the Interstellar Medium (ISM) from the Interplanetary Medium (IPM) though the distinction is complicated on length scales that are appropriate here since the nearest stars are not that far out of the influence of our own solar system in some cases and the merger of the two (ISM and IPM) is spatially complex \cite{Poppe2016}. The distribution of grain sizes is not well understood in detail and represents a challenge for us, but also an opportunity for measurements as we exit the solar system. Further refinement is needed in this area and one which we are actively pursuing. The estimates below are just that. We have tried to err on the side of being conservative, though more data is needed to constrain the grain distribution.
    
\subsection{Interstellar Gas and Dust}

    On average the ISM is about 99\% gas and 1\% dust by mass with the gas consisting primarily of Hydrogen ($\#>90$\%) and Helium ($\#\sim9$\%) and about 0.1\% higher $Z$. Both the dust and the atomic and molecular impacts will erode the sail and spacecraft.
    
\subsection{Interstellar Gas Impacts}

    The gas density and ionization varies significantly depending on the ISM region (warm, cold, cool, H-II, etc.). If we use a nominal value for H density of 1/cm$^3$ then we would encounter approximately $4\times10^{18}$/cm$^2$ over a journey to $\alpha$ Centauri. At a speed of 0.2c this gives a flux of about $6\times10^7$ p/s$\cdot$cm$^2$ with an energy of 19 MeV per H atom/proton or $1.2\times10^7$ J/cm$^2$ or about 20 mW/cm$^2$. The penetration depth of 19 Mev protons is about 0.4 cm water equivalent or a few mm in Be or Si. As discussed below if we face the wafer ``edge on'' with a buffer layer around the edge of the wafer we can mitigate the H/proton hits. If we are ``face on'' the wafer is so thin (can be made $<100$ $\mu$m thick and recent work has resulted in 1 $\mu$m thickness in critical regions) that the protons largely pass through the wafer without depositing much of their energy, though interactions will still occur and we prefer ``edge on.'' For the higher $Z$ species the range (for the same speed) is
    \begin{equation}
        \frac{R_1}{R_2}=\bigg(\frac{Z_2}{Z_1}\bigg)^2\frac{M_1}{M_2}.
    \end{equation}
    Thus, helium with $Z=2$ and 4 times the mass has the same range as a proton. Gas impacts, which are highly anisotropic both spatially and energetically due to spacecraft motion, are much higher than the cosmic ray impacts. Cosmic ray impacts will largely irradiate the bulk of the wafer while gas impacts will predominantly irradiate the forward facing spacecraft edge and are easier to mitigate.
    
\subsection{Interstellar Dust}

    Interstellar dust is estimated to exist over a very large range of sizes from a few molecules to around 500 nm in size. A small fraction is thought to consist of larger refractory material that condensed as the material left by stars. The cumulative dust density in the local interstellar medium of the Local Bubble is estimated to be approximately $10^{-6}$ grains/m$^3$ for sizes down to 10 nm. For a dust grain $\sim100$ nm in size the mass is $10^{-18}$ to $10^{-17}$ kg. The approximate cumulative density for grains larger than 100 nm is about $\textrm{few}\times10^{-8}$ grains/m$^3$ while for grains larger than 1 $\mu$m the cumulative density is about $10^{-9}$ grains/m$^3$ \cite{Poppe2016}. Larger grains are much less prevalent.
    
    The total number of dust grain gits during the hourney on the spacecraft is given by
    \begin{equation}
        N=n\sigma L,
    \end{equation}
    where $n$ is the number density per m$^3$, $\sigma$ is the cross section of the craft (this dominates over the small dust/gas cross section), and $L$ is the journey length in meters. Assuming $n(>1\mu\textrm{m})=10^{-9}$ dust grains/m$^3$ and $L=4\times10^{16}$ m ($\sim4$ ly --- distance to $\alpha$ Centauri), $nL=4\times10^7$ grains/m$^2$ during the journey. This is a large number and thus we see that the number of grain hits during an interstellar journey is not small. However, we may have some control over the spacecraft effective cross section $\sigma$ as well as shielding if needed.
    
    The cross section can be controlled to some extent. We assume an effective $10\times10$ cm spacecraft. If we orient the spacecraft ``face on'' (worst case), then $\sigma\sim10^{-2}$ m$^2$ for a 10 cm wafer. If we orient the spacecraft ``edge on'' with wafer thickness of 100 microns (default), then $\sigma=10^{-5}$ m$^2$. IF we use a long and thin spacecraft (difficult, but possible) with a 1o micron thick wire, then $\sigma=10^{-10}$ m$^2$. Taking these cross sections we can compute the number of hits ($N>1$ $\mu$m) during the journey:
    \begin{align}
    \begin{split}
        \textrm{Face on:}& \hspace{2mm} N=4\times10^5\\
        \textrm{Edge on:}& \hspace{2mm} N=400\\
        \textrm{Long thin rod:}& \hspace{2mm} N=4\times10^{-3}.
    \end{split}
    \end{align}
    The uncertainties are large and the factor of ``4'' is meaningless here. We want to get a rough order of magnitude. One question to be answered of course is ``what is the consequence of a dust grain hit?'' This is one of the items to test and assess as a part of the roadmap. With such high speeds the transit time in the wafer would be short compared to the lattice (phonon) time scales and a punch through may be more likely and less damaging. This is an area where more simulations are needed. No relativistic dust grain accelerators exist for testing, unfortunately.
    
\subsection{Impact Energy}

    The energy per impact for $v=0.3c$ ($10^8$ m/s) for a mass $m_\textrm{grain}=10^{-17}$ kg dust grain ($\sim0.1-1$ micron size depending on structure) is
    \begin{equation}
        \textrm{KE}_I=\frac{1}{2}m_\textrm{grain}v^2\sim0.1\textrm{J}.
    \end{equation}
    The total energy deposition of all grain impact is
    \begin{equation}
        \textrm{KE}_T=N \textrm{KE}_I.
    \end{equation}
    The worse case is the face on case where $N=4\times10^5$. In this case,
    \begin{equation}
        \textrm{KE}_T=4\times10^5\times0.1\textrm{J}=4\times10^4\textrm{J}.
    \end{equation}
    Compare this to the total KE of the wafer spacecraft:
    \begin{align}
    \begin{split}
        \textrm{KE}_{sc}=\frac{1}{2}m_\textrm{wafer}v^2\sim&10^{-3}(\textrm{kg})\times(10^8(\textrm{m/s}))^2\\
        =&10^{13}\textrm{J}.
    \end{split}
    \end{align}
    Thus the ratio of the total dust grain KE to the spacecraft KE is
    \begin{equation}
        \frac{\textrm{KE}_T}{\textrm{KE}_{sc}}\sim\frac{4\times10^4}{10^{13}}=4\times10^{-9}.
    \end{equation}
    Hence the total energy of impacts is negligible compared to the total KE of the wafer. It is NOT negligible in total energy delivered to the wafer (i.e. damage to the wafer), but it is not all delivered at once and the face on case is the worst case.
    
\subsection{Momentum Transfer}

    The total momentum tansfer $P_T$ of the dust grains to the spacecraft (assume worst case of an elastic collusion which is clearly an overestimate) is $P_T=NP_\textrm{grain}$, where the momentum transfer per grain hit is
    \begin{equation}
        P_\textrm{grain}=2m_\textrm{grain}v,
    \end{equation}
    assuming an elastic collision. $v=c/3$ here. For a dust grain of mass $m_\textrm{grain}\sim10^{-17}$ kg, then
    \begin{equation}
        P_T=2\times10^{-17}(\textrm{kg})\times10^8(\textrm{m/s})=2\times10^{-9}\textrm{N$\cdot$s}. 
    \end{equation}
    Compare to the total momentum of the spacecraft, assuming the worst case of the low mass wafersat:
    \begin{equation}
        P_\textrm{sc}=m_\textrm{sc}v\sim10^{-3}(\textrm{kg})\times10^8(\textrm{m/s})=10^5\textrm{N$\cdot$s}.
    \end{equation}
    Thus the ratio of the (worst case, face on and elastic collision) momentum dust grain transfer to the spacecraft momentum is
    \begin{equation}
        \frac{P_T}{P_\textrm{sc}}=\frac{10^{-3}}{10^5}=10^{-8}.
    \end{equation}
    Hence the total momentum transfer of the dust grains $P_T$ to the spacecraft (even for wafers) is negligible. Dust grain hits will not slow it appreciably.
    
    We need to do much more refined studies/simulations of the impact of dust grains on the wafer electronics. For example, what happens during an impact? Can we build redundant electronics so that dust hits do not compromise the system? If we just blow a hole through the wafer it is much less problematic than complete shattering. We discuss mitigation and shielding below.
    
\subsection{Interplanetary Dust}

    The solar system density and size distribution is more of a concern during the initial drive phase. Dust in the solar system is observed as zodiacal scattering and emission and has been studied in a number of ways, primarily in the infrared as well as with in-situ dust detectors. The solar system dust cloud forms a ``pancake'' shaped cloud in the ecliptic plane. The dust grains span a range of sizes with the peak mass flux between 100 and 400 $\mu$m. The overall distribution of the smaller sizes (less than 1 $\mu$m) is not well known. The density is highly dependent on the location in the solar system with highest concentrations in the inner solar system and in the ecliptic plane. This makes statements on density imprecise but we can get a rough estimate from both zodiacal light and in-situ measurements. Cumulative number densities for grains $>1$ $\mu$m are roughly 10$^{-8}$ m$^{-3}$ within approximately 50 AU. Outside of 50 AU, the densities across all grain sizes are believed to slowly decrease with increasing heliocentric distance \cite{Poppe2016}. Dust grains larger than 200 $\mu$m have a cumulative density no larger than 10$^{-12}$ m$^{-3}$. Hitting a 1 mm dust grain in the solar system would likely be catastrophic to the local wafer area affected or possible destroy the wafer. Since the solar system dust distribution is extremely anisotropic and inhomogeneous the path of the spacecraft will be important in estimating the hit probability. A path perpendicular to the ecliptic is preferred in general and is reduced in total impacts by more than a factor of 10 compared to a path in the ecliptic \cite{Poppe2016}.
    
\subsection{Dust Impact Rate}

    The dust impact rate is $R=n\sigma v$, where $v$ is the spacecraft speed. Using the cumulative dust density for grains larger than 1 $\mu$m of $n=10^{-8}$ grains/m$^3$ and $v=0.3c$, we get $nv=1$ m$^{-2}$s$^{-1}$.
    
    For the three cases of the wafersat considered above we get the following rates:
    \begin{align}
    \begin{split}
        \textrm{Face on:}& \hspace{2mm} R=10^{-2}\textrm{Hz}\\
        \textrm{Edge on:}& \hspace{2mm} R=10^{-5}\textrm{Hz}\\
        \textrm{Long thin rod:}& \hspace{2mm} R=10^{-10}\textrm{Hz}
    \end{split}
    \end{align}
    Fortunately, the time spent in the solar system is relatively short for the low mass probes.
    
\subsection{Total Dust Impact While in the Solar System}

    Since the solar system is highly anisotropic any estimates we give will be very dependent on the actual path taken out of the solar system. We can compute a rough estimate in the same way we did the interstellar case but it is important to keep in mind the large degree of variability. The path length is also highly variable and depends on the inclination angle of the trajectory relative to the ecliptic plane.
    
    As we did for the interstellar case the total number of hits during the journey on the spacecraft is given by $N=n\sigma L$. Assuming $n=10^{-8}$ m$^{-3}$ and $L=10^{12}$ m (assume 7 AU as a nominal distance), we have $nL=10^4$ grains/m$^2$. This is small compared to the interstellar case.
    
    As before, we assume an effective $10\times10$ cm spacecraft for the wafer-scale case. We assume the same orientation cases for the interstellar case and the same cross sections:
    \begin{align}
    \begin{split}
        \textrm{Face on:}& \hspace{2mm} N=10^2\\
        \textrm{Edge on:}& \hspace{2mm} N=10^{-1}\\
        \textrm{Long thin rod:}& \hspace{2mm} N=10^{-6}
    \end{split}
    \end{align}
    The number of hits expected in the solar system is relatively low. Since hitting any significantly sized object ($>1$ mm) would be catastrophic, the best strategy is to send many probes to avoid any ``single point failure'' and to send them out with trajectories away from the ecliptic plane. Even if we raise the distance in the solar system from 7 AU to 70 AU we still only increase the impacts by a factor of 10, which is  still well below a critical concern level, at least for the edge forward case. The major uncertainty is the lack of information about the smaller IPM dust grains.
    
\subsection{Dust Impact Mitigation}
    
    Several techniques exist to mitigate dust impacts. Mitigation can be from minimizing spacecraft geometry (cross section), shielding (a thin beryllium shield, for example), and spacecraft topology. As we discussed, a long thin spacecraft with small cross section relative to the velocity vector can result in large reductions in the number of hits. Shielding needs to be explored further. A Be shield that is a few mm thick may allow significant protection but this would dominate  the mass of the lightest spacecraft (WaferSat) significantly. Another solution may lie in a highly redundant topology that allows for a large number of hits and still maintain functionality. It is critical to prevent shattering of the wafer for the low mass case. A wafer could be produced with a ``waffle pattern'' with small contiguous regions and thinned or even legs attachments that would prevent shattering of the adjacent sections. This is an area where design, simulation, and testing with a dust accelerator (currently only good to 100 km/s) is critical to explore.
    
\subsection{Shielded Edge On Design}

    One possible design that combines shielding with an edge on wafer orientation is to make the outer facing (along the velocity vector) edge out of Be or another suitable material that could withstand high speed dust grain impacts. This could also be an ablative or a ceramic composite similar to bullet proof vests. For example in the case of the wafersat with a wafer thickness of approximately 100 microns, the outer forward edge could easily be 1 cm ``wide'' by 100 microns thick or perhaps somewhat thicker than the wafer if needed, and still add only a modest amount to the wafer mass. A design that is simpler but twice the shielding mass, though still a small portion of the wafer, is to make the entire outer rim out of shielding. The optimal choice of shield material will need to be designed and tested but this appears to be a promising option.
    
\subsection{Spacecraft Perturbation from Dust Impacts}

    Dust impacts will have a perturbing effect on the spacecraft that becomes increasingly important for lower mass probes. The worst case will be the wafersat case. We can make some estimates of the overall perturbation by comparing the dust impact applied torque impulse and subsequent angular momentum. We will assume a coordinate system where $x$ and $y$ are in the plane of the wafer and $z$ is the normal. The velocity vector is assumed to be along the $x$-$z$ axis of the wafer with the primary component being along the $x$ axis.
    
    For a thin circular disk of uniform mass with total mass $m$, the moment of inertia along the $x$ or $y$ axis is
    \begin{equation}
        I_{x,y}=\frac{1}{4}mR^2=\frac{1}{16}mD^2,
    \end{equation}
    where $R=D/2$ is the disk radius. Along the $z$ axis it is
    \begin{equation}
        I_z=\frac{1}{2}mR^2=\frac{1}{8}mD^2=2I_{x,y}.
    \end{equation}
    For a square wafer the moments of inertia are $I_{x,y}=\frac{1}{4}mD^2$ and $I_z=\frac{1}{6}mD^2$ where $D$ is the size of the square side. For a $D=10$ cm wafer we have the following moments:
    \begin{align}
    \begin{split}
        \textrm{Square:} \hspace{2mm} I_{x,y}=&2.50\times10^{-6}\hspace{1mm}\textrm{kg$\cdot$m$^2$}\\
        \hspace{2mm} I_z=&1.67\times10^{-6}\hspace{1mm}\textrm{kg$\cdot$m$^2$}\\
        \textrm{Circle:} \hspace{2mm} I_{x,y}=&6.25\times10^{-7}\hspace{1mm}\textrm{kg$\cdot$m$^2$}\\
        \hspace{2mm} I_z=&1.25\times10^{-6}\hspace{1mm}\textrm{kg$\cdot$m$^2$}
    \end{split}
    \end{align}
    
    The total momentum transfer $P_T$ of an interstellar dust grain impact to the spacecraft, assuming the worst case (not realistic) of an elastic collision gives $P_\textrm{grain}=2m_\textrm{grain}v$. For a dust mass of $m_\textrm{grain}\sim10^{-17}$ kg, then $P_\textrm{grain}=2\times10^{-9}$ N$\cdot$s.
    
    We consider two cases:
    \begin{enumerate}
        \item Impact head on, but allow the dust to hit anywhere along the wafer thickness and allow for a wafer plane misalignment with the velocity vector. If the wafer is 0.1 mm thick the applied torsion impulse will be worst if the impact is at the upper or lower edge (0.01 mm off the centerline). The torque applied is $\bm{\tau}=\bm{r}\times\bm{p}$. With a 10 cm wafer (round) we get an impulsive torque of
        \begin{align}
        \begin{split}
            dL=\tau dt=&\frac{1}{2}10^{-4}(\textrm{m})2\times10^{-9}(\textrm{N$\cdot$s})\\
            =&10^{-13}\textrm{Nm$\cdot$s}\\
            =&I\alpha\hspace{.5mm}dt=I\hspace{.5mm}dw,
        \end{split}
        \end{align}
        \begin{align}
        \begin{split}
            dw=\tau\hspace{.5mm}dt/I\sim&\frac{10^{-13}(\textrm{Nm$\cdot$s})}{2.5\times10^{-6}(\textrm{kg$\cdot$m$^2$})}\\
            =&4\times10^{-8}\textrm{rad/sec}.
        \end{split}
        \end{align}
        This is extremely small and can be counteracted by the wafer photon thrusters.
        \item We will assume an initial misalignment of the wafer plane to the velocity vector of 0.001 radians (1 mrad) $\sim200$ arc sec. We can do much better than this if needed. As a quick check we see an angular misalignment of 0.001 radians over a 10cm lever arm give a displacement at the edge of 0.001 rad $\times$ 0.1 m $=10^{-4}$ m. This is the thickness of the wafer or 2 times the offset hit in case 1) and thus still leads to a very small impulsive torgue and very small $dw\sim4\times10^{-8}$ rad/sec and hence can be corrected by the wafer photon thrusters.
        \item In this case we allow the dust grain to hit the radial edge so as to cause the wafer to ``spin up.'' The impulsive torque is $dL=\tau\hspace{.5mm}dt=0.05$m.
    \end{enumerate}
    
\section{Onboard and Beamed Power Options}

    Providing adequate power during the long cruise phase is critical to maintain bidirectional laser communications. This is especially critical with the smaller mass payloads that have extremely limited mass and size for the communications system. We focus on the most difficult case, namely the wafer scale spacecraft. In order to have a reasonable laser communications link from spacecraft to Earth and assuming the main phased array used to drive the spacecraft is also used as a phased array receiver or at least as a light bucket (non-phased array receive mode with ``light bucket'' combining of sub-elements). We baseline a 1 watt (optical) burst mode transmission (0.5\% duty cycle) from the wafersat as the laser comm transmitter. This then requires an average electrical power, assuming 50\% laser comm diode conversion efficiency, of about 10 mW$_{\textrm{electrical}}$. The current baseline is to use a small radio thermal generator (RTG) as the long term on-board power. We derate the laser communications fraction to 0.2\% to allow for other subsystem usage. There are several suitable RTG materials. Pu-238 is one of them. Pu-238 has a half-life of about 87.7 years and produces about 400 mW$_\textrm{thermal}$/g initially. Using a Pu-238 RTG and nanowire TE converters allows for an RTG system that has a mass of about 0.3g. The Pu-238 mass can be reduced as the burst power and or duty cycle are reduced. Energy is stored on the wafer using thin film supercapacitors.  Such a small system, while theoretically possible, has not been built. Small RTG’s were built for early pacemakers so there is some precedent. The cold background aids the efficiency with an assumed conversion efficiency of 7\% or about 30 mW$_\textrm{electrical}$/g (Pu-238). A MEMS or other micromechanical converter could increase the conversion efficiency to above 30\% if developed. We assume the lower 7\% conversion.
    
    Note that the wafer alone receives significant power from the drive laser if illuminated (Figure 32). A narrow bandgap photo converter (PV) is feasible with greater than 50\% efficiency today and this allows the possibility of direct beamed power to the spacecraft for considerable distances. A more advanced option is to coat the drive reflector ($\sim1$ m$^2$) with a narrow bandgap PV which would give nearly 100 times the power of the wafer alone. This could conceivably allow beamed power all the way out to the nearest star if the burst fractional time was reduced (see figure below).
    
\subsection{PV When Reaching Target Star}
    
    As the spacecraft approaches a star the onboard PV could be used as a secondary power source using the light from the nearby star to power the system. If the approach was close enough this could dominate allowing a much greater burst fraction and hence data rate. For example, if we approach within 1 AU of a star similar to our sun we would get 0.14 W/cm$^2$ and with 40\% efficient PV (assuming mild future development, or concentrating with the same optic used for laser communications) we could get about 60 mW/cm$^2$ or 6 W over the 10 cm wafer or 600 W over a 1 m reflector. This would completely dominate the power budget and allow a vastly greater data rate.  Ideally both RTG and PV would be available.
    
    \begin{figure}
        \centering
        \includegraphics[width=0.45\textwidth]{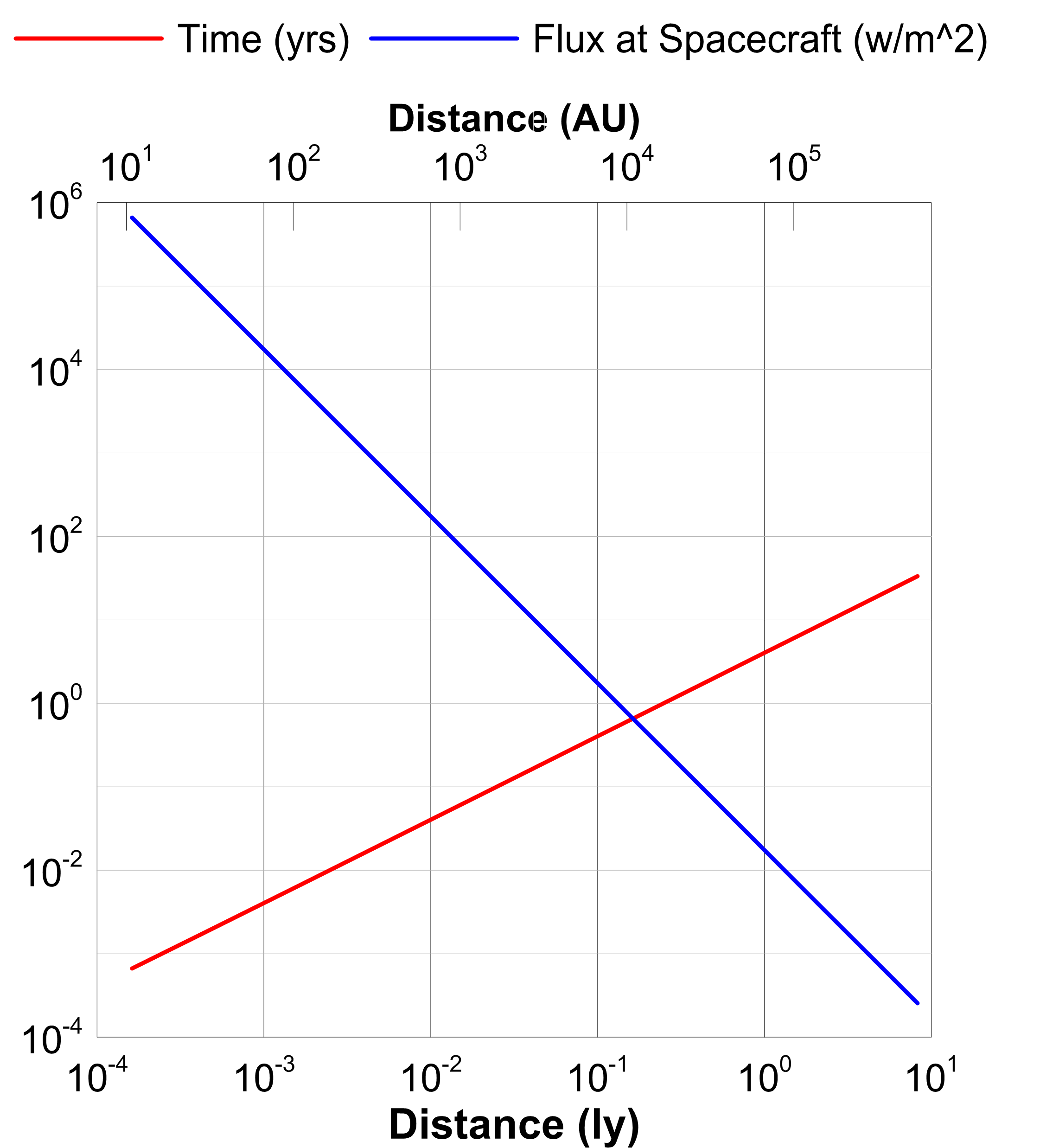}
        \caption{Flux (W/m$^2$) at spacecraft from DE-STAR and time since launch vs distance (ly). Assume DE-STAR 4 driver with 70 GW.}
        \label{fig:fluxatcraft}
    \end{figure}
    
\section{Spacecraft Thermal Issues}

    Being far away from the sun the spacecraft will tend to run very cold unless there is some way to actively heat the payload. As the spacecraft becomes smaller this becomes increasing difficult as the surface area to volume increases as $1/a$ where $a$ is the spacecraft ``size.'' Since the system is in radiative equilibrium with the environment, the power radiated will be proportional (roughly) to the exposed surface area, modified by an effective emissivity factor $\epsilon(\lambda)$ which is a function of the radiating wavelength. In general $\epsilon(\lambda)$ has a mild functional dependence on $\lambda$. Assuming the spacecraft is powered by an RTG, the power available will be roughly proportional to the volume $a^3$ assuming a fixed RTG mass fraction. We can compute a rough estimate for the outer spacecraft temperature by equating the internal RTG power $P_\textrm{rtg}$ with the radiating power 
    \begin{equation}
        P_r\sim A\epsilon(\lambda_p)\sigma T^4,
    \end{equation}
    where $A$ is the total effective (including view factors) radiating surface area, $\epsilon(\lambda_p)$ is the effective emissivity at the peak radiating wavelength and $\sigma\sim5.67\times10^{-8}$ W/m$^2\cdot$K$^4$ is the Stefan-Boltzmann constant. Equating $P_r=P_\textrm{rtg}$ we get 
    \begin{equation}
        T=\bigg(\frac{P_\textrm{rtg}}{A\epsilon(\lambda_p)\sigma}\bigg)^{1/4}\sim65\bigg(\frac{P_\textrm{rtg}}{A\epsilon(\lambda_p)}\bigg)^{1/4}\hspace{0.5mm}\textrm{K}.
    \end{equation}
    The wafer scale system is the most challenging and will run the coldest. If we assume 0.3 g of Pu-238 for the wafer case we get $P_\textrm{rtg}\sim0.12$ W, $A\sim0.02$ m$^2$ for a 10 cm sized wafer (2 sided radiated area), giving $T\sim100$ K assuming unity emissivity. We can lower the effective emissivity, to some extent, which will increase $T$. Even with unity emissivity this operating temperature ($~100$ K) is not that cold and with proper doping, a Si based wafer will generally operate well. If we used a thin layer of low emissivity coating (``superinsulation'') we should be able to have $\epsilon(\lambda_p)\sim0.1$ or lower, yielding $T\sim180$ K. Si-Ge, GaAs or other semiconductors are also options for the wafer. Si and Si-Ge have the great advantage of ease and cost of fabrication. Other compounds will be needed for specialized parts of the wafer (laser comm for example), but the mild operating temperature ($T\sim100$ K) is encouraging. Doped Si for IR ROIC's have been produced that operate well below 10 K so the operating temperature does not appear to be a critical issue. For larger payloads the operating temperature will be larger assuming a fixed fractional mass allocation to the RTG. We do note that the case of the long cylindrical payload geometry for reducing the effective cross section to minimize dust hits, as opposed to the disk baseline geometry, will decrease the average operating $T$. Due to the RTG heating thermal issues look to be quite reasonable.
    
\section{Photon Thrusters}
    
    All of the payload designs require some active attitude control capability for pointing during laser communications as well as for pointing during imaging and to assume spacecraft orientation to minimize dust impacts and correct orientation after a dust impact. Small trajectory corrections are also desirable. The baseline is to use photon thrusters to provide the active pointing and guidance capability. Photons give a thrust of P/c or about 3.3 nN/W. Photon thrusters can be accomplished in two ways. One of them uses the on-board electrical power to drive small edge mounted laser diodes or LED's. A second way is thermal photons. Assuming the RTG baseline for on-board power there is also the possibility of using the RTG heat, which is ultimately radiated as photons. The effect of the RTG thermal photon thrusting is inevitable even if undesired. Assuming a 7\% conversion of thermal to electrical for the RTG, it is clear the thermal effect dominates. A possible way to minimize the RTG thermal effect is to symmetrize the thermal emission of the RTG but it would be useful to take advantage of the much larger RTG thermal photon thruster than to minimize it. Utilizing the RTG thermal photons for thruster control is complicated by the need to control the direction of emission. This could be done with a suitable optical design but it is not a trivial problem. A switchable heat pipe is another option. From the RTG calculations we get about 400 mW$_\textrm{thermal}$/g (Pu-238) or about 30 mW$_\textrm{electrical}$/g and ignoring the thermal photons we would get a photon thruster force of about $P/c=10^{-10}$ N/g $=0.1$ nN/g (Pu-238). Note that the electrical power is not only converted into light but also into heat in the laser/LED and this could all be effectively used if designed properly. Hence we can assume near unity conversion efficiency for electrical power to photon thrust. From the calculations of the dust impacts for the edge impacts requiring reorientation we need about $10^{-13}$ Nm$\cdot$s of torque impulse per dust grain hit. For the edge hit wafer spin up mode we need about $10^{-10}$ Nm$\cdot$s of torque impulse per dust grain hit. Both of these required torque impulses are easily obtained with the photon thrusters even in the wafersat assuming about 0.1 - 0.3 g (Pu-238) for the hit rates above. A secondary option we will explore would be to use a small field emission type of thruster that utilizes a small amount of mass. This is more power efficient but does require some mass for ejection. This must be charge neutral upon emission or the system will become space charge limited. This has the potential to be $\sim10^{4-5}$ more thrust to power efficient if an ISP between chemical (300) and ion engine (3000) can be achieved. It is an option to be further explored in a trade study. 
    
\subsection{Pointing with Photon Thrusters}
    
    If we assume a 10 mW photon thruster we get a force (as above) of 3.3 nN/W $\times10^{-2}$ W $=3.3\times10^{-11}$ N or 33 pN. For reorientation of the wafer required for imaging or laser communications back to Earth we can compute the accelerations and thus times to move by a given angle. For the 10 cm wafer, as above, we have $I_{x,y} = 2.50\times10^{-6}$ kg$\cdot$m$^2$ for a square and $I_{x,y}=1.25\times10^{-7}$ kg$\cdot$m$^2$ for a circle. This gives angular accelerations of 
    \begin{align}
    \begin{split}
        \textrm{Square:}\hspace{2mm}\alpha_{x,y}=&\frac{\tau}{I_{x,y}}=\frac{3.3\times10^{-12}\hspace{1mm}\textrm{N$\cdot$m}}{2.5\times10^{-6}\hspace{1mm}\textrm{kg$\cdot$m$^2$}}\\
        =&1.32\times10^{-6}\hspace{1mm}\textrm{rad/s$^2$},\\
        \textrm{Circle:}\hspace{2mm}\alpha_{x,y}=&\frac{\tau}{I_{x,y}}=\frac{3.3\times10^{-12}\hspace{1mm}\textrm{N$\cdot$m}}{6.25\times10^{-7}\hspace{1mm}\textrm{kg$\cdot$m$^2$}}\\
        =&5.28\times10^{-6}\hspace{1mm}\textrm{rad/s$^2$}.
    \end{split}
    \end{align}
    In  a time of $t=10^3$ seconds this gives angular rates of
    \begin{align}
    \begin{split}
         \textrm{Square:}\hspace{2mm}\omega_{x,y}=&\alpha_{x,y}t\\
         =&(1.32\times10^{-6}\hspace{1mm}\textrm{rad/s$^2$})(10^3\hspace{1mm}\textrm{s})\\
         =&1.32\times10^{-3}\hspace{1mm}\textrm{rad/s},\\
         \textrm{Circle:}\hspace{2mm}\omega_{x,y}=&\alpha_{x,y}t\\
         =&(5.28\times10^{-6}\hspace{1mm}\textrm{rad/s$^2$})(10^3\hspace{1mm}\textrm{s})\\
         =&5.28\times10^{-3}\hspace{1mm}\textrm{rad/s},
    \end{split}
    \end{align}
    and a $\Delta\theta_{x,y}$ of
    \begin{align}
    \begin{split}
        \textrm{Square:}\hspace{2mm}\Delta\theta_{x,y}=&\alpha_{x,y}t^2=1.32\hspace{1mm}\textrm{rad},\\
        \textrm{Circle:}\hspace{2mm}\Delta\theta_{x,y}=&\alpha_{x,y}t^2=5.28\hspace{1mm}\textrm{rad}.
    \end{split}
    \end{align}
    Thus very large angular changes can be made in relatively short times.
    
\subsection{Flyby Imaging}

    As another example consider the case of a $v=0.3c$ mission flying by an object of interest at a distance of $r=1$ AU. This would yield a worst case (perpendicular) angular rate of $\omega = v/r =108$ (m/s)/$1.5\times10^{11}$ m $\sim7\times10^{-4}$ rad/s $\sim140$ arcsec/s. This again is the worst case. We would traverse a distance of 1 AU in $1.5\times10^{11}$ m/$108$ (m/s) or 1500 s.  To stabilize the image we would need to ``slew'' at this rate either in whole body spacecraft frame or electronically in the sensor. Recall that for a 10 mW photon thruster with $t=1000$ s we get $\omega_{x,y} =1.32\times10^{-3}$ rad/s. From above we could conceivably stabilize in the spacecraft frame if we ``spun up'' the wafer for $10^3$ s and take an image as we fly by at 1 AU even at 0.3c. For a diffraction optical system with $D=10$ cm (assume square) we have 
    \begin{equation}
        \theta_{DL}=\frac{2\lambda}{D}=\frac{2\times10^{-6}}{0.1}=2\times10^{-5}\textrm{rad}\sim4\hspace{0.5mm}\textrm{arcsec}.
    \end{equation}
    For a circular aperture replace $D$ by $D/\alpha$ where $\alpha=1.22$. At $r=1$ AU this gives a resolution of 
    \begin{align}
    \begin{split}
        r\theta_{DL}=&1.5\times10^{11}(\textrm{m})\cdot2\times10^{-5}\\
        =&3\times10^6\hspace{0.5mm}\textrm{m}=3000\hspace{0.5mm}\textrm{km}.
    \end{split}
    \end{align}
    This would resolve a planet the size of the Earth easily but not give much substructure. If we could get to within 0.1 AU we would get 300 km resolution, as an example. This would require faster slewing which could be done by spinning the spacecraft up longer or and using a ``streak camera'' approach, though electronic pixel shifting might be a better approach particularly since we want pointing control as we go by. 
    
\subsection{Magnetic Torquer}

    Other possible non-expendable orientation control include a magnetic torquer that acts against the interstellar magnetic field. The interstellar magnetic field is highly anisotropic and inhomogeneous. A rough estimate is about 1 micro-Gauss (0.1 nT) based on the average galactic magnetic field. As an example, such a magnetic torquer would be a single 10 cm loop with 1 amp of current. Using a 0.1 nT estimate for the interstellar magnetic field gives a torque of about $10^{-12}$ N$\cdot$m (iAB). It is possible to use a thin film current loop with a large number of turns to amplify this effect. 100 turns would not be unreasonable. Using pulsed current from the RTG power source could allow such a system. Since the spacecraft will run very cold during the cruise phase high $T_c$ superconducting films might be an option. Getting 3D control would require a 3-axis torquer. In the case of the wafersat this would be a challenge, though a ``pop-up'' additional coil may be feasible to allow 3-axis torquing. This same technique could be used upon entry to the stellar system where a larger magnetic field may be present and where the much larger PV power from the star could be used to allow for more rapid slewing for imaging (compared to the RTG and photon thrusters case we assumed above).
    
\subsection{Trajectory Modification}

    The ability to make some modest trajectory modifications would be extremely useful. We assume the use of photon thrusters. We can get an estimate for this but assuming only the use of the electrical portion and not the thermal photon component. The thrust/RTG active mass is $10^{-10}$ N/g $=0.1$ nN/g (Pu-238). Over the course of 1 year this would yield a $\Delta P$ (momentum) of about $3\times10^7$ s/yr $\times10^{-10}$ N/g $=3\times10^{-3}$ N$\cdot$s/g$\cdot$yr. Assuming a 1 gram wafer this would give an angular change of $\phi=\Delta P/P$ where $P$ is the total momentum with $P=mv\sim10^{-3}$ (kg)$\times10^8$ m/s $=10^5$ N$\cdot$s. Thus,
    \begin{equation}
        \phi=\frac{\Delta P}{P}=\frac{3\times10^{-3}\hspace{0.5mm}\textrm{N$\cdot$s/g$\cdot$yr}}{10^5\hspace{0.5mm}\textrm{N$\cdot$s}}=3\times10^{-8}\hspace{0.5mm}\textrm{rad/g$\cdot$yr}.
    \end{equation}
    Assuming a 20 year mission with would yield $\phi_\textrm{total}=6\times10^{-7}$ rad/g. This yields a total displacement $D$ at a distance $L$ of $\alpha$ Centauri (4.4 ly $\sim4.4\times10^{16}$ m) of  
    \begin{equation}
        D=L\phi_\textrm{total}\sim3\times10^{10}\hspace{0.5mm}\textrm{m/g},
    \end{equation}
    or about 0.2 AU/g. Using the thermal photons this would increase about 16 times to 3 AU/g. If we assume 0.3 g of Pu-238 we get about 1 AU of displacement over a 20 year mission if we use the thermal photons. This is small but possibly useful.
    
\section{Ground vs Space Deployment}

    It would be far simpler and less expensive if we could deploy the main DE driver on the ground vs in space. In \cite{Pelton2015} we discuss the issue of ground, airborne, and space deployment options for our related technology work on DE planetary defense \cite{Pelton2015}\cite{Hughes2014}. The primary concern for ground deployment is the perturbations (seeing) of the atmosphere. With typical seeing at ``good'' mountain top sites of a bit better than 1 arc second ($\sim 5$ micro-rad) this is far from the 0.1 nrad required. Even the best adaptive optics system fall far short of this. With the upcoming 30 m class telescopes we hope to be able to get to decent Strehl ratios with multi-AO systems but at much larger diffraction limited values than we need. Ground based interferometry in the visible is done with 10 m class telescopes with modest success and this is encouraging. The two Keck telescopes on Mauna Kea are about 85 m apart, while the VLT's can be up to 200 m apart and the NPOI (Navy Precision Optical Interferometer) has a 440 m baseline, for example. This is encouraging. The key will be high fractional encircled energy, Strehl. \textbf{The ground based option is a possibility to explore, as the phased array is ideal for adaptive optics use, since by its very nature it is an adaptive optics system.} It is a part of the roadmap we propose but it is not clear we can get good Strehl (power on target) with the kind of system we propose for ground deployment even with the most complex AO we can imagine and it is extremely far from any current systems or imagined systems. We have also considered hybrid system using ground based deployment for the critical DE components and phase conjugating orbital beam directors that correct the atmospheric distortions in each sub element and then ``redirect'' to the spacecraft. It is not clear that this is feasible for the very fine targeting we need. It also remains an option to explore in optimizing cost and deployment. Airborne (aircraft, balloons, etc.) are an intermediate approach and also a part of the deployment roadmap for testing of the smaller systems but full scale deployment on such platforms is extremely problematic due to the scale required. The optimization of where to deploy in space is also a part of a longer term analysis (LEO, GEO, lunar, L2 etc.). These are not only very cost sensitive optimizations but suitability for maintenance is also a significant factor. \textbf{During the development and test phase of the roadmap we will explore the limits of ground based deployment to better quantify this.} Smaller arrays, say 0.01-1 km, should be built for ground use before going to space. Ground does offer much lower cost and the option of much higher areal power density to offset the reduced array size. \textbf{This will also help us understand ground deployment of the PAT mode.} In addition the ground-based solution is complicated by the atmospheric air glow and non-thermal processes such as OH lines in particular \cite{Lubin2016}. An onboard local oscillator to tune the laser if needed could be commanded from the ground using the laser array to transmit a command to the wafer. Uplinking commands is feasible (modulo TOF) using the laser array to transmit. Weather is also an issue, as is water vapor fluctuations \cite{Suen2015}\cite{Suen2013}. The atmospheric emission is mitigated by using a very narrow linewidth laser for communications since the data rates are low. Uplinking commands is feasible (modulo TOF) using the laser array to transmit. Once the spacecraft is far away the TOF will be complicated and tracking and beacon locking will be challenging in all cases whether in space or on the ground. In order to minimize backgrounds for reception it is highly desirable to fully synthesize the received beam as discussed above. This requires knowledge of the location of the spacecraft and knowledge of both the astrometry and ephemeris on (sub) nrad levels. This is not trivial. Ground deployment should be explored before the space option primarily due to the dramatic reduction in cost and ease of maintenance and expansion.
    
\section{Microwave and Millimeter vs Laser}

    Our analysis is wavelength agnostic but guides us. It is important to understand the issues related to wavelength dependence, power, and array size as it affects the spacecraft speed. From the above (and appendix) the final speed with continued illumination for a power $P_0$, array size $d$, and wavelength $\lambda$ is
    \begin{equation}
         v_{\textrm{max-}\infty}=\bigg(\frac{P_0(1+\epsilon_r) d}{c\lambda\alpha}\bigg)^{1/2}(\xi h\rho m_0)^{-1/4}.
    \end{equation}
    For the same spacecraft speed we need to maintain the ratio of $P_0 d/\lambda$. The Earth's atmosphere is extremely opaque in the far IR (essentially blocked from about 30 to 300 microns) \cite{Denny2014}\cite{Suen2015}\cite{Suen2013}. If we imagine a millimeter wavelength (gyrotron for example) approach with a wavelength of 1 mm we immediately see that to achieve the same spacecraft speed we would have to increase the array size by a factor of 1000 (1 mm/1 micron) thus requiring an array about the size of the entire Earth and we will still have atmospheric phase perturbations to deal with. In fact, the atmospheric transmisison at 1 micron is superior to that at 1 mm due to molecular water vapor absorption. At sea level the atmosphere is essentially opaque at 1 mm while relatively transparent at 1 micron. An alternative approach is to increase the power by a factor of 1000 to 100 TW and keep the array size fixed at 10 km. This is greater than the entire Earth's current power generation. Consider the SKA as an option. The high frequency portion (\textbf{no longer currently baselined}) of the array will be at a wavelength of about 1 cm ($10^4$ $\mu$m) or a frequency of 30 GHz and have an inner ``close packed' array of approx 3000 ten meter class telescopes. If the high frequency portion of the SKA were completely close packed it would have an equivalent diameter of about $d=0.6$ km.  \textbf{For the same power $P_0$ the speed is proportional to $(d/\lambda)^{1/2}$ and hence an entire SKA high frequency array at 1 cm wavelength would be equal to a 6 cm telescope at 1 $\mu$m wavelength.} From another perspective, using a small 1 m laser array at 1 micron is equivalent to a 10 km array operating at 1 cm or one that is 100 times the area of the planned SKA at a wavelength even shorter than the planned SKA or to a hypothetical 1 km array operating at 1 mm (300 GHz).  One might imagine using a sparse array (widely separated optics) but this causes large sidelobe issues resulting in large loss of power on the spacecraft. \textbf{If the areal packing fraction of the sparse array is $\epsilon_\textrm{frac}$ then the fraction of power that is in the main beam at the spacecraft is $\epsilon_\textrm{frac}P_0$ where $P_0$ is the total power.} In the equation for the speed above we would replace $P_0$ with $\epsilon_\textrm{frac}P_0$. For $N$ antennas of size $D^{*}$ (assume square) the packing fraction is $\epsilon_\textrm{frac} = N D^{*2}/d^2$ and $P_0$ is replaced by $P_0 N D^{*2}/d^2$ \textbf{and thus the speed is less if we try to spread out the array as more power goes into sidelobes.} A filled array is critical to high performance.
    \begin{align}
    \begin{split}
        v_{\textrm{max-}\infty}=&\bigg(\frac{\epsilon_\textrm{frac}P_0(1+\epsilon_r)d}{c\lambda\alpha}\bigg)^{1/2}(\xi h\rho m_0)^{-1/4}\\
        =&\bigg(\frac{ND^{*2}P_0(1+\epsilon_r)d}{d^2c\lambda\alpha}\bigg)^{1/2}(\xi h\rho m_0)^{-1/4}\\
        =&\bigg(\frac{ND^{*2}P_0(1+\epsilon_r)d}{dc\lambda\alpha}\bigg)^{1/2}(\xi h\rho m_0)^{-1/4}.
    \end{split}
    \end{align}
    Note the final speed drops as the array size $d$ increases (array becomes sparser)  ($v\sim d^{-1/2}$) for a fixed number $N$ of antennas (or optics) of size $D^{*}$.  This is because of the reduction of power in the main beam, due to sidelobe losses, as the array becomes sparser ($d$ increases). Here $d>D^{*}N^{1/2}$ as the packing fraction cannot exceed unity. Microwave and millimeter wave systems do not seem to be a fruitful direction here.
    
\section{Comparison to Matter-Antimatter Annihilation Propulsiom}

\begin{figure}
        \centering
        \includegraphics[width=0.45\textwidth]{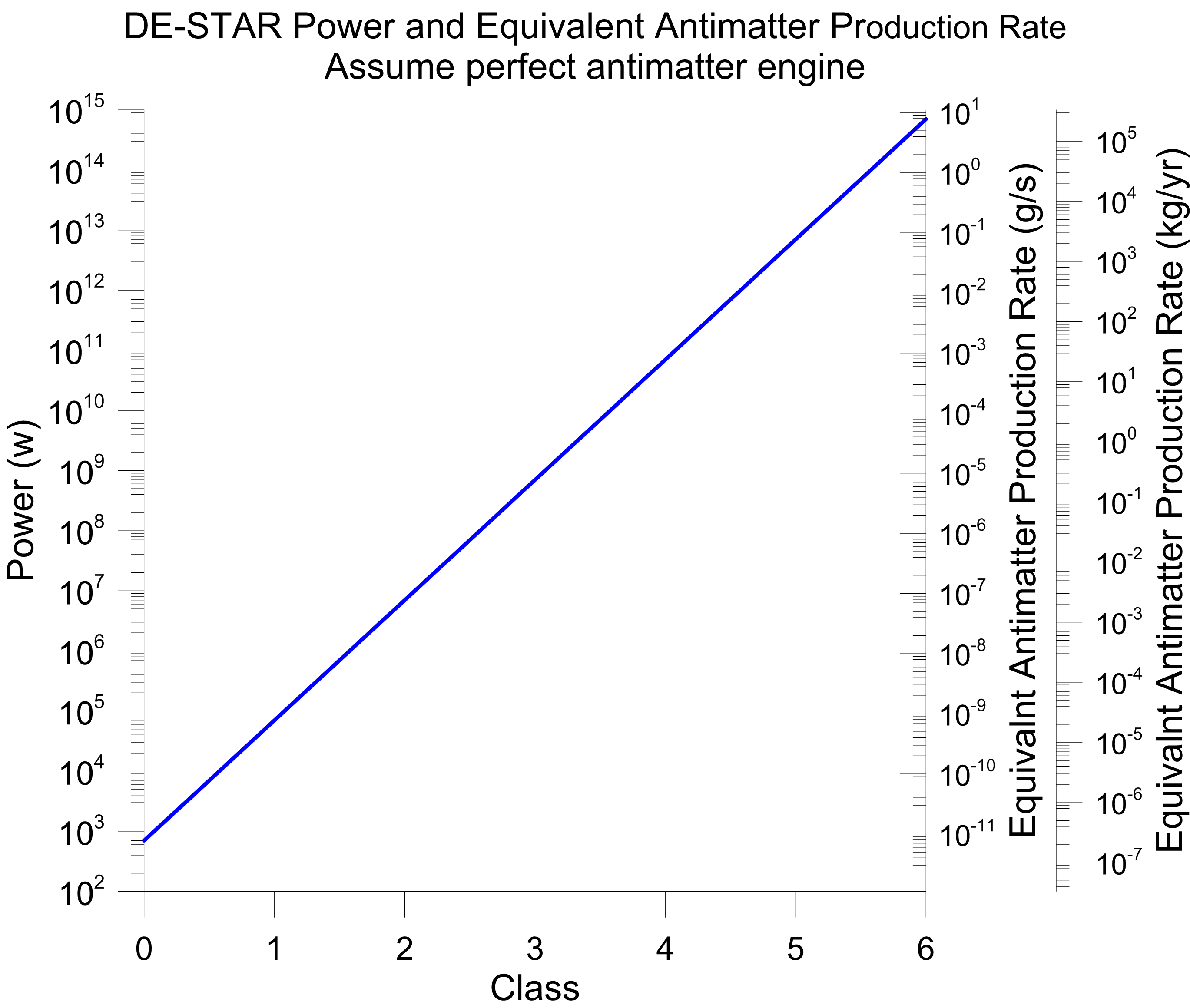}
        \caption{Relationship between directed energy propulsion and a hypothetical perfect antimatter-matter annihilation engine in terms of equivalent antimatter production rates compared to DE driver class. This is the most optimistic case for antimatter and assumes perfect conversion to massless particles, zero exhaust divergence, and no confinement or reaction chamber mass. A real antimatter based system would be far worse. Class is the log of the array size when the size is specified in meters. Class 0 is a 1 m array and class 4 is a 10 km array for example. From Lubin 2016.}
        \label{fig:antimatter}
    \end{figure}

    In \cite{Lubin2016} we compare directed energy systems to hypothetical antimatter based propulsion systems. Antimatter is often invoked in the context of the ``ultimate propellant'' for enabling interstellar flight \cite{Forward1984}. Even ignoring all the complexities of producing, storing, and utilizing annihilation based engines (which are formidable and which do not currently have a feasible path forward), and ignoring the large auxiliary masses needed for storage and reaction (which greatly reduce the final speed), and even assuming perfect conversion efficiency to massless particles with zero divergence angle out the exhaust (i.e. a perfect antimatter engine, AME) we still conclude that a directed energy based system is more efficient, in terms of  power to thrust alone, by a factor of 2 since the DE photons bounce off the reflector (Figure \ref{fig:antimatter}). When the realities of the extremely low production efficiencies of antimatter production and extremely large effective cost of production, the large (and unknown path to) confinement masses for trapping of anti-H$_2$ or other antimatter candidates, and the large mass for the reaction chamber to get focusing of the annihilation products to enable good conversion efficiency are taken into account it is not at all clear that antimatter is consistent with the goal of interstellar flight. We would like to have a perfect AME, and it would be very useful for flight, particularly for maneuvering and slowing down, but we do not know how to do so and even if we did the directed energy system would still be advantageous, let alone being a feasible and much lower cost path forward. If we can accomplish some degree of photon recycling the contrast becomes even more favorable for DE systems.
    
\section{Payload Sizes}

    Once a suitable laser driver is built the payloads can be any size from miniature relativistic probes, such as the wafer scale one for interstellar flight we have discussed, to large spacecraft capable of transporting humans in the solar system. Some examples of the many mission scenarios possible are shown below. Note the a single laser driver can be used to launch sequentially or in parallel any number of spacecraft and thus the system enables and is amortized over a large mission space. As the speed scales as a mild function of payload mass in the optimized case ($v\sim m_0^{-1/4}$) there is significant choice in mission planning. The difference in speed when increasing the mass by a factor of 10 reduces the speed by a factor of about 1.8, while increasing the mass by a factor of 100 reduces speed by about 3.2, and increasing the mass by $10^4$ reduces speed by a factor of 10.
    
    The following gives a selected set of possible missions. It is assumed that the reflector mass is equal to the base spacecraft mass (i.e. system mass not including reflector, or total system mass is twice the base spacecraft mass). The reflector is assumed to be 1 micron thick and the reflector density is assumed to be 1.4 g/cc. The laser array is assumed to be 10 km on a side and the reflector is assumed to be square. If the ground based option were viable the speeds would be comparable for the same sized array, but the illumination time available would have to be carefully considered for most siting option and would likely limit ground based sites (except near the poles) to very small mass payloads with short acceleration times. The mass given is the base spacecraft mass and hence the reflector mass. For thinner reflectors the speed multipliers discussed above can be easily used, though the times and $L_0$ change. For the smallest payloads, as the speeds become relativistic the full relativistic solution needs to be used for higher precision calculations. The laser power is assumed to be 70 GW. No photon recycling is assumed for reasons of both complexity over long distances as well as being relatively ineffective at relativistic speeds \cite{Kulkarni2018}. As discussed, no simple braking mechanism exists except for a duplicate system at the target or ejecting a retro reflector for short range systems within our solar system.
    
    \begin{itemize}
        \item[$-$]1 g: Wafer-scale spacecraft with 0.85 m reflector capable of significant relativistic flight.
            \begin{itemize}
                \item[-] Time to when laser diffraction spot equals reflector size ($t_0$): 186 s
                \item[-] Distance to when laser diffraction spot equals reflector size ($L_0$): $4.01\times10^9$ m
                \item[-] Speed when laser diffraction spot equals reflector size ($v_0$): $4.31\times10^7$ m/s
                \item[-] Beta when laser diffraction spot equals reflector size ($\beta_0$): 0.14
                \item[-] Speed with continued illumination ($v_{\textrm{max}-\infty}$): $6.10\times10^7$ m/s
                \item[-] Beta with continued illumination ($\beta_{\textrm{max}-\infty}$): 0.20
                \item[-] Acceleration when reflector is fully illuminated ($a_0$): $2.37\times10^4$ ``g''
            \end{itemize}
        \item[$-$]10 g: Multiple wafer-scale systems with 2.7 m reflector.
            \begin{itemize}
                \item[-] $t_0$: 1050 s
                \item[-] $L_0$: $1.27\times10^{10}$ m
                \item[-] $v_0$: $2.43\times10^7$ m/s
                \item[-] $\beta_0$: 0.081
                \item[-] $v_{\textrm{max}-\infty}$: $3.43\times10^7$ m/s
                \item[-] $\beta_{\textrm{max}-\infty}$: 0.11
                \item[-] $a_0$: $2.37\times10^3$ ``g''
            \end{itemize}
        \item[$-$]100 g: Multiple wafer and sub-class CubeSat systems with 8.5 m reflector.
            \begin{itemize}
                \item[-] $t_0$: 5880 s
                \item[-] $L_0$: $4.01\times10^{10}$ m
                \item[-] $v_0$: $1.36\times10^7$ m/s
                \item[-] $\beta_0$: 0.046
                \item[-] $v_{\textrm{max}-\infty}$: $1.93\times10^7$ m/s
                \item[-] $\beta_{\textrm{max}-\infty}$: 0.064
                \item[-] $a_0$: 237 ``g''
            \end{itemize}
        \item[$-$]1 kg: CubeSat-class system with 27 m reflector.
            \begin{itemize}
                \item[-] $t_0$: $3.32\times10^4$ s
                \item[-] $L_0$: $1.27\times10^{11}$ m
                \item[-] $v_0$: $7.67\times10^6$ m/s
                \item[-] $\beta_0$: 0.026
                \item[-] $v_{\textrm{max}-\infty}$: $1.08\times10^7$ m/s
                \item[-] $\beta_{\textrm{max}-\infty}$: 0.036
                \item[-] $a_0$: 23.7 ``g''
            \end{itemize}
        \item[$-$]10 kg: Significant imaging capability with 85 m reflector.
            \begin{itemize}
                \item[-] $t_0$: $1.86\times10^5$ s
                \item[-] $L_0$: $4.01\times10^{11}$ m
                \item[-] $v_0$: $4.31\times10^6$ m/s
                \item[-] $\beta_0$: 0.014
                \item[-] $v_{\textrm{max}-\infty}$: $6.10\times10^6$ m/s
                \item[-] $\beta_{\textrm{max}-\infty}$: 0.02
                \item[-] $a_0$: 2.37 ``g''
            \end{itemize}
        \item[$-$]100 kg: Significant robotic mission with multi-mission capability and 270 m reflector.
            \begin{itemize}
                \item[-] $t_0$: $1.06\times10^5$ s
                \item[-] $L_0$: $1.27\times10^{12}$ m
                \item[-] $v_0$: $2.43\times10^6$ m/s
                \item[-] $\beta_0$: 0.0081
                \item[-] $v_{\textrm{max}-\infty}$: $3.46\times10^6$ m/s
                \item[-] $\beta_{\textrm{max}-\infty}$: 0.011
                \item[-] $a_0$: 0.237 ``g''
            \end{itemize}
        \item[$-$]1,000 kg: Smallest sized human ``shuttle craft'' system with 850 m reflector.
            \begin{itemize}
                \item[-] $t_0$: $5.88\times10^6$ s
                \item[-] $L_0$: $4.01\times10^{12}$ m
                \item[-] $v_0$: $1.36\times10^6$ m/s
                \item[-] $\beta_0$: 0.0046
                \item[-] $v_{\textrm{max}-\infty}$: $1.93\times10^6$ m/s
                \item[-] $\beta_{\textrm{max}-\infty}$: 0.0064
                \item[-] $a_0$: 0.0237 ``g''
            \end{itemize}
        \item[$-$]10,000 kg: Medium human-capable or cargo craft for interplanetary travel with 2.7 km reflector.
            \begin{itemize}
                \item[-] $t_0$: $3.32\times10^7$ s
                \item[-] $L_0$: $1.27\times10^{13}$ m
                \item[-] $v_0$: $7.67\times10^5$ m/s
                \item[-] $\beta_0$: 0.0026
                \item[-] $v_{\textrm{max}-\infty}$: $1.08\times10^6$ m/s
                \item[-] $\beta_{\textrm{max}-\infty}$: 0.0036
                \item[-] $a_0$: $2.37\times10^{-3}$ ``g''
            \end{itemize}
        \item[$-$]100,000 kg: Large human-capable or cargo craft for interplanetary travel with 8.5 km reflector.
            \begin{itemize}
                \item[-] $t_0$: $1.86\times10^8$ s
                \item[-] $L_0$: $4.01\times10^{13}$ m
                \item[-] $v_0$: $4.31\times10^5$ m/s
                \item[-] $\beta_0$: 0.0014
                \item[-] $v_{\textrm{max}-\infty}$: $6.10\times10^5$ m/s
                \item[-] $\beta_{\textrm{max}-\infty}$: 0.0020
                \item[-] $a_0$: $2.37\times10^{-4}$ ``g''
            \end{itemize}
    \end{itemize}
    
\section{Roadmap}

    Should we begin on this path, and if so how should we begin?  Like any long journey it is easy get discouraged and not take the first steps. There are thousands of reasons not to begin. It is too hard, we are not technologically ready, we will not live to see the final journey to the stars... Most of these could be said about any profound endeavor. One difference on this journey is we have a very large scale of masses that are relevant to propel to extremely high speed rather than trying to propel a human, and while one of the long term goals is to send a probe to a nearby star and return data, this is not the only objective. Part of the starting efforts will be to scope a more complete roadmap from desktop to orbital with an emphasis on understanding the technology readiness level (TRL) of each element and what is required to increase it for future missions. Given the large range between our current chemical propellant propulsion and our goals of relativistic speeds and the range of useful masses from sub-gram to large systems, we have an enormous parameter space to work in. All of these are along the path, particularly since this system is modular, scalable, and on a very rapid development path and thus lends itself to a roadmap. With laser efficiencies near 50\% the rise in efficiency will not be one of the enabling elements along the roadmap, but free space phase control over large distances during the acceleration phase will be. This will require understanding the optics, phase noise and systematic effects of the combined on-board metrology and off-board phase servo feedback. Reflector stability during acceleration will also be on the critical path as will increasing the TRL of the amplifiers for space use. For convenience we break the roadmap into several steps. One of the critical development items for space deployment is greatly lowering the mass of the radiators. While this sounds like a decidedly low tech item to work on, it turns out to be one of the critical mass drivers for space deployment. Current radiators have a mass to radiated power of 25 kg/kW, for radiated temperatures near 300 K. This is an area where some new ideas are needed. Our current Yb fiber baseline laser amplifier mass to power is 5 kg/kW (with a very realistic, and in place, 5 year roadmap to 1 kg/kW) and current space photovoltaics of less than 7 kg/kW with very real possibilities to greatly reduce this. The roadmap is precisely that: a roadmap. It is a map to the future where radical transformations will be enabled, but one that is methodical and one that feeds back onto itself as we proceed. This is not a trivial project but it is one where the rewards are truly profound in their consequences for all humanity.\\
    
\subsection{Technology Maturation}

    \begin{itemize}
        \item[$-$]\textbf{Laser and Phased Array}
        \begin{itemize}
            \item[-]Increase TRL of laser amplifiers to at least TRL 6
            \item[-]Test of low mass, thin film optics as an option
            \item[-]Reduce SBS effect to lower bandwidth and increase coherence time/length
            \item[-]Optimize multiple lower power amplifiers vs fewer higher power units --- SBS/coherence trades
            \item[-]Maturation and miniaturization of phase control elements for phased arrays
            \item[-]Phase tapping and feedback on structure
            \item[-]Structural metrology designs
            \item[-]Study of optimized Kalman filters as part of phase control and servo targeting loop
            \item[-]Study and test near field phase feedback from small free-flyer elements
            \item[-]Study beam profiling and methods to smooth beam on reflector
            \item[-]Beam randomization techniques to flatten beam
        \end{itemize}
        \item[$-$]\textbf{Reflector}
        \begin{itemize}
            \item[-]Study multilayer dielectric coating to minimize loss and maximize reflectivity --- trade study
            \item[-]Study materials designs for minimal mass reflectors --- plastics vs glasses
            \item[-]Shape designs for reflector stability --- shaping
            \item[-]Study designs with varying thicknesses and dielectric layers
            \item[-]Study designs with low laser line absorption and high thermal IR absorption (emission)
            \item[-]Study broader band reflectors to deal with relativistic wavelength shift with speed
            \item[-]Study self-stabilizing designs
            \item[-]Simulations of reflector stability and oscillations during acceleration phase --- shape changes
            \item[-]Study spinning reflector to aid stability and randomization of differential force and heating
            \item[-]Study techniques for reflector to laser active feedback
            \item[-]Study techniques to keep beam on reflector using ``beacons''
        \end{itemize}
        \item[$-$]\textbf{Wafer-Scale Spacecraft}
        \begin{itemize}
            \item[-]Study materials for lowest power and high radiation resistance and compatibility with sensors
            \item[-]Determine power requirements
            \item[-]Study onboard power options --- RTG, beta converter, beamed power
            \item[-]Design narrow bandgap PV for beamed power phase
            \item[-]Design on-wafer laser communications
            \item[-]Design optical and IR imaging sensor
            \item[-]Star tracker and laser lock modes
            \item[-]Study swarm modes including intercommunications
            \item[-]Design watchdog timers and redundant computational and sensor/power topologies
            \item[-]Test in beam line to simulate radiation exposure
            \item[-]Study high speed dust impacts on the wafer and design in fault tolerance
            \item[-]Design onboard or thin film ``pop-up'' optics
            \item[-]Design fiber optic or similar cloaking to mitigate heating during laser exposure
            \item[-]Simulate thermal management both during laser exposure and during cruise phase
            \item[-]Simulate radiation exposure during cruise phase
            \item[-]Simulate imaging of target optics
            \item[-]Study use of WaferSat for planetary and terrestrial probes
        \end{itemize}
        \item[$-$]\textbf{Communications}
        \begin{itemize}
            \item[-]Optimize wafer-only laser communications
            \item[-]Study feasibility of using acceleration reflector as part of laser communications
            \item[-]Study feasibility of using reflector as thin film optics for laser communications and imaging
        \end{itemize}
        \item[$-$]\textbf{System Level}
        \begin{itemize}
            \item[-]Detailed design studies including mass trade-offs and costing vs system size
            \item[-]Develop cost roadmaps identifying critical elements as impediments to deployment vs size
            \item[-]Design, build, and test ground based structures with metrology feedback system
            \item[-]Design and simulate orbital structures of various sizes (fixed vs sub-element free-flyer)
            \item[-]Study orbital trade-offs and project launcher feasibility vs time
            \item[-]Study LEO, GEO, Lagrange points, and lunar options
            \item[-]Simulations of proper orbital trajectories including any Earth blockage effects
            \item[-]Work with space PV designers to optimize efficiency and minimize mass
            \item[-]Develop PV roadmap for mass, efficiency, radiation resistance, and minimizing mass
            \item[-]Develop roadmap to reducing radiator mass by $10\times$ as a goal
            \item[-]Study target selection of possible exoplanet systems
            \item[-]Study solar system targets
            \item[-]Study multi-mode use including space debris, beamed power, SPS, planetary defense, etc.
            \item[-]Begin discussion of geopolitical concerns such a system may evoke
        \end{itemize}
    \end{itemize}
    
\subsection{Operational Maturation and Steps}
    \noindent
    \begin{itemize}[leftmargin=0mm]
        \item[] \textbf{Step I} --- Ground based -- Small phased array ($\sim1$ m), beam targeting and stability tests -- 10 kW.
        \item[] \textbf{Step II} --- Ground based -- Target levitation and lab-scale beam line acceleration tests -- 10 kW.
        \item[] \textbf{Step III} --- Ground based -- Beam formation at large array spacing (10 m - 10 km) with sparse array.
        \item[] \textbf{Step IV} --- Ground based -- Scale to 100 kW with arrays in the 1-3 m size -- Possible suborbital tests.
        \item[] \textbf{Step V} --- Ground based -- Scale to 1 MW with 10-100 m optics. Explore 1 km ground option.
        \item[] \textbf{Step VI} --- Orbital testing with small 1-3 class arrays and 10-100 kW power -- ISS possibility.
        \item[] \textbf{Step VII} --- Orbital array assembly tests in 10 m class array.
        \item[] \textbf{Step VIII} --- Orbital assembly with sparse array at 100 m level.
        \item[] \textbf{Step IX} --- Orbital filled 100 m array.
        \item[] \textbf{Step X} --- Orbital sparse 1 km array.
        \item[] \textbf{Step XI} --- Orbital filled 1 km array.
        \item[] \textbf{Step XII} --- Orbital sparse 10 km array.
        \item[] \textbf{Step XIII} --- Orbital filled 10 km array.
    \end{itemize}
    
\section{Other Benefits}

    As we outline in our papers, the same basic system can be used for many purposes including both stand-on and stand-off planetary defense from virtually all threats with rapid response, orbital debris mitigation, orbital boosting from LEO to GEO for example, future ground to LEO laser assisted launchers, standoff composition analysis of distant objects (asteroids for example) through molecular line absorption, active illumination (LIDAR) of asteroids and other solar system bodies, beamed power to distant spacecraft, among others. The same system can also be used for beaming power down to the Earth via micro or mm waves for selected applications. \textbf{This technology will enable transformative options that are not possible now and allows us to go far beyond our existing chemical propulsion systems with profound implications for the future} \cite{Lubin2013}\cite{Lubin2014}\cite{Pelton2015}\cite{Riley2014}\cite{Zhang2015a}.\\

\section{\label{sec:conclusion}Conclusions}

    It is now feasible to seriously discuss, plan, and execute a program to use directed energy to propel spacecraft to relativistic speeds allowing the possibility of realistic interstellar flights for the first time.  There has been a dramatic change in the practical possibilities of using directed energy brought about by a revolution in photonics that is on an exponential rise. While photonic propulsion has been spoken about for a very long time, it has largely been confined to dreaming and to the realm of science fiction in terms of practical application for spaceflight and particularly relativistic spaceflight. This has now changed to the point where a serious program can begin to enable a future no longer constrained by low speed chemical and ion propulsion. We outline a roadmap to that future with a logical series of steps and milestones. One that is modular and scalable to any sized system. The same system has many other applications and spinoffs and this will greatly aid in the cost amortization. While the roadmap to the future of directed energy propulsion is extremely challenging it is nonetheless a feasible roadmap to begin. The difficulties are many but the rewards and long term consequences are not only profound but will be transformative in their scope.

\section{Acknowledgments}

We gratefully acknowledge funding from NASA NIAC 2015 NNX15AL91G and the NASA California Space Grant NASA NNX10AT93H as well as a generous gift from the Emmett and Gladys W. fund in support of this research.

\appendix

\section{Laser Sail --- Non-Relativistic Solution}

    Assumes the speed is low enough that the Doppler shift, relativistic mass increase, and time dilation effects are small. This is reasonably valid for $\beta<0.2$. Assume a sail with thickness $h$, size $D$, density $\rho$, and bare (no sail) payload mass $m_0$. Assume an array of size $d$ and assume perfect reflectivity $\epsilon_r=1$. The distance from the laser to the sail is $L$ and the spot size at distance $L$ is $D_s$.

\subsection{Circle vs. Square Array}

    For a square, the full diffraction angle to first minimum is $\theta=2\lambda/d$. If the DE array is circular, of diameter $d_c$, then the diffraction angle to the first minimum is determined by the first zero in the Bessel function of the first kind, $J_1$. The full diffraction angle $\theta_c$ is then given by:
    \begin{equation}
        \theta_c=\frac{2\alpha\lambda}{d}=\frac{2\lambda}{d_e},
    \end{equation}
    where $\alpha\sim1.22$ and $d_e$ is the effective size (or diameter) of the area, with $d_e=d/\alpha$. In everything for the square aperture we then simply replace $d$ by the effective size $d_e$ to obtain expressions for circular arrays. We therefore have $\alpha=1$ for a square array and $\alpha=1.22$ for a circular array. Note that for a circular DE array it makes more sense to then have a circular sail to slightly lower the sail mass for the same size (diameter) $D$ compared to a square sail. The sail mass is then lower by $\pi/4$. The two effects of a smaller area for a circular array of the same size will tend to be nearly cancelled by the larger sized (circular) sail to match. The difference is speed is only about 3.8\% between the two cases of square array on square sail (slightly faster) compared to circular array on circular sail.

\subsection{General Case of Square or Circular Array and Square or Circular Sail}

\subsubsection{Definitions}
    \begin{equation}
        \textrm{Square sail:}\hspace{1mm}m_\textrm{sail-sq}=D^2h\rho
    \end{equation}
    \begin{equation}
        \textrm{Circular sail:}\hspace{1mm}m_\textrm{sail-cir}=\frac{\pi}{4}D^2h\rho=\frac{\pi}{4}m_\textrm{sail-sq}
    \end{equation}
    We can write the mass of the sail in general as
    \begin{equation}
        m_\textrm{sail}=\xi D^2h\rho,
    \end{equation}
    where $\xi=1$ for a square sail and $\xi=\pi/4$ for a circular sail.
    \begin{equation}
        \textrm{Spot size:}\hspace{1mm}D_s=L\theta=\frac{2L\lambda\alpha}{d}
    \end{equation}
    \begin{align}
    \begin{split}
        \textrm{Force (thrust on sail):}\hspace{1mm}F=&\frac{2P_0}{c}\hspace{1mm}\textrm{for}\hspace{1mm}D_s<D\\
        F=&\frac{2P_0}{c}\bigg(\frac{D}{D_s}\bigg)^2=\frac{P_0}{c}\frac{d^2D^2}{L^2\lambda^2\alpha^2}\hspace{1mm}\textrm{for}\hspace{1mm}D_s>D
    \end{split}
    \end{align}
    
\subsubsection{Case 1), wherein $D_s<D$}

    Let $L_0$ be the distance where the spot size is equal to the sail size. We have:
    \begin{equation}
        L_0\theta=D=\frac{2\lambda\alpha L_0}{d}\rightarrow L_0=\frac{dD}{2\lambda\alpha}
    \end{equation}
    \begin{equation}
        F=\frac{2P_0}{c}\hspace{1mm}\textrm{is constant while $D_s<D$.}
    \end{equation}
    The energy to distance $L_0$ is given by:
    \begin{equation}
        \textrm{KE}=FL_0=\frac{1}{2}mv^2
    \end{equation}
    \begin{equation}
        v=\sqrt{\frac{2FL_0}{m}}=\sqrt{2aL_0},\hspace{2mm}a=\frac{F}{m},\hspace{2mm}m=m_\textrm{sail}+m_0,\hspace{2mm}m_\textrm{sail}=\xi D^2h\rho
    \end{equation}
    \begin{equation}
        \textrm{KE}=FL_0=\frac{2P_0}{c}\frac{dD}{2\lambda}=\frac{P_0 dD}{c\lambda}
    \end{equation}
    \begin{equation}
        v(L)=\sqrt{\frac{4P_0L}{c(\xi D^2h\rho+m_0)}},\hspace{2mm}\textrm{$v=v_0$ when $L=L_0$}
    \end{equation}
    \begin{equation}
        v_0=\bigg(\frac{4P_0L_0}{c(\xi D^2h\rho+m_0)}\bigg)^{1/2}=\bigg(\frac{2P_0dD}{c\lambda\alpha(\xi D^2h\rho+m_0)}\bigg)^{1/2}
    \end{equation}
    \begin{equation}
        a=\frac{F}{m}=\frac{2P_0}{c(\xi D^2h\rho+m_0)}=\textrm{constant while $D_s<D$}.
    \end{equation}
    \begin{equation}
        t(L)=\frac{v(L)}{a}=\bigg(\frac{Lc(\xi D^2h\rho+m_0)}{P_0}\bigg)^{1/2}.
    \end{equation}
    The time to $v=v_0$ is given by:
    \begin{equation}
        t_0=\frac{v_0}{a}=\bigg(\frac{L_0c(\xi D^2h\rho+m_0)}{P_0}\bigg)^{1/2}=\bigg(\frac{cdD(\xi D^2h\rho+m_0)}{2P_0\lambda\alpha}\bigg)^{1/2}
    \end{equation}
    If $m_0=0$ (no payload mass), then:
    \begin{equation}
        m=m_\textrm{sail}=\xi D^2h\rho
    \end{equation}
    \begin{equation}
        v(L)=\bigg(\frac{4P_0L}{c\xi D^2h\rho}\bigg)^{1/2},\hspace{2mm} v_0=\bigg(\frac{4P_0L_0}{c\xi D^2h\rho}\bigg)^{1/2}=\bigg(\frac{2P_0dD}{c\lambda\alpha\xi D^2h\rho}\bigg)^{1/2}\propto h^{-1/2},P_0^{1/2},d^{1/2},D^{-1/2},\lambda^{-1/2},\rho^{-1/2}
    \end{equation}
    \begin{equation}
        a=\frac{2P_0}{\xi D^2h\rho c}\propto P_0,D^{-2},h^{-1},\rho^{-1}
    \end{equation}
    \begin{equation}
        t(L)=\frac{v(L)}{a}=\bigg(\frac{Lc\xi D^2h\rho}{P_0}\bigg)^{1/2}
    \end{equation}
    \begin{equation}
        t_0=\frac{v_0}{a}=\bigg(\frac{L_0c\xi D^2h\rho}{P_0}\bigg)^{1/2}=\bigg(\frac{cdD\xi D^2h\rho}{2P_0\lambda\alpha}\bigg)^{1/2}=\bigg(\frac{cd\xi D^3h\rho}{2P_0\lambda\alpha}\bigg)^{1/2}\propto P_0^{-1/2},d^{1/2},D^{3/2},h^{1/2},\lambda^{-1/2},\rho^{1/2}
    \end{equation}
    Note that $v_0\propto d^{1/2},D^{-1/2},h^{-1/2}$ implies that a smaller reflector is faster. Why is a smaller reflector faster?
    \begin{equation}
        v_0=\bigg(\frac{2\textrm{KE}}{m}\bigg)^{1/2}=\bigg(\frac{2FL_0}{m}\bigg)^{1/2}
    \end{equation}
    Since $m\propto D^2$, $L_0\propto D$, and $F$ is constant, we have:
    \begin{equation}
        v_0\propto\frac{1}{\sqrt{D}}
    \end{equation}
    Hence, if we want high $v$, we must make $D$ small. Note that a smaller sail is faster IF $m_0=0$. Therefore, we must make the sail as small as possible IF highest speed is the metric and if $m_0=0$.
    
    With continued illumination, beyond when the laser spot exceeds the reflector size, the speed increases by a factor of $\sqrt{2}$:
    \begin{equation}
        v_0\rightarrow v_\infty\equiv v(L=\infty)=\sqrt{2}v_0
    \end{equation}
    
\subsubsection{Case 2), wherein $D_s>D$}

    \begin{equation}
        F=\frac{2P_0}{c}\bigg(\frac{D}{D_s}\bigg)^2=\frac{2P_0}{c}\frac{d^2D^2}{L^2\lambda^2\alpha^2}=\frac{2P_0}{c}\bigg(\frac{L_0}{L}\bigg)^2
    \end{equation}
    Let KE$_1$ be the kinetic energy from $L=0$ to $L_0$ (kinetic energy for $D_s<D$). $L_0$ is then the distance at which $D=D_s$. We have:
    \begin{equation}
        \textrm{KE}_1=FL_0=\frac{2P_0}{c}L_0=\frac{P_0dD}{c\lambda\alpha}
    \end{equation}
    Let KE$_2$ be the kinetic energy from $L=L_0$ to $L=\infty$. We have:
    \begin{align}
    \begin{split}
        \textrm{KE}_2=&\int_{L_0}^\infty F dL=\frac{P_0d^2D^2}{2c\lambda^2\alpha^2}\int_{L_0}^L \frac{dL}{L^2}=\frac{P_0d^2D^2}{2c\lambda^2\alpha^2}\bigg(\frac{1}{L_0}-\frac{1}{L}\bigg)\\
        =&\textrm{KE}_1\frac{dD}{2\lambda\alpha}\bigg(\frac{1}{L_0}-\frac{1}{L}\bigg)=\textrm{KE}_1 L_0\bigg(\frac{1}{L_0}-\frac{1}{L}\bigg)=\textrm{KE}_1\bigg(1-\frac{L_0}{L}\bigg)
    \end{split}
    \end{align}
    Therefore as $L\rightarrow\infty$, $\textrm{KE}_1=\textrm{KE}_2$.
    \begin{equation}
        \textrm{KE}_\textrm{total}=\textrm{KE}_1+\textrm{KE}_2=\textrm{KE}_1\bigg(2-\frac{L_0}{L}\bigg)=\frac{2P_0}{c}L_0\bigg(2-\frac{L_0}{L}\bigg)
    \end{equation}
    As $L\rightarrow\infty$, we have:
    \begin{equation}
        \textrm{KE}_\textrm{total}(L\rightarrow\infty)=2\textrm{KE}_1=\frac{2P_0dD}{c\lambda\alpha}=\frac{4P_0}{c}\frac{dD}{2\lambda\alpha}=\frac{4P_0}{c}L_0=\frac{2P_0}{c}2L_0=2L_0F=\frac{1}{2}mv^2
    \end{equation}
    \begin{equation}
        v(L)=\bigg[\frac{2P_0dD}{mc\lambda\alpha}\bigg(2-\frac{L_0}{L}\bigg)\bigg]^{1/2}=\bigg[\frac{4P_0}{mc}L_0\bigg(2-\frac{L_0}{L}\bigg)\bigg]^{1/2}
    \end{equation}
    \begin{equation}
        v_0=v(L=L_0)=\bigg(\frac{2P_0dD}{mc\lambda\alpha}\bigg)^{1/2}=\bigg(\frac{4P_0}{mc}L_0\bigg)^{1/2}=(2a_0L_0)^{1/2},\hspace{2mm}\textrm{where}\hspace{2mm}a_0=\frac{F}{m}=\frac{2P_0}{mc}
    \end{equation}
    \begin{equation}
        v(L)=\bigg[\frac{2P_0dD}{mc\lambda\alpha}\bigg(2-\frac{L_0}{L}\bigg)\bigg]^{1/2}=v_0\bigg(2-\frac{L_0}{L}\bigg)^{1/2}
    \end{equation}
    As before,
    \begin{equation}
        v_\infty=v(L=\infty)=\sqrt{2}v_0
    \end{equation}
    \begin{equation}
        L_0=\frac{dD}{2\lambda\alpha}
    \end{equation}
    \begin{equation}
        a(L)=\frac{F(L)}{m}=\frac{P_0}{2mc}\bigg(\frac{dD}{L\lambda\alpha}\bigg)^2=\frac{2P_0}{mc}\bigg(\frac{L_0}{L}\bigg)^2=a_0\bigg(\frac{L_0}{L}\bigg)^2
    \end{equation}
    \begin{equation}
        v(L)=\bigg(\frac{2P_0dD}{c\lambda\alpha(m_\textrm{sail}+m_0)}\bigg)^{1/2}\bigg(2-\frac{L_0}{L}\bigg)^{1/2}=\bigg(\frac{2P_0dD}{c\lambda\alpha(\xi D^2h\rho+m_0)}\bigg)^{1/2}\bigg(2-\frac{L_0}{L}\bigg)^{1/2},\hspace{2mm}\textrm{for}\hspace{2mm}L>L_0
    \end{equation}
    
\subsection{Maximizing Speed of Laser Driven System}

    Let $v=v(L)$. For $L=L_0$, $v(L_0)\equiv v_0=v_0(D)$. As before,
    \begin{equation}
        L_0=\frac{dD}{2\lambda\alpha}
    \end{equation}
    where $D$ is the sail size and $L=L_0$ when the spot size is equal to $D$.
    \begin{equation}
        v_0(D)=\bigg(\frac{2P_0d}{c\lambda\alpha}\bigg)^{1/2}\bigg(\frac{D}{\xi D^2h\rho+m_0}\bigg)^{1/2}=\bigg(\frac{2P_0d}{c\lambda\alpha}\bigg)^{1/2}\bigg(\frac{D}{m_\textrm{sail}+m_0}\bigg)^{1/2}
    \end{equation}
    where $m_\textrm{sail}=\xi D^2h\rho$ and $m_0$ is the payload mass without the sail. We have the following limiting cases for $v(D)$:
    \begin{equation}
        v_0=v(D=0)=0,\hspace{2mm}\lim_{D\rightarrow\infty}v_0(D)=0
    \end{equation}
    Hence, we can find the $D$ which gives a maximum $v_0$. To do so, we set $dv/dD=0$:
    \begin{align}
    \begin{split}
        \frac{dv}{dD}=&\frac{1}{2}\bigg(\frac{2P_0d}{c\lambda\alpha}\bigg)^{1/2}\bigg(\frac{D}{\xi D^2h\rho+m_0}\bigg)^{1/2}\bigg(\frac{\xi D^2h\rho+m_0-D(2\xi Dh\rho)}{(\xi D^2h\rho+m_0)^2}\bigg)\\
        =&\frac{1}{2}\bigg(\frac{2P_0d}{c\lambda\alpha}\bigg)^{1/2}\bigg(\frac{D}{m_\textrm{sail}+m_0}\bigg)^{1/2}\bigg(\frac{m_0-m_\textrm{sail}}{(m_\textrm{sail}+m_0)^2}\bigg)
    \end{split}
    \end{align}
    Therefore, maximum $v$ ($dv/dD=0$) occurs when $m_\textrm{sail}=m_0$, or when the mass of the sail is equal to the mass of the payload.
    \begin{equation}
        v_{0,\textrm{max}}=\bigg(\frac{2P_0d}{c\lambda\alpha}\bigg)^{1/2}\bigg(\frac{D}{2m_0}\bigg)^{1/2}=\bigg(\frac{P_0d}{c\lambda\alpha}\bigg)^{1/2}(\xi h\rho m_0)^{-1/4}
    \end{equation}
    Since $D=(m_0/\xi h\rho)^{1/2}$ for $m_\textrm{sail}=m_0$, we have:
    \begin{equation}
        v_{0,\textrm{max}}=\bigg(\frac{P_0d}{c\lambda\alpha}\bigg)^{1/2}(\xi h\rho D)^{-1/2}=\bigg(\frac{P_0}{c\lambda\xi h\rho}\bigg)^{1/2}\bigg(\frac{d}{\alpha D}\bigg)^{1/2}=c\bigg(\frac{P_0}{P_1}\bigg)^{1/2}\bigg(\frac{d}{\alpha\xi D}\bigg)^{1/2}
    \end{equation}
    where $P_1\equiv c^3\lambda h\rho\approx2.7\times10^{16}$(W)$\times\lambda$($\mu$m)$\times h$($\mu$m)$\times\rho$(g/cc).
    \begin{equation}
        \lim_{L\rightarrow\infty}v(L)\equiv v_\infty=\sqrt{2}v_0
    \end{equation}
    \begin{equation}
        \frac{dv}{dD}(L=\infty)=0=\frac{dv}{dD}(L=L_0)\sqrt{2}
    \end{equation}
    Therefore, we have the same condition for maximum $v$. When $m_\textrm{sail}=m_0$,
    \begin{equation}
        \rightarrow\lim_{L\rightarrow\infty}v(L)\equiv v_\infty=\sqrt{2}v_0=\bigg(\frac{2P_0d}{c\lambda\alpha}\bigg)^{1/2}(\xi h\rho m_0)^{-1/4}
    \end{equation}
    When $m_\textrm{sail}=m_0$, $m=m_\textrm{sail}+m_0=2m_\textrm{sail}=2m_0$. Therefore, at $v=v_{0,\textrm{max}}$ we have:
    \begin{equation}
        a=\frac{F}{m}=\frac{F}{2m_0}=\frac{P_0}{m_0c}=\frac{P_0}{c\xi D^2h\rho}
    \end{equation}
    \begin{equation}
        t_0=\frac{v_{0,\textrm{max}}}{a}=\bigg(\frac{cd\xi D^3h\rho}{P_0\lambda\alpha}\bigg)^{1/2}=\bigg(\frac{cdD\xi D^2h\rho}{P_0\lambda\alpha}\bigg)^{1/2}=\bigg(\frac{cdDm_0}{P_0\lambda\alpha}\bigg)^{1/2}=\bigg(\frac{2cL_0m_0}{P_0}\bigg)^{1/2}=\bigg(2\frac{m_0c^2}{P_0}\frac{L_0}{c}\bigg)^{1/2}
    \end{equation}
    \begin{equation}
        t_0=\bigg(\frac{cdDm_0}{P_0\lambda\alpha}\bigg)^{1/2}=\bigg(\frac{cd}{P_0\lambda\alpha}\bigg)^{1/2}\bigg(\frac{{m_0}^3}{\xi h\rho}\bigg)^{1/4}
    \end{equation}
    \begin{equation}
        D=\bigg(\frac{m_0}{\xi h\rho}\bigg)^{1/2}
    \end{equation}
    \begin{equation}
        L_0=\frac{dD}{2\lambda\alpha}
    \end{equation}
    where, as before, $t_0$ is the time when the spot size $D_s$ is equal to the sail size $D$.
    
\subsection{Effect of Imperfect Reflector}

    If the reflector has non-unity reflection coeffient $\epsilon_r$, then $P_0$ above is replaced by $P_0(1+\epsilon_r)/2$.

\bibliographystyle{unsrt}  
\bibliography{references} 
\nocite*






\end{document}